\documentclass[twocolumn]{aastex631}
\usepackage{graphicx}
\usepackage{natbib}
\usepackage{xspace}
\usepackage{multirow}
\usepackage{booktabs}

\def\msun{$M_{\odot}$\xspace}
\def\rsun{$R_{\odot}$\xspace}
\def\stella{\texttt{STELLA}\xspace}
\def\snec{\texttt{SNEC}\xspace}
\def\mesa{\texttt{MESA}\xspace}
\def\nifs{$^{56}$Ni\xspace}

\def\bethe{$\times10^{51}\mathrm{ergs}$\xspace}
\def\tablenotemark#1{{\normalfont\textsuperscript{\normalsize\rm #1}}}

\shorttitle{Short Title}
\shortauthors{Short Authors}

\graphicspath{{./}{figures2/}}

\begin{document}

\title{The effects of Thomson scattering and chemical mixing on early-time light curves of double peaked type IIb supernovae}

\email{rogersh0125@snu.ac.kr, scyoon@snu.ac.kr}

\author[0000-0001-7488-4337]{Seong Hyun Park}
\affiliation{Department of Physics and Astronomy, Seoul National University, Seoul 08826, Korea}

\author{Sung-Chul Yoon}
\affiliation{Department of Physics and Astronomy, Seoul National University, Seoul 08826, Korea}
\affiliation{SNU Astronomy Research Center, Seoul National University, Seoul 08826, Korea}

\author{Sergei Blinnikov}
\affiliation{KKTEP, NRC Kurchatov Institute, Moscow 123182, Russia}
\affiliation{Keldysh Institute of Applied Mathematics, Miusskaya pl., 4, 125047, Moscow, Russia}

\begin{abstract}
Previous numerical simulations of double-peaked SNe IIb light curves have demonstrated
that the radius and mass of the hydrogen-rich envelope of the progenitor star
can significantly influence the brightness and timescale of the early-time
light curve around the first peak.  In this study, we investigate how Thomson
scattering and chemical mixing in the SN ejecta affect the optical light curves
during the early stages of the SNe IIb using radiation hydrodynamics
simulations.  By comparing the results from two different numerical codes
(i.e., \stella{} and \snec{}), we find that the optical brightness of the first
peak can be reduced by more than a factor of 3 due to the effect of Thomson
scattering that causes the thermalization depth to be located below the
Rosseland-mean photosphere, compared to the corresponding case where this
effect is ignored. We also observe a short-lived plateau-like
feature lasting for a few days in the early-time optical light curves of our
models, in contrast to typical observed SNe IIb that show a quasi-linear
decrease in optical magnitudes after the first peak. A significant degree of
chemical mixing between the hydrogen-rich envelope and the helium core in SN
ejecta is required to reconcile this discrepancy between the model prediction
and observation. Meanwhile, to properly reproduce the first peak, a significant mixing of \nifs{} into the hydrogen-rich outermost layers should be restricted.
Our findings indicate that inferring the SN IIb progenitor structure from a
simplified approach that ignores these two factors may
introduce substantial uncertainty.  
\end{abstract}

\section{Introduction} \label{sec:intro}
    
    Type IIb supernovae (SNe IIb) are a class of core-collapse SNe that undergo a transition from Type II to Type Ib over time. They are initially classified as Type II due to the presence of hydrogen lines in their spectra during the early stages and these hydrogen lines become weaker or disappear at later times. This implies that their progenitors retain only a small amount of hydrogen-rich (H-rich) envelope ($M_{\mathrm{env}}\lesssim0.5$\msun). Stripping of the H-rich envelope from a massive star can occur in various ways, including stellar winds, episodic mass eruptions, and binary interactions \citep[e.g.,][]{Podsiadlowski92, Podsiadlowski93, Stancliffe09, Yoon10, Claeys11, Smith14, Meynet15, Yoon17, Eldridge18}. 
    
    \begin{figure}
        \includegraphics[width=\linewidth]{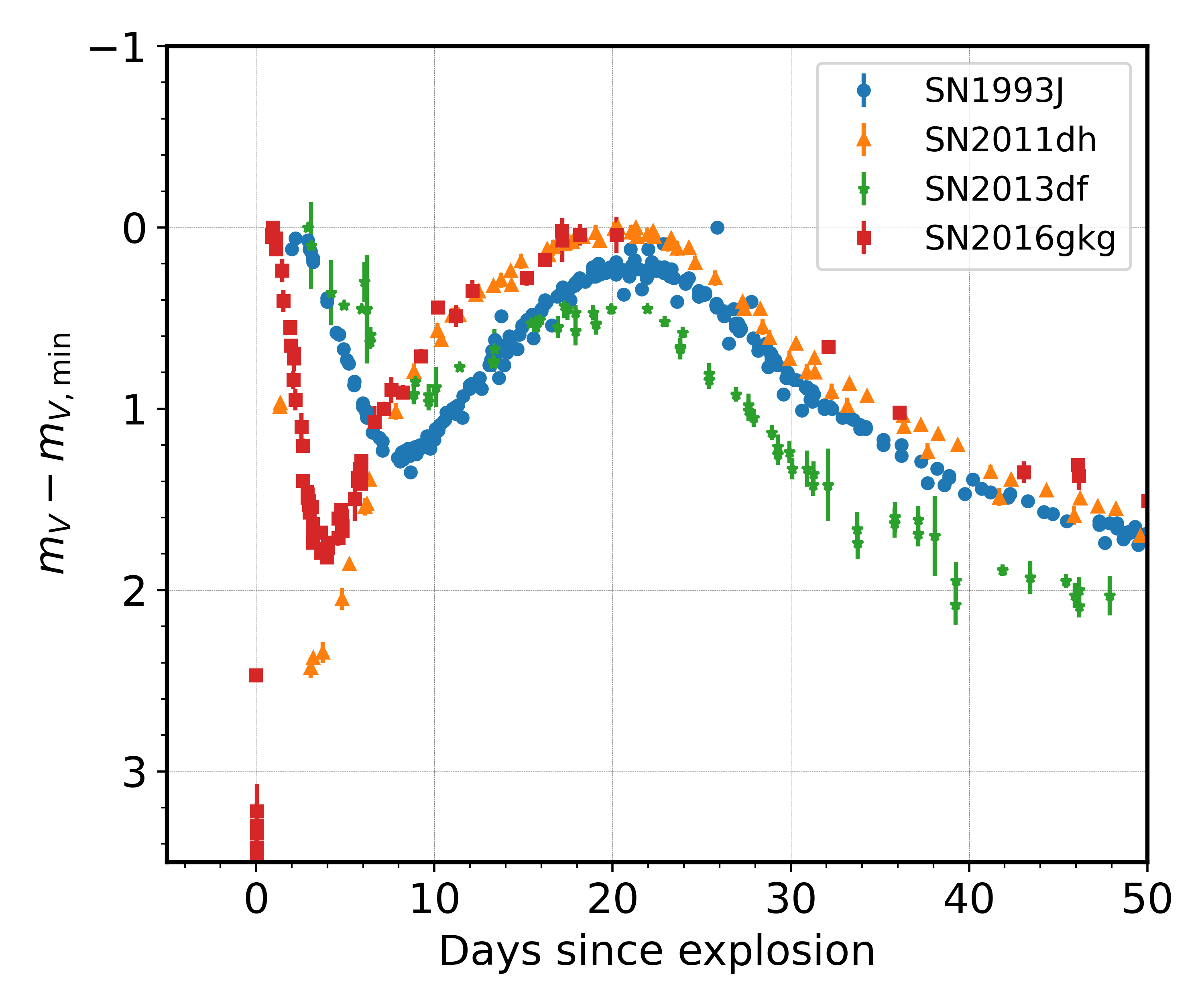}
        \caption{$V$-band light curves of selected samples of observed SNe IIb (SN 1993J; SN 2011dh; SN 2013df; SN 2016gkg). The light curves are normalized with respect to the inferred explosion dates and the peak magnitude of each SN. The photometry data are taken from \citet{Guillochon17}.}
        \label{fig:observed}
    \end{figure}    
    
    Some SNe IIb show two distinct peaks in their optical light curves. The $V$-band light curves of SN 1993J, SN 2011dh, SN 2013df, and SN 2016gkg are presented in Figure \ref{fig:observed} as examples of such SNe IIb \citep{Richmond94,Arcavi11,MoralesGaroffolo14,Tartaglia17,Bersten18,Nakaoka19}. The second peak of SNe IIb, also known as the main peak, is powered by radioactive decay of \nifs{} produced by nucleosynthesis during the explosion. The process of the primary peak formation through radioactive \nifs{} heating shares similarities with the case of SNe Ib/Ic. On the other hand, the first peak is produced by thermal radiation of the hot, highly-ionized, H-rich and extended envelope that has been heated by shock waves during the explosion. Properties of the first peak including the peak luminosity and the rise time are greatly affected by the properties of the pre-SN progenitor such as the mass of the H-rich envelope and the radius. In previous studies, various numerical and analytical methods were employed for either finding relations between the progenitor properties and the first peak properties or applying these relations to infer the progenitor properties using observational data \citep[e.g.,][]{Bersten12,Bersten18,NakarPiro14,NagyVinko16,Piro17,Dessart18,Fremling19,Armstrong21,Pellegrino23}. 

    Previous studies on SNe IIb employed various numerical codes or physical assumptions for modelling the hydrodynamical evolution of the ejecta after the explosion and the resulting optical light curves.  \citet{Woosley94} modelled the SN ejecta using the one-dimensional (1D) hydrodynamics code \texttt{KEPLER} and performed a multi-group radiative transport calculations using the \texttt{EDDINGTON} code in order to compute optical light curves. In \citet{Dessart18}, the radiation-hydrodynamics (RHD) code \texttt{v1d} and the non-LTE radiative transport code \texttt{CMFGEN} were respectively used for modelling the evolution of the ejecta and obtaining the optical light curves. Several studies adopted the \stella code \citep{Blinnikov98,Blinnikov06,Blinnikov11}, a 1D RHD code that implicitly calculates the radiative transfer over multiple frequency bands, in order to generate spectral energy distributions (SEDs) and obtain optical light curve models \citep[cf.][]{Tsvetkov09,Tsvetkov12,Nakaoka19,Balakina21}.  Some studies assume that the SEDs from a SN are black-body spectra of the effective temperature at the photosphere (usually defined by the Rosseland-mean opacity) or a color temperature obtained from an analytical approximation \citep{Bersten12,Bersten18,Piro17,Eldridge18}. In this case, the choice of characteristic temperatures (e.g. $T_\mathrm{eff}$, effective temperature; $T_\mathrm{color}$, color temperature)  may significantly affect the resulting optical light curves. The H-rich envelope heated by the shock during the explosion is highly ionized at the epoch of the first optical peak, making the contribution of Thomson electron scattering to the SED formation significant. In this regard, the impact of Thomson scattering and light dilution on radiative transfer \citep{Mihalas78,Mihalas84,Baschek91,Ensman92}, which may not be adequately captured in the black-body approximation, would be important to properly understand the first peak of SNe IIb in the optical.

    In addition to the effects of Thomson scattering, another factor that may significantly affect the optical light curves of SNe IIb is the chemical mixing within the ejecta during the explosion, which can be induced by the Rayleigh-Taylor instability \citep[RTI; cf.][]{Shigeyama90,Hammer10,Paxton18,Yoon19,Utrobin21}. In a yellow or red supergiant SN IIb progenitor, the SN shock must be decelerated when its front moves through the boundary between the helium (He) core and the H-rich envelope, and the resulting reverse shock can make the ejecta susceptible to the RTI. Although the chemical mixing induced by the RTI might not be as strong as in the case of SN 1987A or SN IIP given the relatively less massive H-rich envelope \citep{Woosley94,NOMOTO1995173,Iwamoto97}, the consequences of chemical mixing for SN IIb optical light curves still need further investigation as we discuss below in Section \ref{sec:chem}. In particular, we argue that a significant chemical mixing across the He core and the H-rich envelope and rather a limited mixing of \nifs{} in the inner ejecta are required to explain the double-peaked SNe IIb light curves.

    In this study, we investigate how the early-time light curves of double peaked SNe IIb in the optical are affected by these two factors: the effects of Thomson scattering (Section \ref{sec:snec}) and chemical mixing (Section \ref{sec:chem}). In Section \ref{sec:model}, we explain the SN IIb progenitor models and the numerical codes used in this work. In Section \ref{sec:snec}, we compare optical light curves generated from \snec and \stella codes and demonstrate how the Thomson scattering in the radiative transfer has crucial impacts on the optical first peaks. In Section \ref{sec:chem}, the effects of chemical mixing on the internal opacity structure and the shape of optical light curve models are discussed. In Section \ref{sec:obs}, we compare our light curve models to the light curves of SN 1993J, a typical example of double-peaked SNe IIb. We then discuss the implications of using the early-time light curves to infer the characteristics of the progenitor structure. In Section \ref{sec:con}, we present a brief summary and conclude this paper. 

        \begin{deluxetable*}{lcccccccccc}
        \centering
        \tablecaption{Pre-SN progenitor properties of the models and observed SNe \label{tab:presnparams}}
        \tablehead{
        \colhead{Model/SNe Name} & \colhead{$M_{\mathrm{f}}$} & \colhead{$R_{\mathrm{f}}$} & \colhead{$M_{\mathrm{H,env}}$} & \colhead{$T_{\mathrm{eff}}$} & \colhead{$L_{\mathrm{f}}$} & \colhead{$M_{\mathrm{Fe}}$} & \colhead{$m_{\mathrm{H}}$} & \colhead{$m_{\mathrm{He}}$} & \colhead{$X_{\mathrm{surf}}$} & \colhead{References}
        } 
        \startdata
        Tm11p200 & 3.35 & 290.5 & 0.098 & 3.67 & 4.58 & 1.37 & 0.016 & 1.64 & 0.18 & \\
        Sm11p400 & 3.45 & 412.4 & 0.138 & 3.60 & 4.58 & 1.45 & 0.033 & 1.67 & 0.28 & \\
        Sm11p600 & 3.64 & 547.3 & 0.189 & 3.56 & 4.68 & 1.38 & 0.077 & 1.75 & 0.49 & \\
        \hline
        SN 1993J   & 2.8 -- 4.5 & 430 -- 720  & $\sim0.2$    & $3.63\pm0.05$ & $5.1\pm0.3$   & & & & & (1) \\
        SN 2011dh  &  3 -- 4    & $\sim200$   & $\sim0.1$    & 4.90 -- 4.99  & $3.78\pm0.02$ & & & & & (2) \\
        SN 2013df  &  2 -- 3.6  & $545\pm65$  & 0.2 -- 0.4   & $3.63\pm0.01$ & $4.94\pm0.06$ & & & & & (3) \\
        SN 2016gkg & 4.8 -- 5.0 & 180 -- 320  & 0.01 -- 0.02 & $5.10^{+0.17}_{-0.19}$ & $3.86\pm0.05$ & & & & & (4) \\
        \enddata
        \tablecomments{$M_{\mathrm{f}}$: total mass in units of \msun, $R_{\mathrm{f}}$: radius in units of \rsun, $M_{\mathrm{H,env}}$: H-rich envelope mass in units of \msun, $T_{\mathrm{eff}}$: effective temperature in units of K in logarithmic scale, $L_{\mathrm{f}}$: luminosity in units of $L_{\odot}$ in logarithmic scale, $M_{\mathrm{Fe}}$: iron core mass in units of \msun, $m_{\mathrm{H}}$: integrated mass of hydrogen in units of \msun, $m_{\mathrm{He}}$: integrated mass of He in units of \msun, $X_{\mathrm{surf}}$: abundance of hydrogen at the stellar surface}
		\tablerefs{(1) \citet{Aldering94}, \citet{Woosley94}, \citet{Blinnikov98}, \citet{Maund04} (2) \citet{Maund11}, \citet{VanDyk11}, \citet{Bersten12}, \citet{Jerkstrand15}, \citet{Ergon22} (3) \citet{MoralesGaroffolo14}, \citet{VanDyk14}, (4) \citet{Piro17}, \citet{Tartaglia17}, \citet{Bersten18}}
        \end{deluxetable*}

\section{Modelling} \label{sec:model}

    \subsection{Pre-SN progenitor models} \label{sec:modelpresn}
\begin{figure}
            \includegraphics[width=\linewidth]{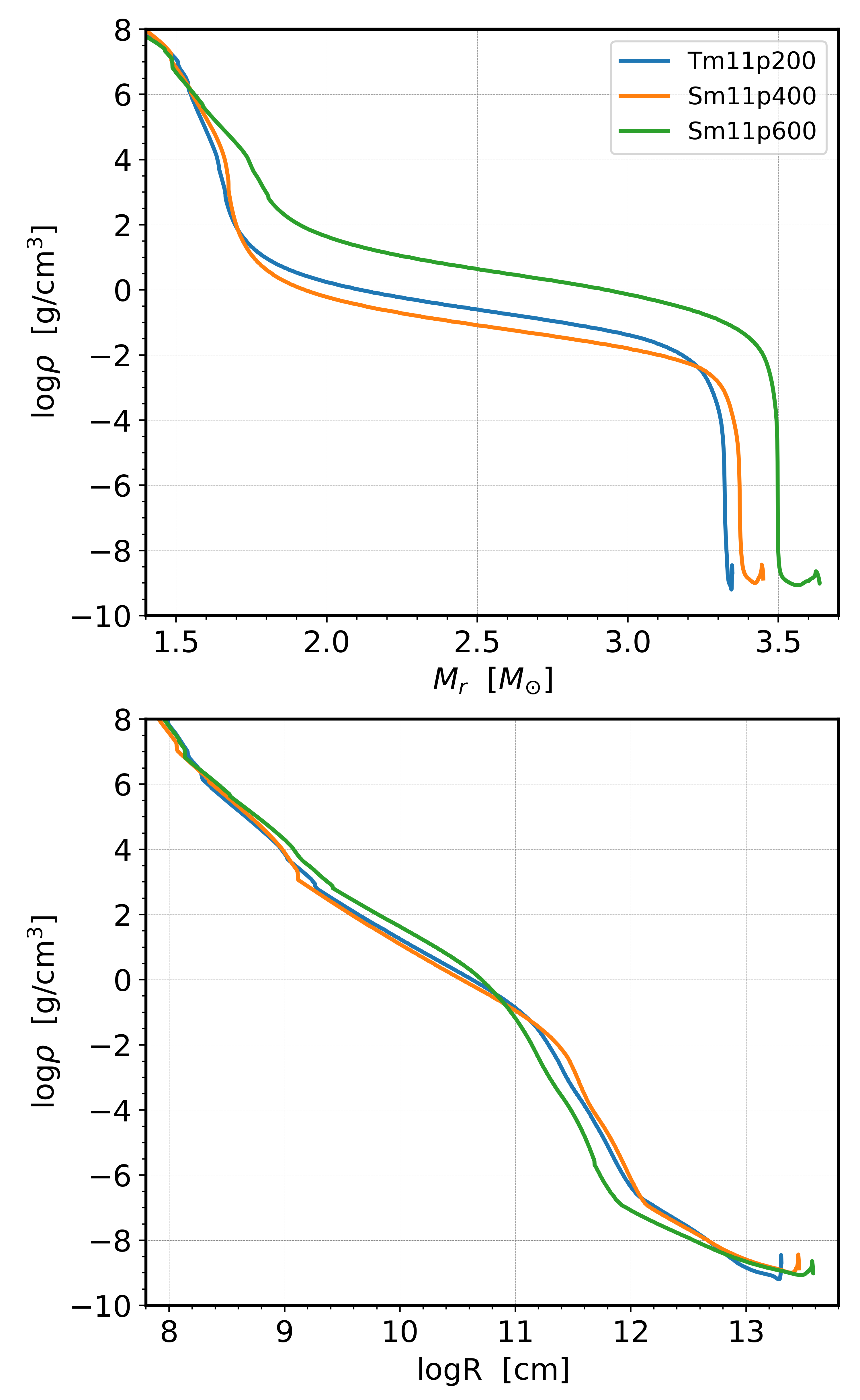}
            \caption{Upper panel: Mass coordinate-density structures of three SNe IIb progenitor models, Tm11p200, Sm11p400, and Sm11p600, with the y-axis in a logarithmic scale. Lower panel: corresponding radius-density structures with both axes in a logarithmic scale.}
            \label{fig:rhostructure}
        \end{figure}

\begin{figure}
            \includegraphics[width=\linewidth]{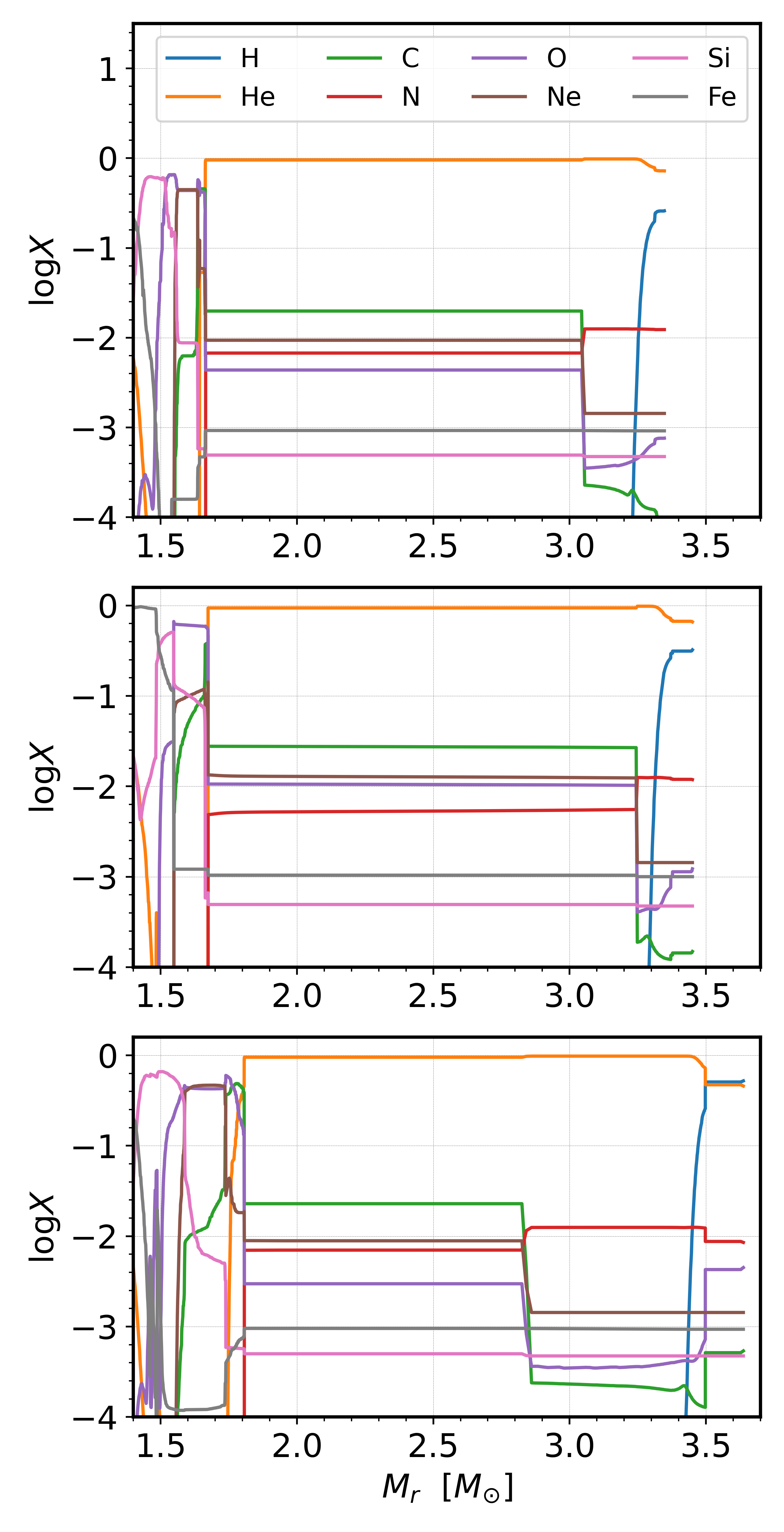}
            \caption{Internal chemical composition of the pre-SN SNe IIb progenitor models Tm11p200 (top), Sm11p400 (middle), and Sm11p600 (bottom).}
            \label{fig:chemstructure}
        \end{figure}

	Our progenitor models are selected from the yellow supergiant (YSG) and red supergiant (RSG) models presented in \citet{Yoon17} who calculated binary star evolution at solar and LMC metallicities in a parameter grid consisting of various initial orbital periods and primary star masses for a fixed mass ratio of 0.9, using the \mesa code \citep{Paxton11,Paxton13,Paxton15,Paxton18,Paxton19}. The Sm11p600 pre-SN model is a RSG model evolved from a primary star having an initial mass of $M=11$\msun and an initial metallicity of $Z=0.02$ in a binary system with an initial period of 600 days. The initial conditions for the Sm11p400 model are the same, but the initial binary period is 400 days. The Tm11p200 model is a YSG model sharing the same initial primary mass, mass ratio, and metallicity, but its initial binary period is 200 days, and a reduced mass-loss rate by a factor of two compared to the standard rate given by the Dutch scheme for hot star winds in the \mesa{} code is employed during the evolution. In \citet{Yoon17}, the evolution of these models is stopped when the central temperature reaches $T_{\mathrm{c}}=10^9$~K. For this study, we evolve these models further until the infall velocity of the iron core reaches 1000 km/s (See \citealp{Yoon17} for detailed information on our fiducial pre-SN models.). Pre-SN properties of the models are presented in Table \ref{tab:presnparams}, along with the pre-SN progenitor properties inferred from SNe IIb observations for comparison. Some of the values presented in Table \ref{tab:presnparams} slightly differ from those in Tables 2 and 3 of \citet{Yoon17} because of different terminal conditions. In Figures \ref{fig:rhostructure} and \ref{fig:chemstructure}, we show the internal density profiles and chemical structures of the models, respectively. 
        
    \subsection{Numerical methods} \label{sec:modelcodes}

        In this study, we employ three different numerical codes - \stella{}, \mesa{}, and \snec{} - to simulate SNe explosions and the resulting light curves. 

\stella{} is a one-dimensional (1D) RHD code capable of implicitly calculating radiative transfer in multi-frequency bands \citep{Blinnikov98,Blinnikov00,Blinnikov06,Blinnikov11}. SN explosion is treated as a thermal bomb applied at a location near the inner boundary of the ejecta. Boltzmann-Saha distributions in local thermodynamic equilibrium (LTE) are assumed in calculating ionization and level populations of atoms. The effects of monochromatic Thomson scattering are fully considered in solving radiative transfer equations, which means that the source function does not correspond to the Planck function. 
%
%

The adopted opacity includes contributions from bound-free and free-free absorptions, lines, and electron scattering.  As for the line opacity, the current version of \stella{} assumes pure absorption to account for the non-LTE effects of fluorescence. The validity of this approximation has been discussed by \citet{Blinnikov98}, \citet{Kozyreva20}, and \citet{Potashov21}. \citet{Blinnikov98} compared SNe IIb light curve models calculated with \stella{} and the non-LTE code \texttt{EDDINGTON} \citep{Eastman93} to conclude that treating the line interactions as pure absorption (i.e., the so-called thermalization parameter $\epsilon_\mathrm{th}$ corresponds to 1.0) yields the best agreement between the two calculations. Although a purely absorptive line opacity was assumed for \texttt{EDDINGTON} in \citet{Blinnikov98}, \citet{Pinto00} compares SN Ia light curve models calculated with the \texttt{EDDINGTON} code adopting realistic fluorescence effects and the pure line absorption assumption and find a good agreement between the two results. \citet{Kozyreva20} compared the \stella{} calculations with those of the non-LTE code \texttt{ARTIS} \citep{Kromer09} for SNe Ia light curves and
conclude that  the choice of $\epsilon = 0.9$ results in the most consistent results between the two numerical codes. These authors 
also compared the \stella{} models with an observed sample of  various types of SNe (i.e., SN Ia: SN 2005cf; SN IIP: SN 1999em; SN II-pec: SN 1987A) and found the best agreement between the \stella{} models and the observed supernova sample for $\epsilon = 0.9$. However, the results with $\epsilon_\mathrm{th} =$ 0.9 and 1.0 are not significantly different in \citet{Kozyreva20} and we use $\epsilon_\mathrm{th}=1.0$ following \citet{Blinnikov98}. In spite of the uncertainty regarding the non-LTE effects, the focus of this work is on the light curves in the photospheric stages during which the non-LTE effects are relatively unimportant \citep{Baklanov13}. 
See the references cited above for more detailed discussions.
        
 The open source code \snec{} introduced by \citet{Morozova15} solves 1D Lagrangian RHD equations and calculates bolometric light curves.  Instead of using multi-frequency bands, the Rosseland-mean opacity is used for the photon diffusion process. The effects of Thomson scattering are not considered and the adopted source function corresponds to the Planck function ($S_{\nu}=B_{\nu}$).  Therefore, \snec{} assumes that the SED of the SNe follows black-body radiation of the effective temperature at the photosphere defined by the Rosseland-mean opacity.  \citet{Piro17} and \citet{Eldridge18} have used \snec{} for modelling SNe IIb light curves. 

As \citet{Morozova15} acknowledged, the methodology of \snec{} code is largely similar to that of \citet{Bersten11}. These authors solve the same 1D RHD equations with those of \snec, but instead of adopting the black-body spectrum of the photospheric temperature, they calculate multi-color light curves by determining the thermalization depth using the analytical method of \citet{Ensman92}, indirectly including the effects of Thomson scattering in their models. In Section \ref{sec:snec}, we compare the multi-band light curves obtained with \snec and \stella to investigate the effects of Thomson scattering and also discuss the validity of the analytical approach introduced by \citet{Ensman92}. 
        
         \mesa{} is a 1D stellar astrophysics code widely used for numerically computing stellar evolution and is also capable of simulating hydrodynamic explosion of core-collapse SNe \citep[CCSNe;][]{Paxton15}. \mesa recently incorporates the effects of RTI by 1D approximation, employing the method introduced by \citet{Duffell16} who treated matter mixing induced by RTI as a diffusion process. For the discussion of the effects of chemical mixing in Section \ref{sec:chem}, \mesa{} is employed for calculating the shock propagation after the explosion including the effects of RTI for different degrees of chemical mixing until the shock reaches the stellar surface. We then hand over the model to \stella, numerically solving multi-group radiative transfer equations to obtain broad-band light curves and investigate the effects of chemical mixing on the shape of the first peak.
        
	\begin{deluxetable*}{ccccccccc}
            \centering
            \tablecaption{SN explosion parameters and properties of the optical peaks \label{tab:snparams}}
            \tablehead{
            \colhead{Model Name} & \colhead{$M_{\mathrm{cut}}$} & \colhead{$E_{\mathrm{kin}}$} & \colhead{Band} & \colhead{$t_{\mathrm{peak}}$} & \colhead{$M_{\mathrm{peak}}$} & \colhead{$L_{\mathrm{peak}}$} & \colhead{$T_{\mathrm{peak,ph}}$} & \colhead{$T_{\mathrm{peak,color}}$}
            } 
            \startdata
            \multirow{3}{*}{Tm11p200\_STELLA\_SCA} & \multirow{3}{*}{1.40} & \multirow{3}{*}{2.15} & $B$ & 1.57 & -16.62 & 42.92 & 4.22 & 4.37\\
            & & & $V$ & 1.66 & -16.46 & 42.87 & 4.20 & 4.35\\
            & & & $R$ & 1.67 & -16.43 & 42.86 & 4.19 & 4.35\\
            \hline
            \multirow{3}{*}{Sm11p400\_STELLA\_SCA} & \multirow{3}{*}{1.48} & \multirow{3}{*}{2.32} & $B$ & 2.94 & -17.44 & 43.08 & 4.15 & 4.29 \\
            & & & $V$ & 3.13 & -17.30 & 43.03 & 4.13 & 4.27 \\
            & & & $R$ & 3.44 & -17.28 & 42.95 & 4.09 & 4.23 \\
            \hline
            \multirow{3}{*}{Sm11p600\_STELLA\_SCA} & \multirow{3}{*}{1.41} & \multirow{3}{*}{2.14} & $B$ & 4.47 & -17.75 & 43.10 & 4.10 & 4.24 \\
            & & & $V$ & 5.09 & -17.64 & 43.00 & 4.05 & 4.19 \\
            & & & $R$ & 5.30 & -17.63 & 42.96 & 4.04 & 4.16 \\
            \enddata
            \tablecomments{SN explosion parameters, i.e. the excised core mass at the explosion and the kinetic energy of the ejecta, are presented. Light curve and SN properties at the first peaks of $B$-, $V$-, and $R$-band light curves of SNe models are also presented. $M_{\mathrm{core}}$: excised core mass in units of \msun, $E_{\mathrm{kin}}$: kinetic energy of the ejecta in units of $10^{51}$ ergs, Band: optical band of which the first peak of the light curve is measured, $t_{\mathrm{peak}}$: the time of the first peak in units of days, $M_{\mathrm{peak}}$: the absolute magnitude at the time of the first peak, $L_{\mathrm{peak}}$: the bolometric luminosity at the time of the first peak in units of ergs in logarithmic scale, $T_{\mathrm{peak,ph}}$: gas temperature at the Rosseland-mean photosphere in units of K in logarithmic scale, $T_{\mathrm{peak,color}}$: the color temperature at the time of the first peak in units of K in logarithmic scale.}
        \end{deluxetable*}
        
\section{The effects of Thomson scattering} \label{sec:snec}
    \subsection{Setup} \label{sec:snec-setup}
        
        In this section, we compare light curve models computed using different methodologies and discuss the importance of Thomson scattering in determining optical light curves of SNe IIb. 
In order to simulate the SN explosion after the core collapse, we apply the thermal bomb by arbitrarily increasing the temperature at the location slightly above the outer boundary of the iron core. The iron core mass and the adopted mass cuts are presented in Tables \ref{tab:presnparams} and \ref{tab:snparams}, respectively. We only focus on the first peak and do not consider the main peak powered by the radioactive decay of \nifs. Therefore, we do not include radioactive \nifs in the SN ejecta in our simulations, except for some test models discussed in Section \ref{sec:obs}. This leaves the thermal radiation of SNe ejecta heated by the SN shock as the only luminosity source. 

        For each progenitor model,  we apply a weak chemical mixing  using the boxcar smoothing method (See Section~\ref{sec:chem} for the extensive discussion on chemical mixing) and calculate three different SN light curve models using the following three different approaches:  1) standard \stella, 2) \stella with the opacity regarded as fully absorptive within the ejecta, and 3) \snec. The first approach is the standard setup for \stella calculations as explained in Section \ref{sec:modelcodes}, which includes the effects of Thomson scattering. For the second approach, we still use the \stella{} code but the treatment of opacity differs. All the terms with Thomson cross-section give the same extinction as in scattering but are treated as pure absorption. This opacity treatment is similar to that of \snec. The model names for three different cases are respectively denoted by suffixes  `\_STELLA\_SCA', `\_STELLA\_ABS', and `\_SNEC'. 
        
    \subsection{Results} \label{sec:snec-results}

\begin{figure}
        \gridline{\fig{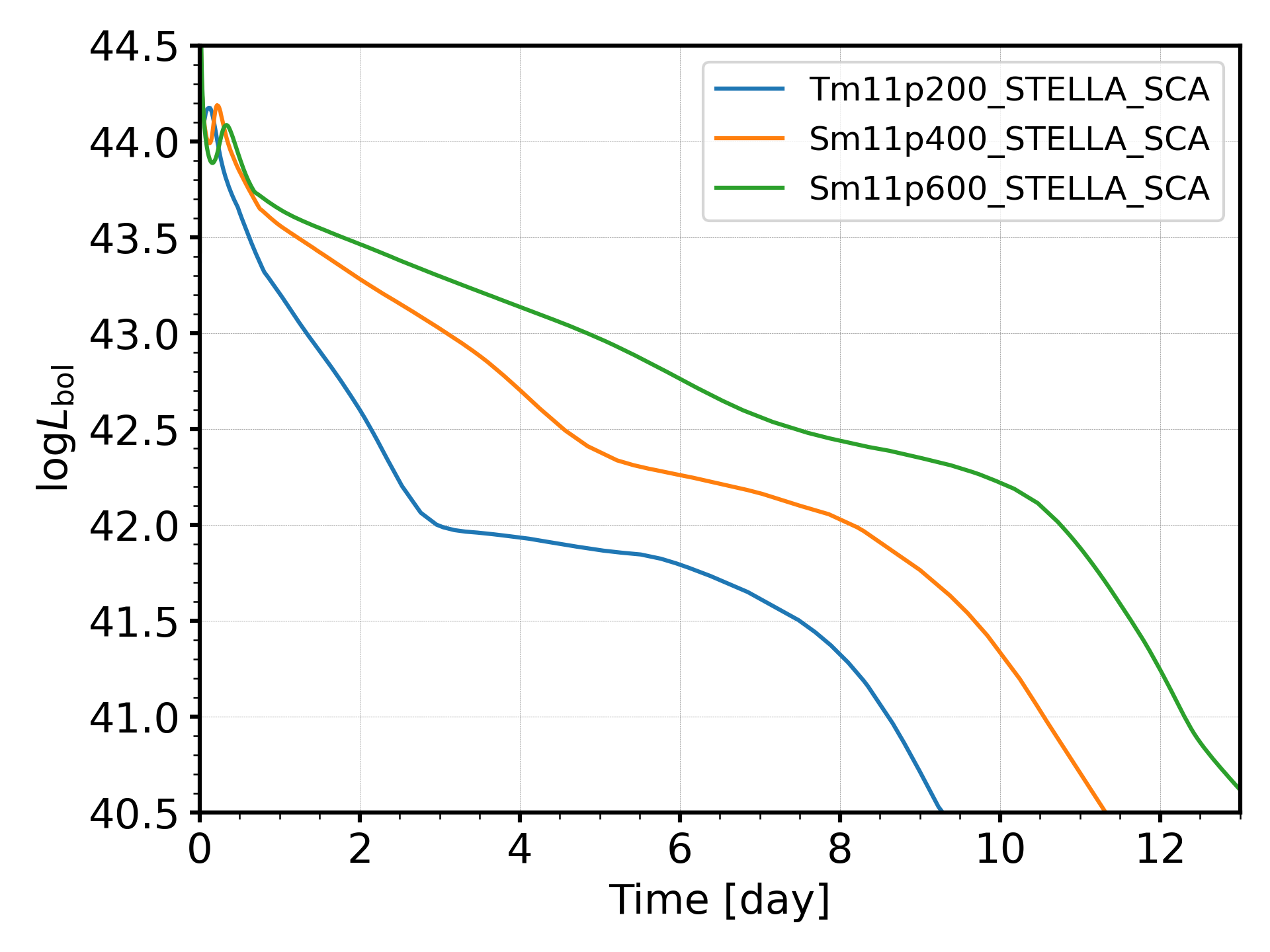}{0.47\textwidth}{}}
        \gridline{\fig{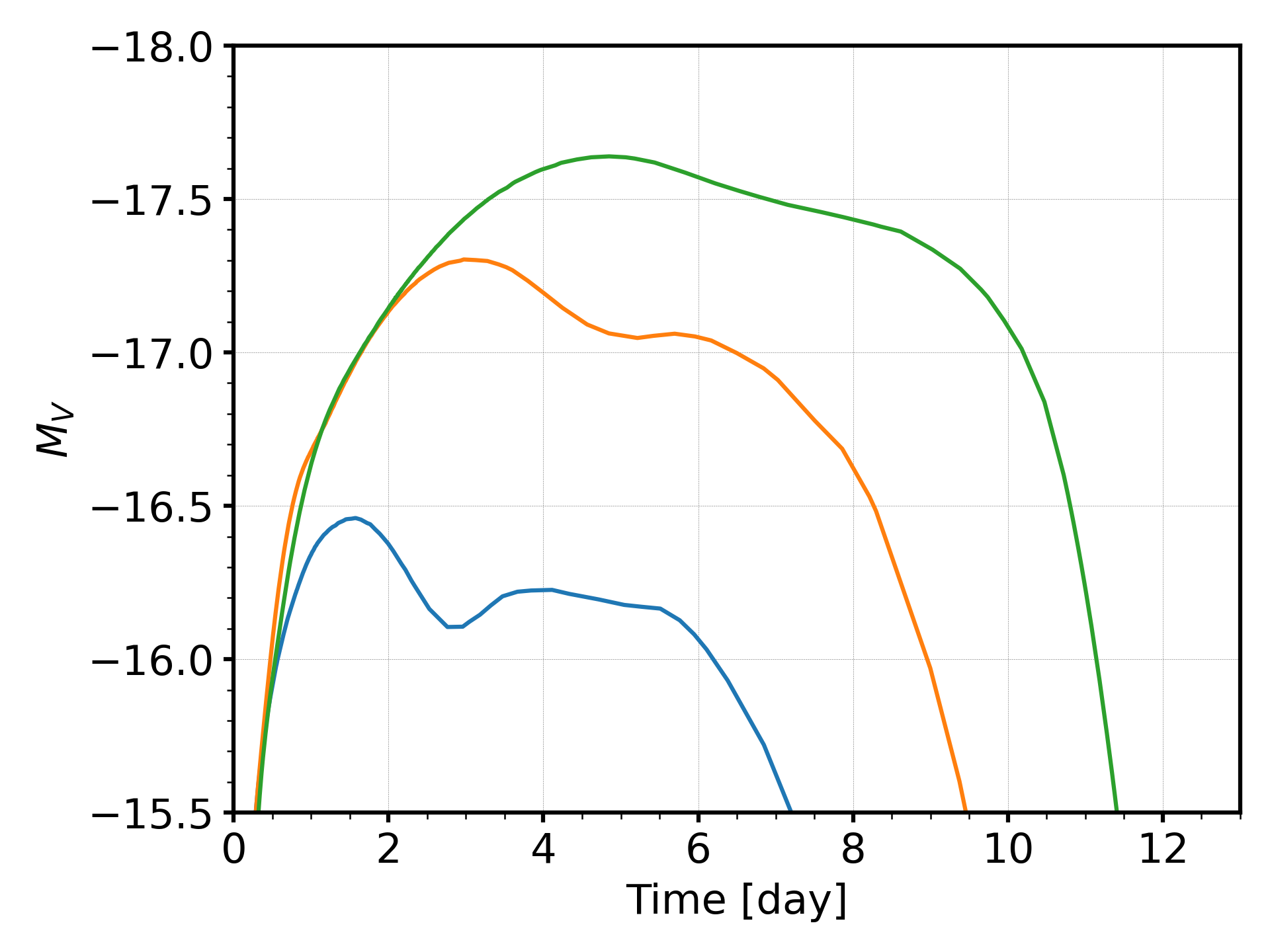}{0.47\textwidth}{}}
        \gridline{\fig{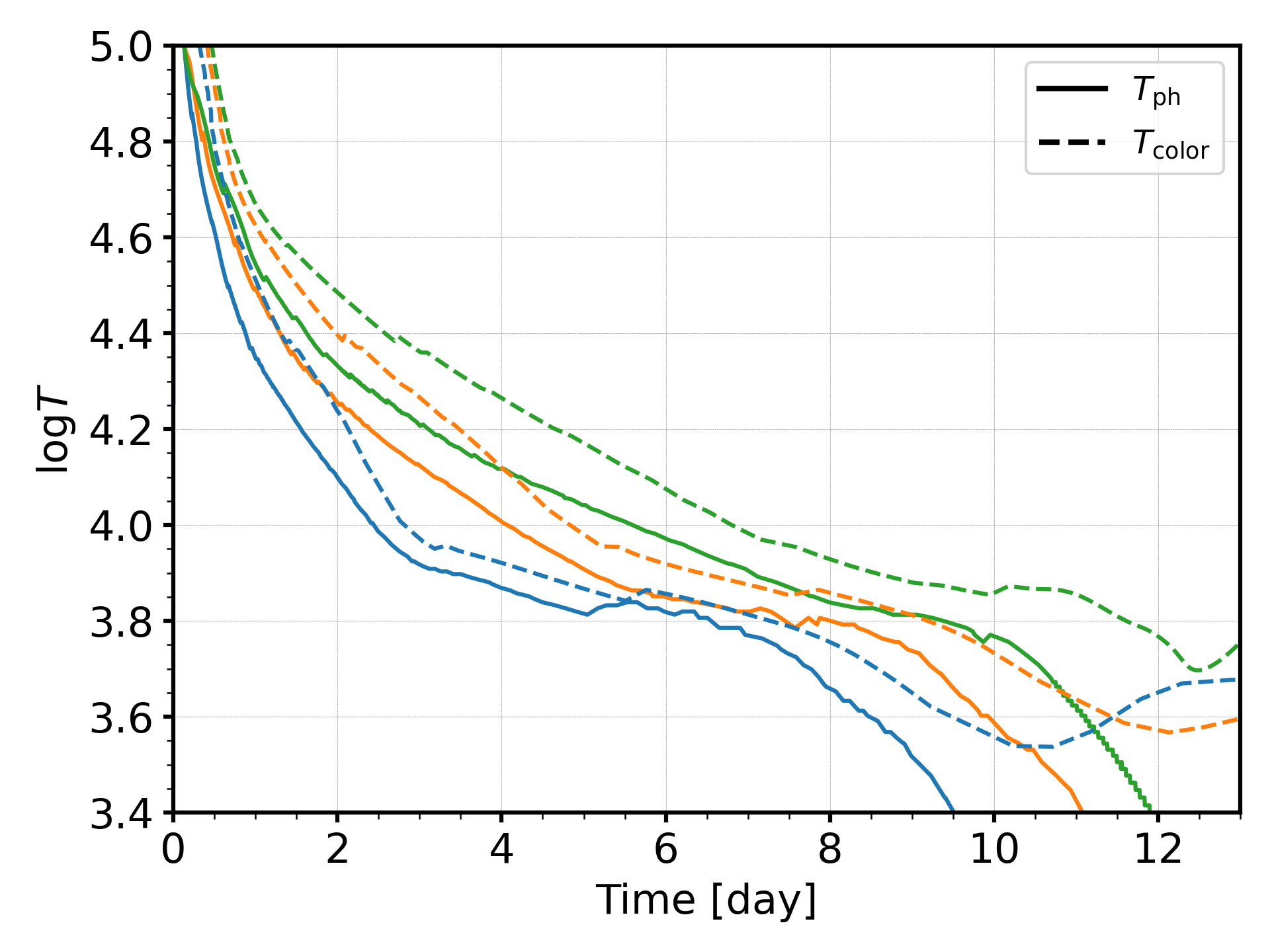}{0.47\textwidth}{}}
        \caption{Top panel: bolometric luminosities of the STELLA\_SCA models. Middle panel: $V$-band light curves of the models. Bottom panel: temporal evolution of the gas temperature at the Rosseland-mean photosphere ($T_{\mathrm{ph}}$; solid line) and the color temperature, that is the temperature that gives the best black-body fit to the SED ($T_{\mathrm{color}}$; dashed line). See Table \ref{tab:snparams} for the physical parameters of the models. 
        \label{fig:Henv_analysis}}
        \end{figure}

        In Figure \ref{fig:Henv_analysis}, we present bolometric and $V$-band light curves of STELLA\_SCA models for three different progenitors (Tm11p200, Sm11p400, \& Sm11p600) having a similar kinetic energy of $\sim$2\bethe (see Table \ref{tab:snparams}), along with the temporal evolution of the color temperature ($T_{\mathrm{color}}$), gas temperature at the photosphere ($T_{\mathrm{ph}}$, and the mass coordinate of the photosphere divided by the pre-SN total mass. Here, the photosphere is defined by the Rosseland-mean opacity and the color temperature is determined by the temperature that gives the best black-body fit to the SN spectrum.
        
        After the shock breakout, bolometric luminosities of the STELLA\_SCA models show three distinct evolutionary phases. First, the initial rapid decline lasts for $\sim$3 days in Tm11p200, $\sim$5 days in Sm11p400, and 7$\sim$8 days in Sm11p600. Then, a plateau-like feature appears for about 3 days before the luminosity again declines rapidly. The decline rate of the light curve during the initial decline phase is lower for a higher H-rich envelope mass ($M_{\mathrm{H,env}}= 0.098, 0.138, \& 0.198$\msun for Tm11p200, Sm11p400, \& Sm11p600). As a result, the slope change between the first and second phases appears to be less prominent for a larger H-rich envelope mass. 
        
        The luminosity decline rate and the ejecta cooling rate are determined by several factors such as initial radius, ejecta mass, initial density profile, chemical composition, and ejecta kinetic energy \citep{Arnett80}. There is a positive correlation between the H-rich envelope mass and the radius of the progenitor models selected for this study (see \ref{tab:presnparams}), and the more rapid for a more compact and less massive H-rich envelope is consistent with the analytical solutions of \citet{Arnett80}. The semi-plateau phase is reached once the ejecta is cooled down to $T\lesssim8000$ K. From this point, the hydrogen recombination becomes important and the photosphere recedes into the inner layers of the ejecta closely following the recombination front. During the semi-plateau phase, the photospheric gas temperature decline rate is significantly decreased for a few days as we see in the bottom panel of Figure \ref{fig:Henv_analysis}. The semi-plateau phase and its relation to the chemical structure of the ejecta are discussed in Section \ref{sec:chem}.
        
        
        As can be seen in the middle panel of Figure \ref{fig:Henv_analysis}, the $V$-band peak is brighter and the rise to the peak is slower for a more massive H-rich envelope. Initially, the emitted radiation is mostly in the short-wavelength region, i.e. X-ray and EUV. As the ejecta cools down, the SED shifts nearer to the optical wavelength region, increasing the optical luminosity. In Table \ref{tab:snparams}, we present rise times to the optical peaks in $B$-,$V$-, and $R$-bands, peak magnitudes, bolometric luminosities, gas temperatures at the photosphere, and color temperatures at the time of the optical peaks. The color temperatures of \stella models are determined by the temperature that gives the best black-body fit to the SED. The $V$-band peaks are reached at 1.5, 3.3, and 5.1 days after the explosion for Tm11p200, Sm11p400, and Sm11p600, respectively. Different ejecta cooling rates make the difference in the rise time to the $V$-band peaks, with models having a less massive H-rich envelope reaching the peak earlier. In all cases, the bolometric luminosities at the time of the optical peaks are similar ($L_{\mathrm{bol}}\simeq10^{43}$ ergs/s), but the $V$-band peak magnitudes are -16.4, -17.2, -17.6 mags for Tm11p200, Sm11p400, and Sm11p600 models, respectively. The bottom panel of Figure \ref{fig:Henv_analysis} shows that both the photospheric gas temperature and the color temperature during the initial phase decline fastest in Tm11p200 and slowest in Sm11p600. At the time of the $V$-band peak, the photospheric gas temperatures are 17000K, 12600K, 11200K and the color temperatures are 25100K, 17800K, 15500K for Tm11p200, Sm11p400, and Sm11p600 models, respectively. This explains why the $V$-band peaks become brighter in the order of Tm11p200, Sm11p400, and Sm11p600, despite that the corresponding bolometric luminosities are similar. 

\begin{figure}
        \gridline{\fig{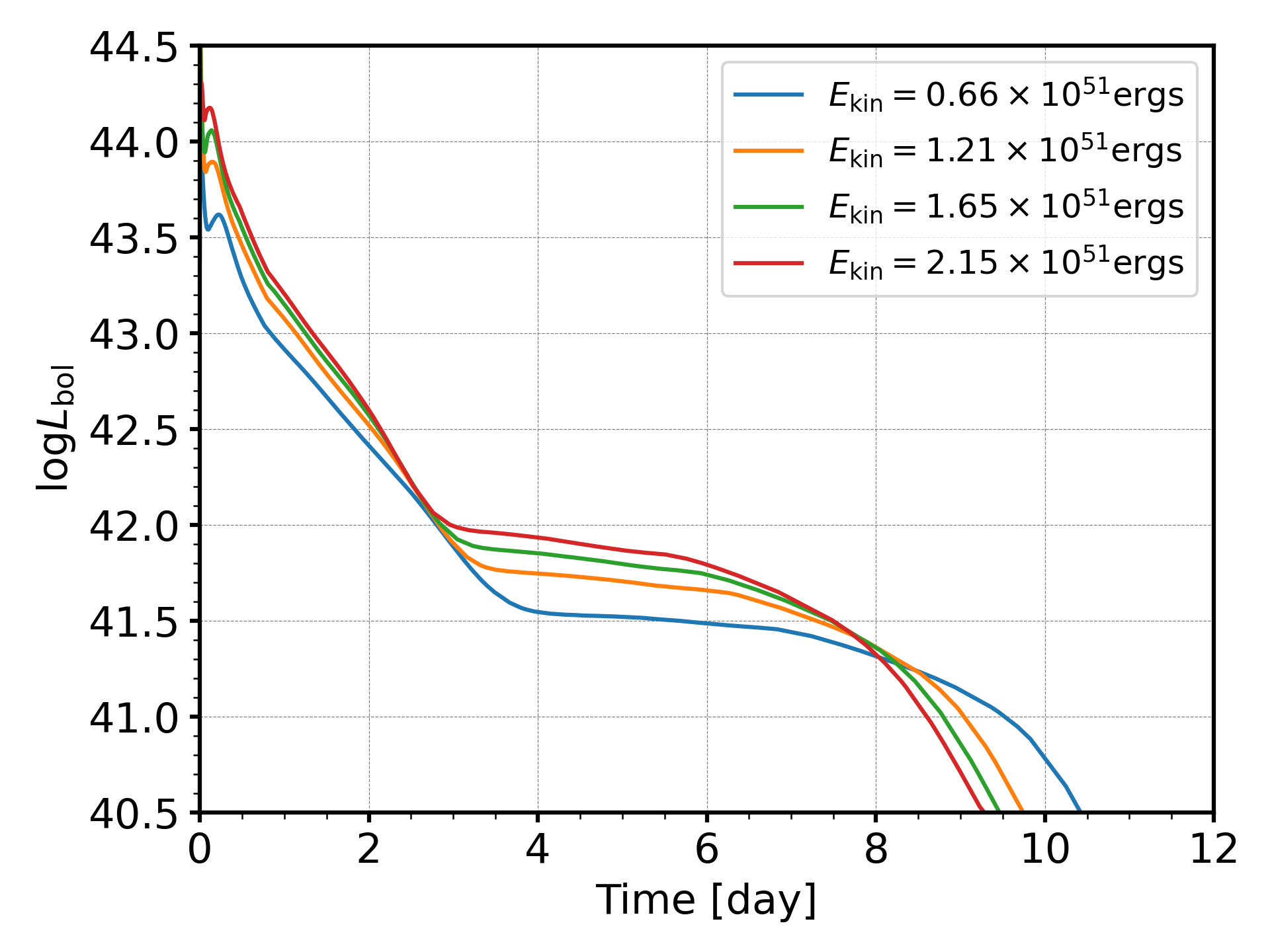}{0.47\textwidth}{}}
	\gridline{\fig{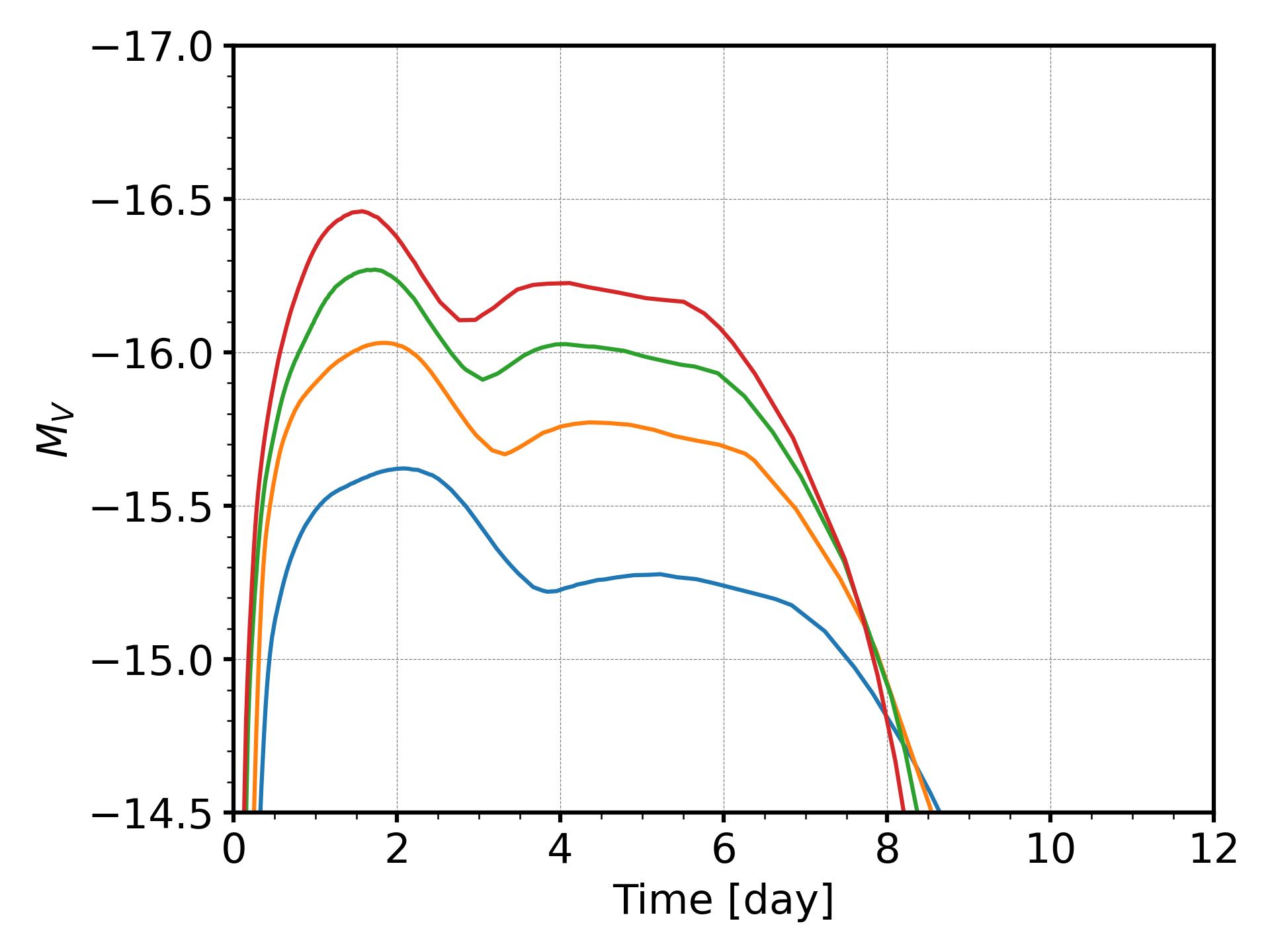}{0.47\textwidth}{}}
        \caption{Upper panel: bolometric luminosities of SNe models from the Tm11p200 progenitor model with different ejecta kinetic energies as indicated 
by the labels.  Here the model with $E_\mathrm{kin} = 2.15$\bethe{} corresponds to the Tm11p200\_STELLA\_SCA model. Lower panel: $V$-band light curves of the same models.
        \label{fig:p400_energy_lcs}}
        \end{figure}

        After the optical peak, the optical brightness declines until the semi-plateau phase is reached. In the semi-plateau phase, $V$-band luminosities of the STELLA\_SCA models remain more or less constant. In Figure \ref{fig:p400_energy_lcs}, we present SN models calculated with the Tm11p200 progenitor model for different ejecta kinetic energies: 0.83\bethe, 1.33\bethe, and 2.32\bethe. For a higher kinetic energy, the width of the optical light curve is narrower because the temperature and density decrease faster and the semi-plateau phase becomes shorter in duration. The optical peak becomes brighter with a higher kinetic energy because the thermal energy stored in the H-rich envelope is greater and the energy is released on a shorter timescale.
        
\begin{figure*}
        \gridline{\fig{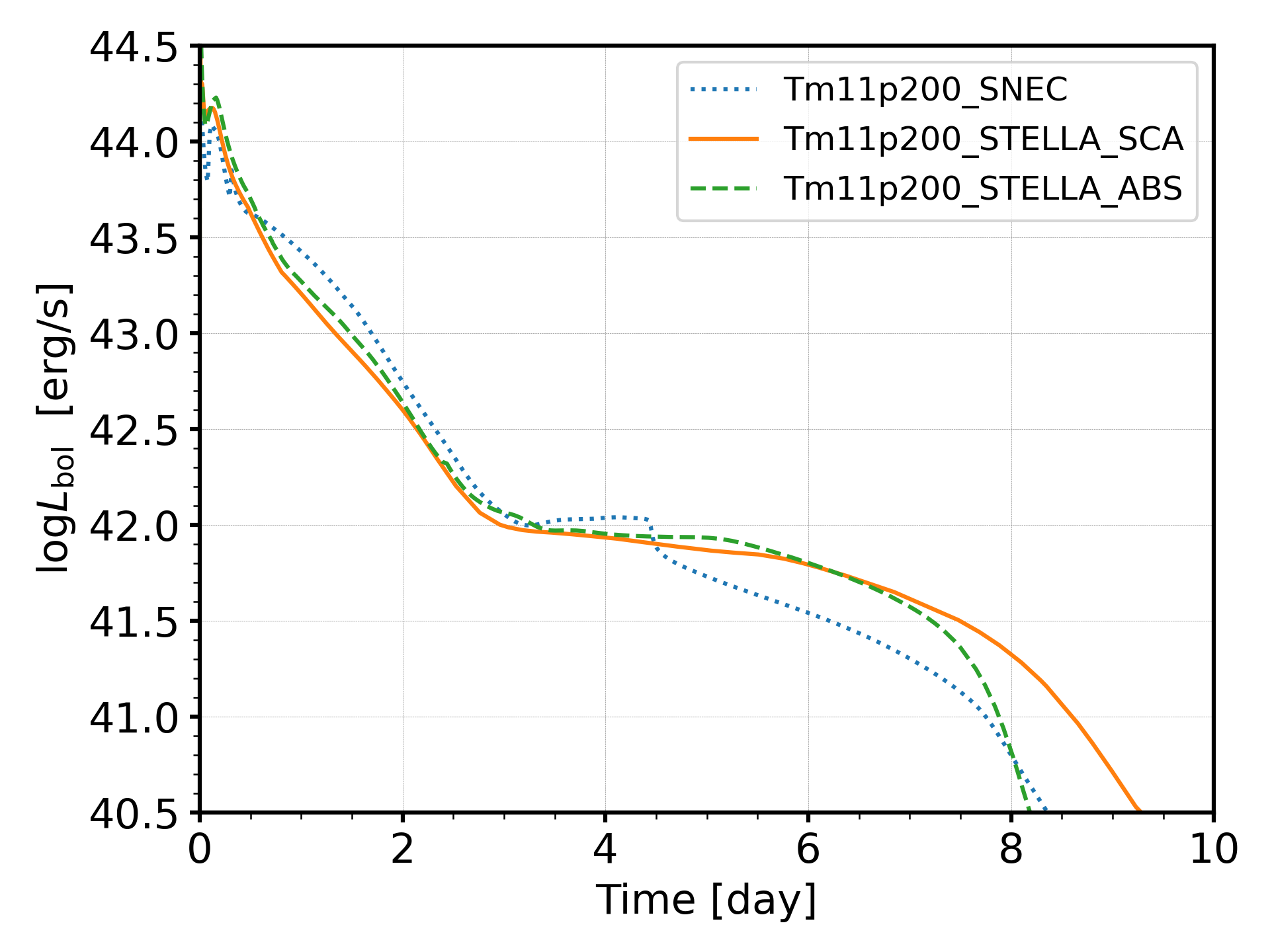}{0.47\textwidth}{}
                  \fig{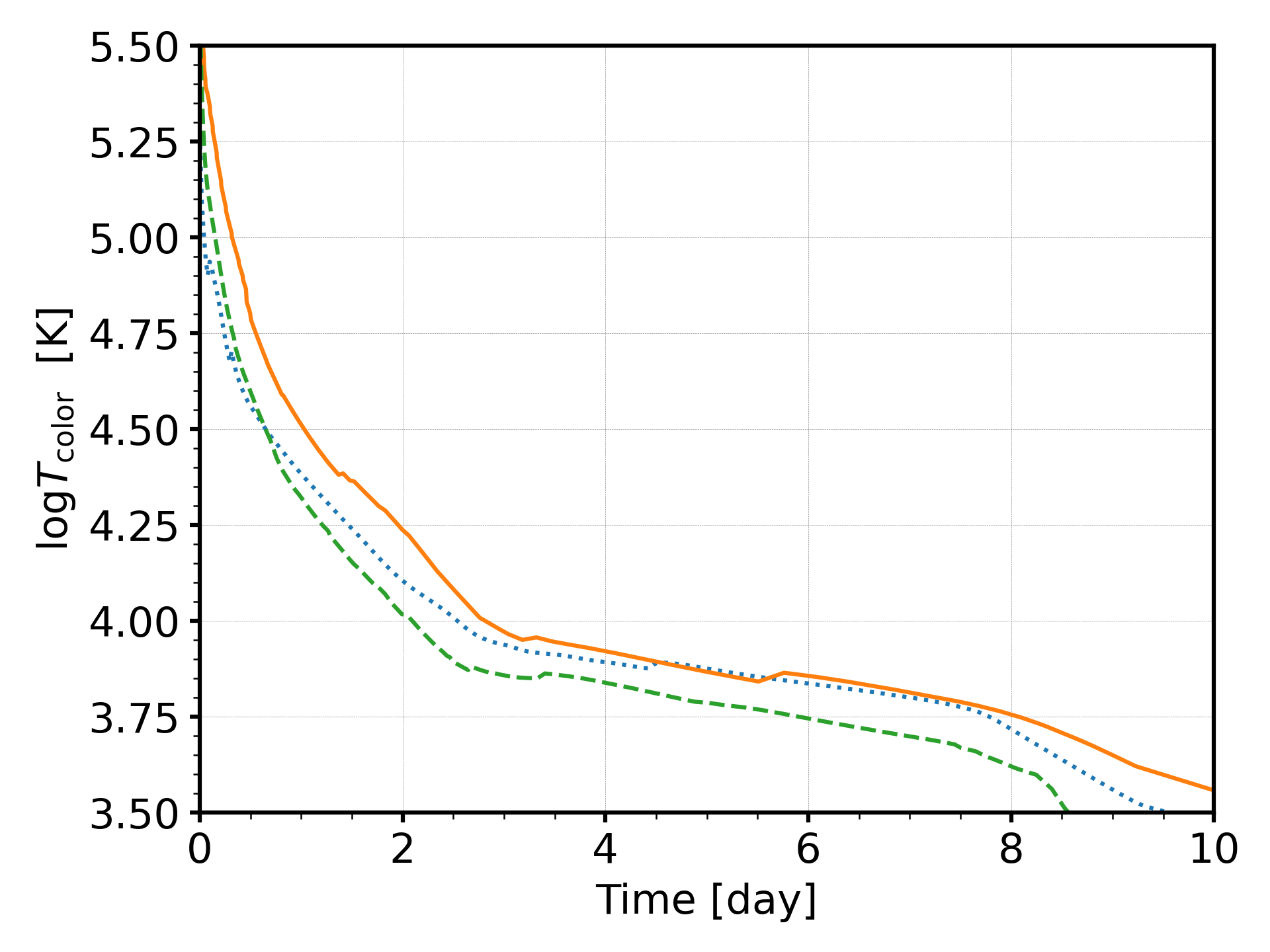}{0.47\textwidth}{}
                  }
        \gridline{\fig{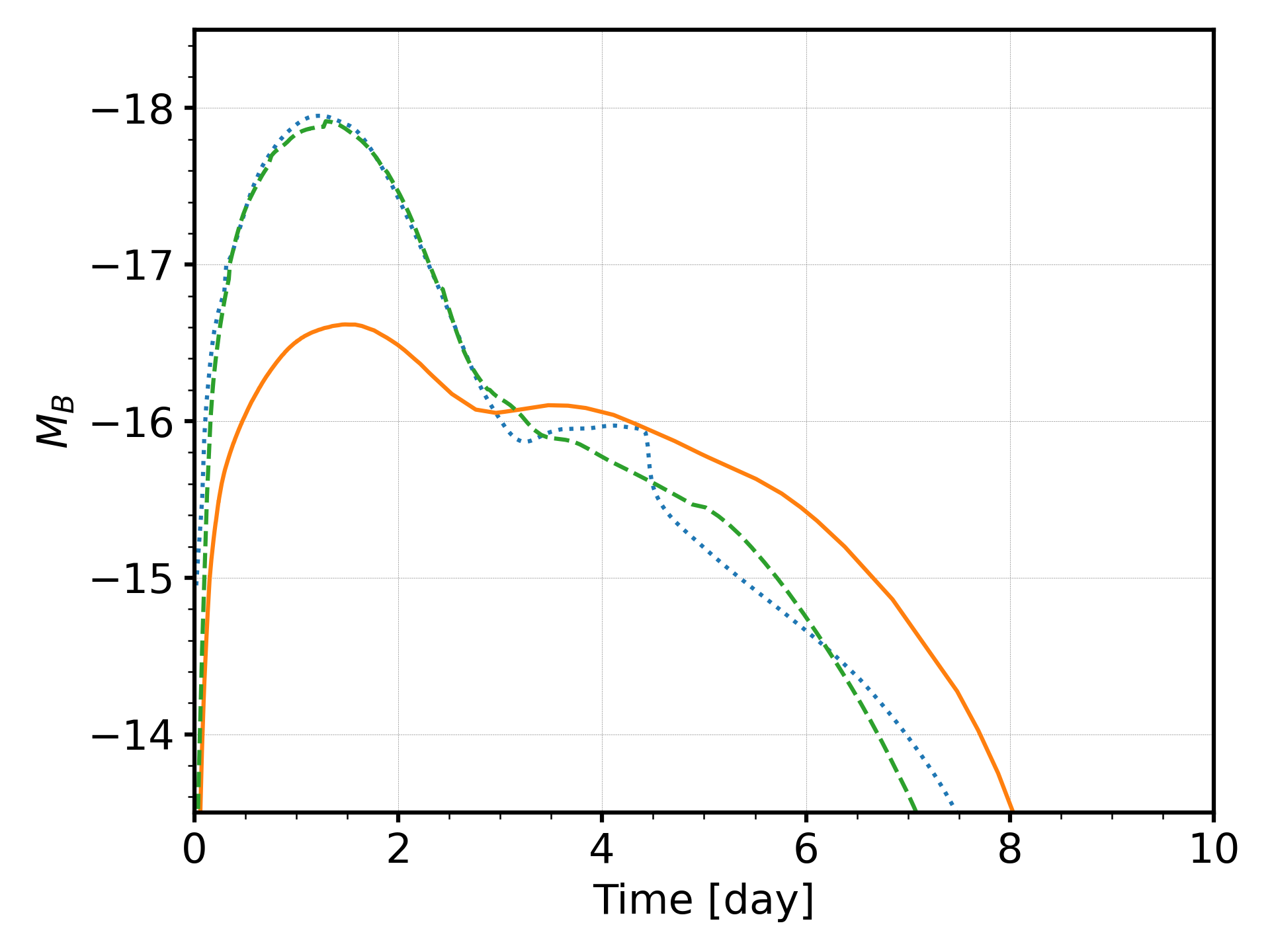}{0.47\textwidth}{}
                  \fig{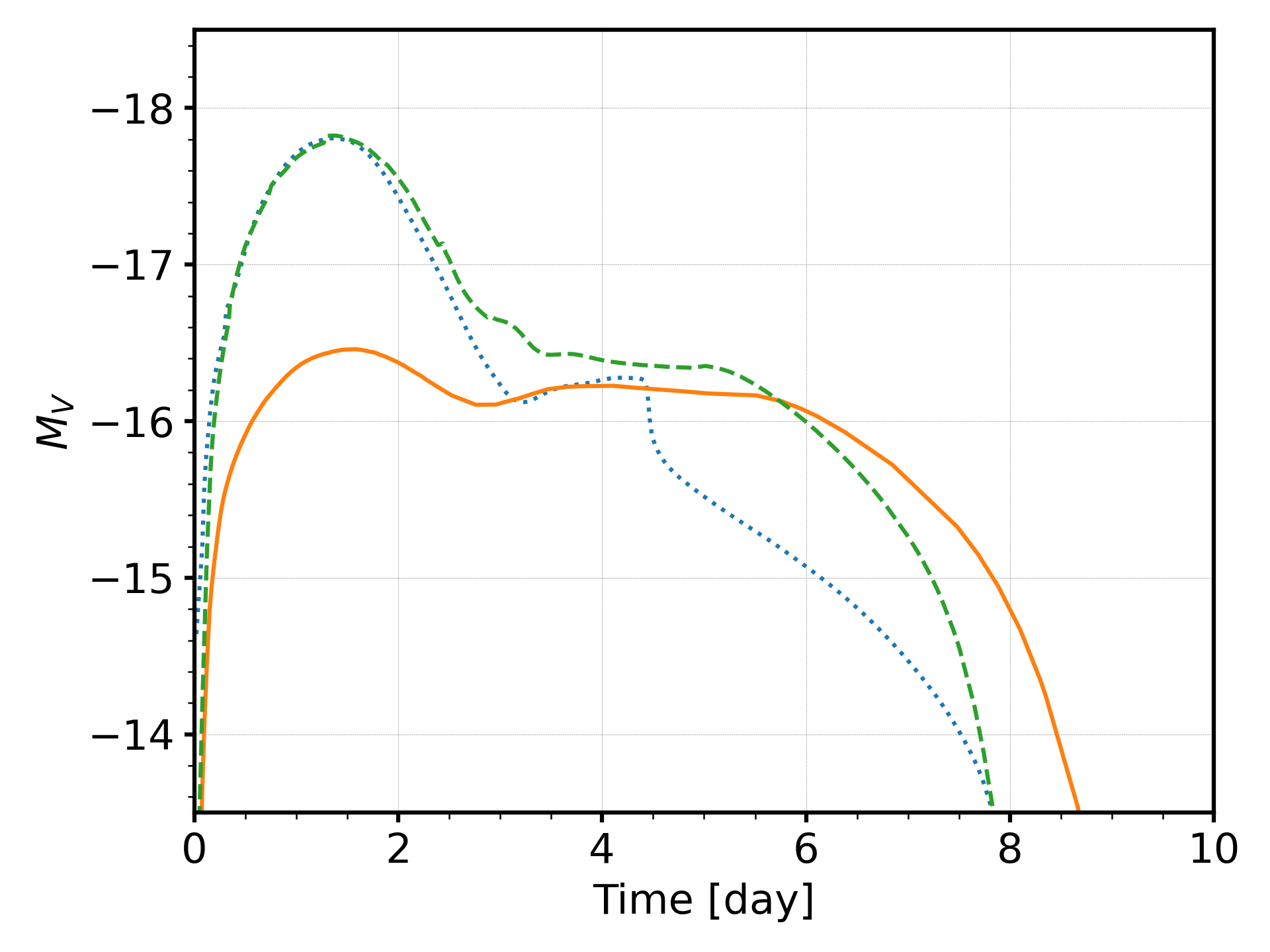}{0.47\textwidth}{}
                  }
        \gridline{\fig{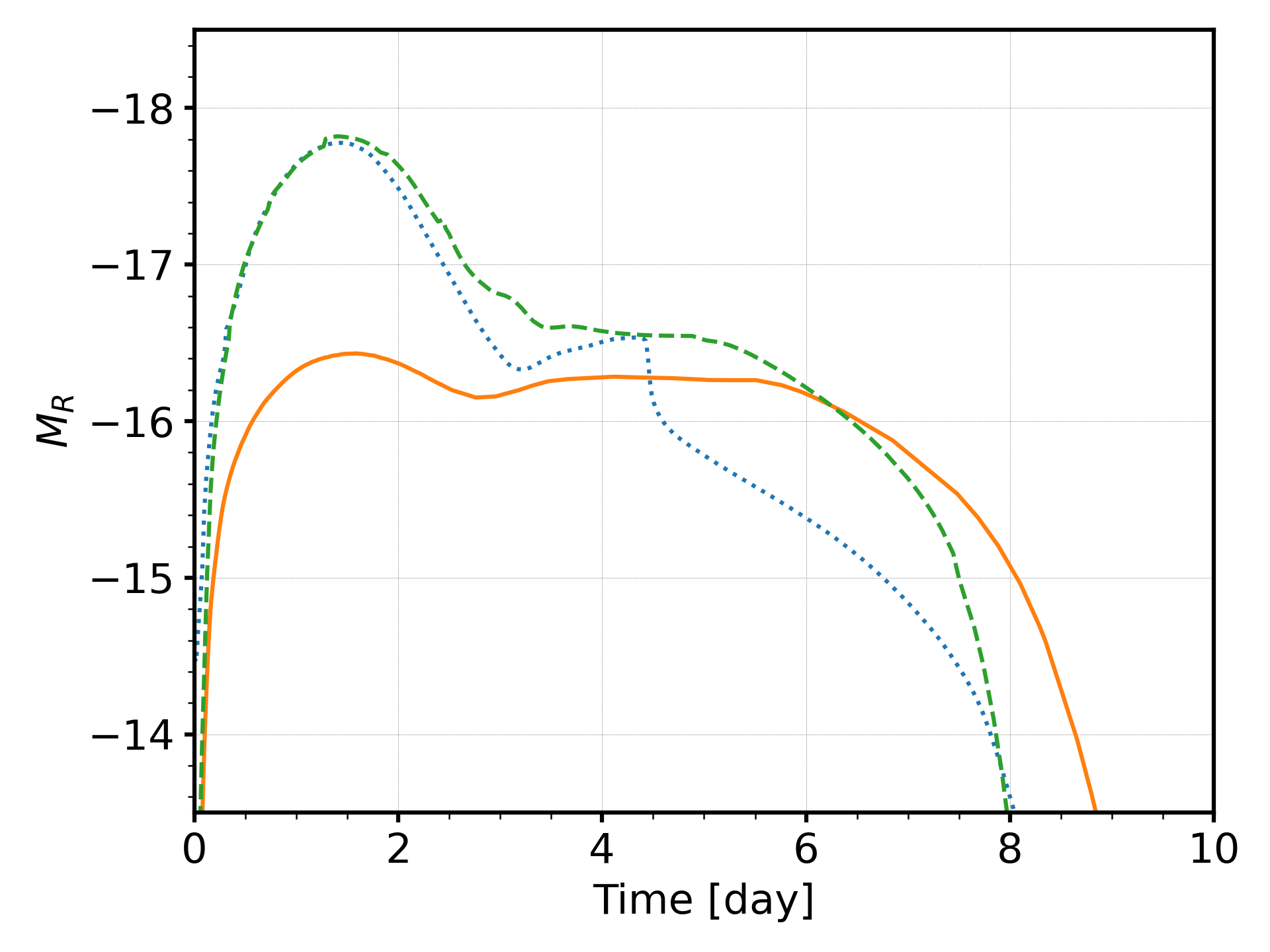}{0.47\textwidth}{}
                  \fig{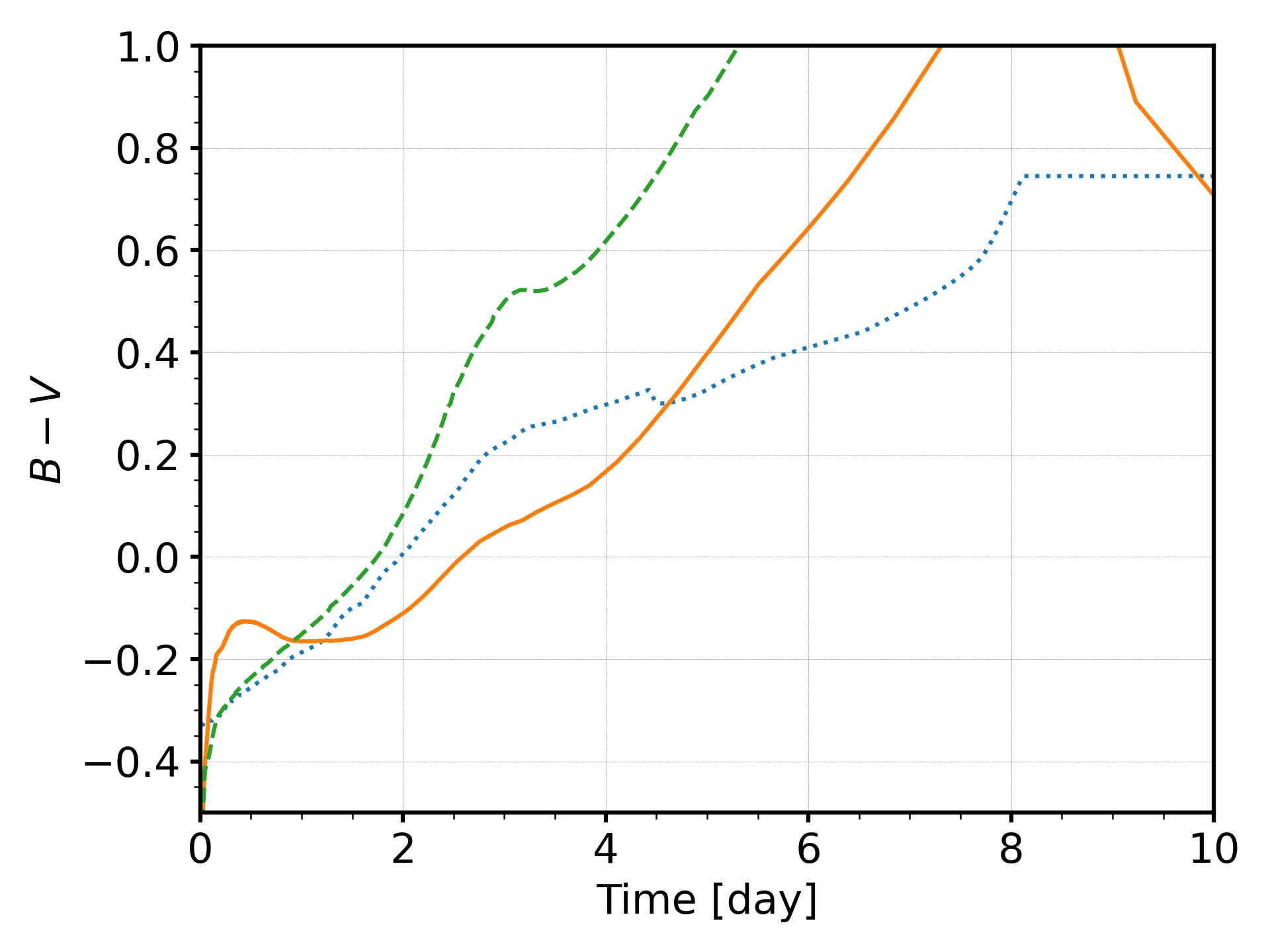}{0.47\textwidth}{}
                  }
        \caption{SNe models based on the progenitor model Tm11p200 calculated with three different methods: 1) \snec (blue dotted), 2) \stella with Thomson scattering (orange solid), 3) \stella assuming pure absorption (green dashed). The panels are showing bolometric luminosities (top left), color temperatures (top right), $B$-band (middle left), $V$-band (middle right), $R$-band light curves (bottom left), and $(B-V)$ color evolution (bottom right). See the text for the definition of the color temperature. The core mass cut of the models is $1.40$\msun, and the ejecta kinetic energy is $2.16$\bethe.
        \label{fig:p200snecresults}}
        \end{figure*}

\begin{figure*}
        \gridline{\fig{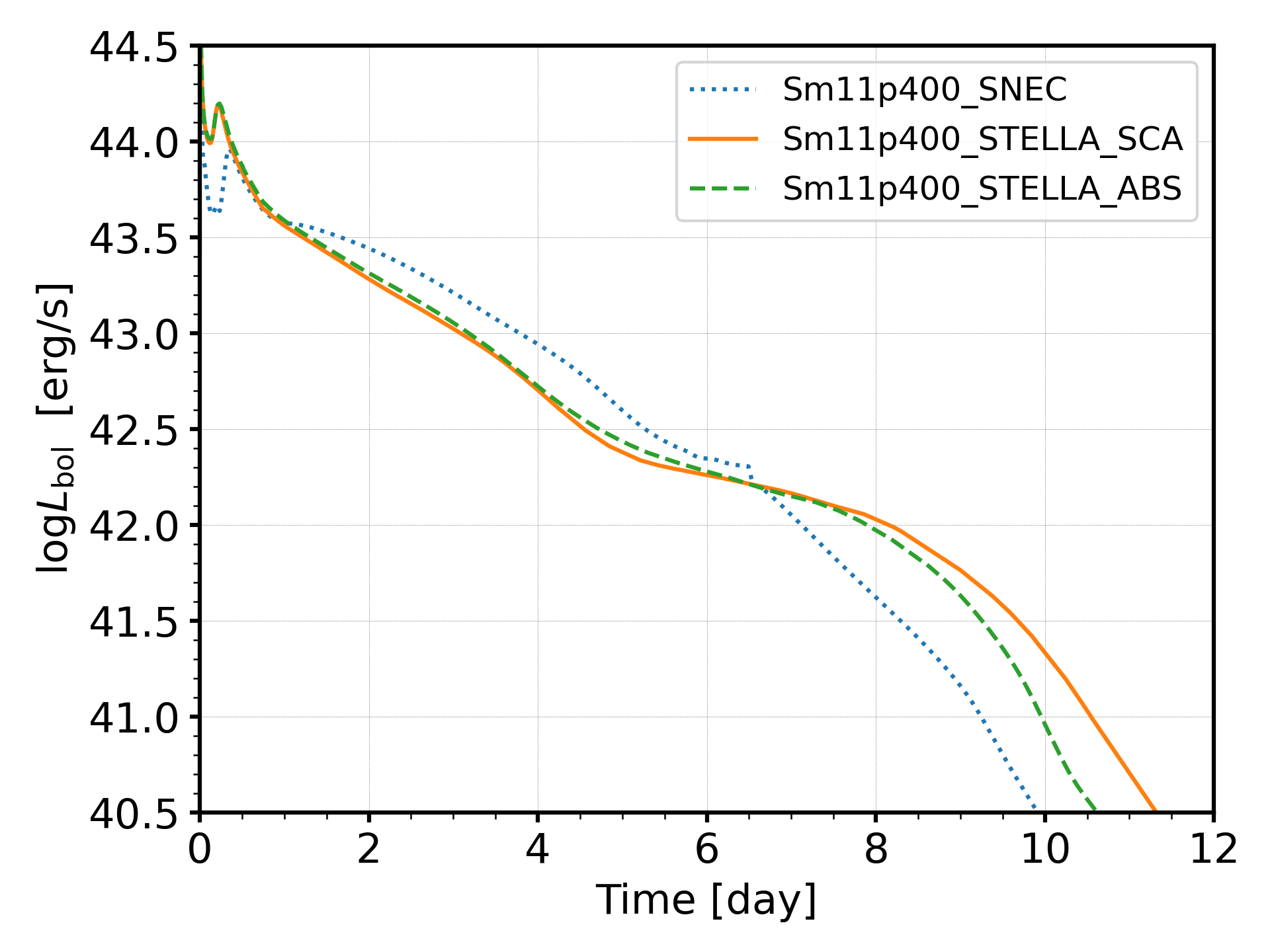}{0.47\textwidth}{}
                  \fig{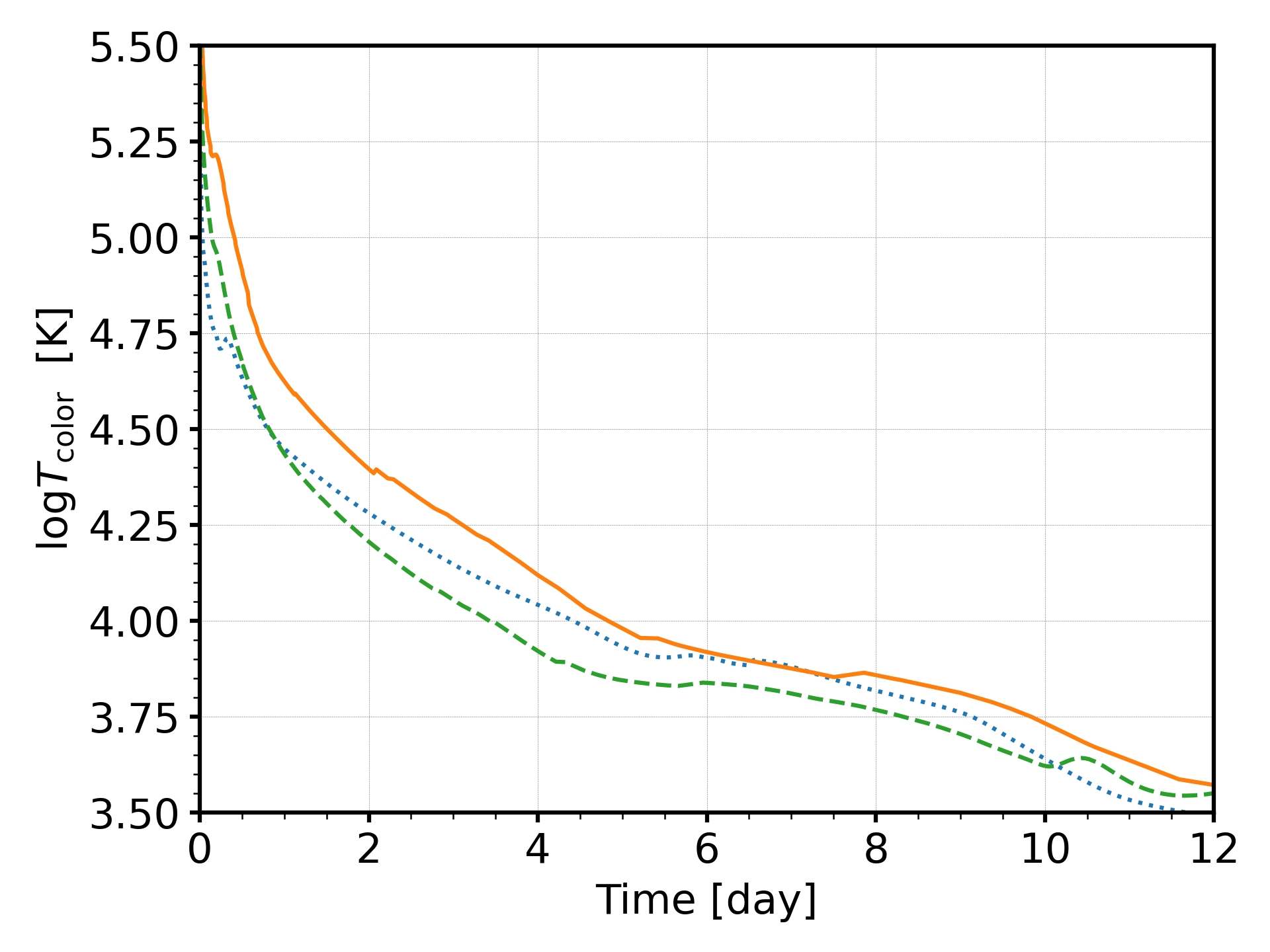}{0.47\textwidth}{}
                  }
        \gridline{\fig{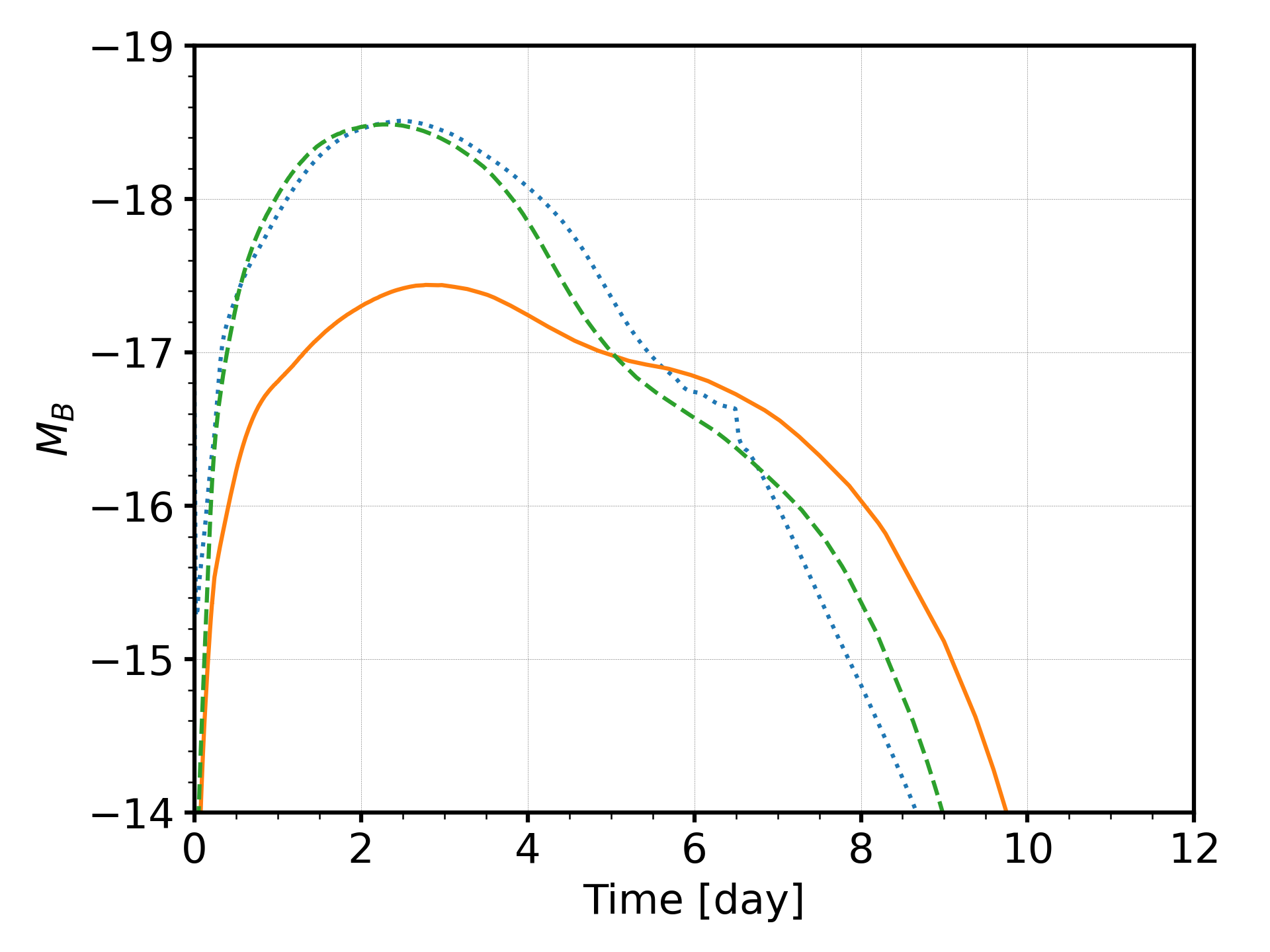}{0.47\textwidth}{}
                  \fig{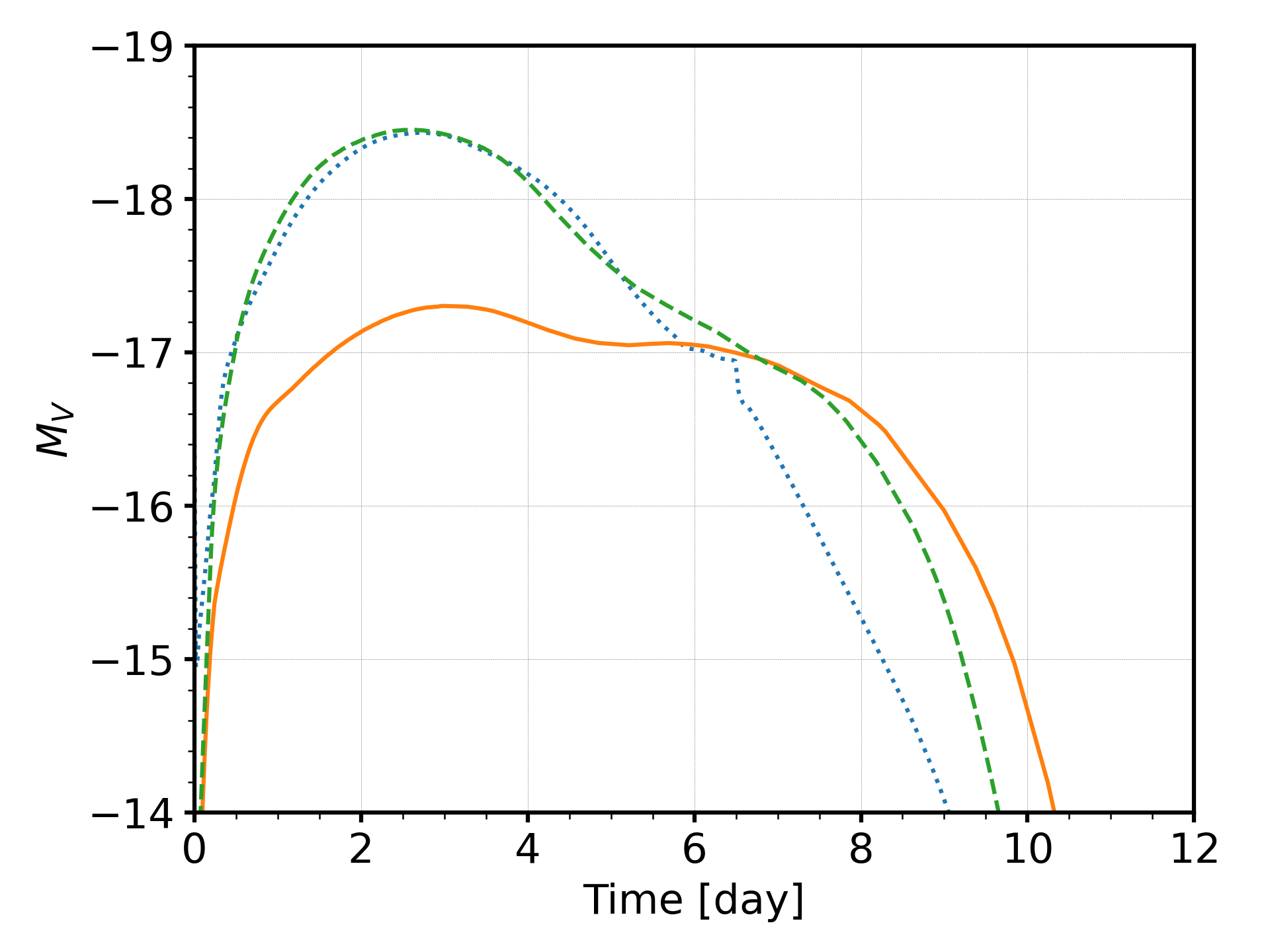}{0.47\textwidth}{}
                  }
        \gridline{\fig{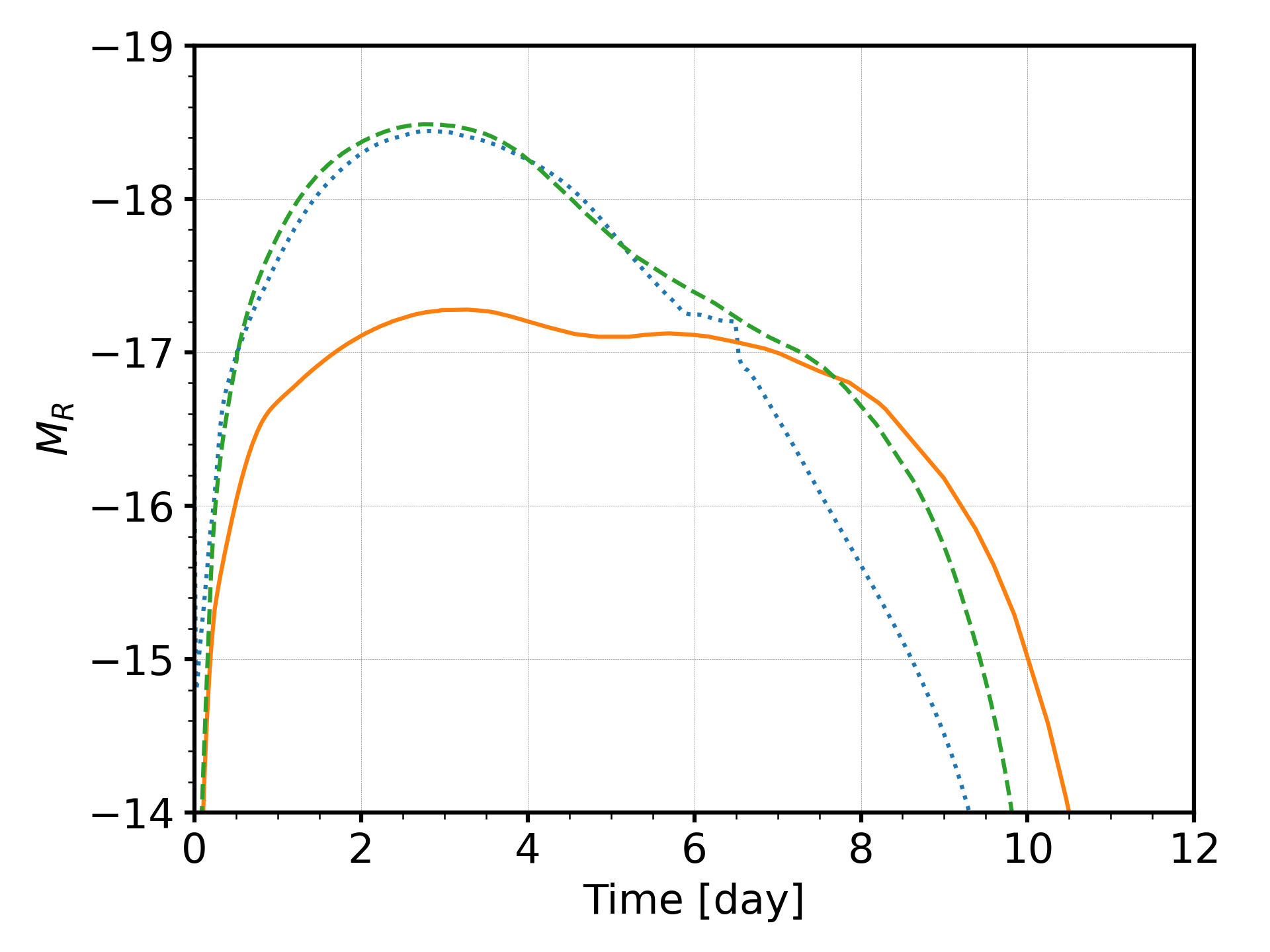}{0.47\textwidth}{}
                  \fig{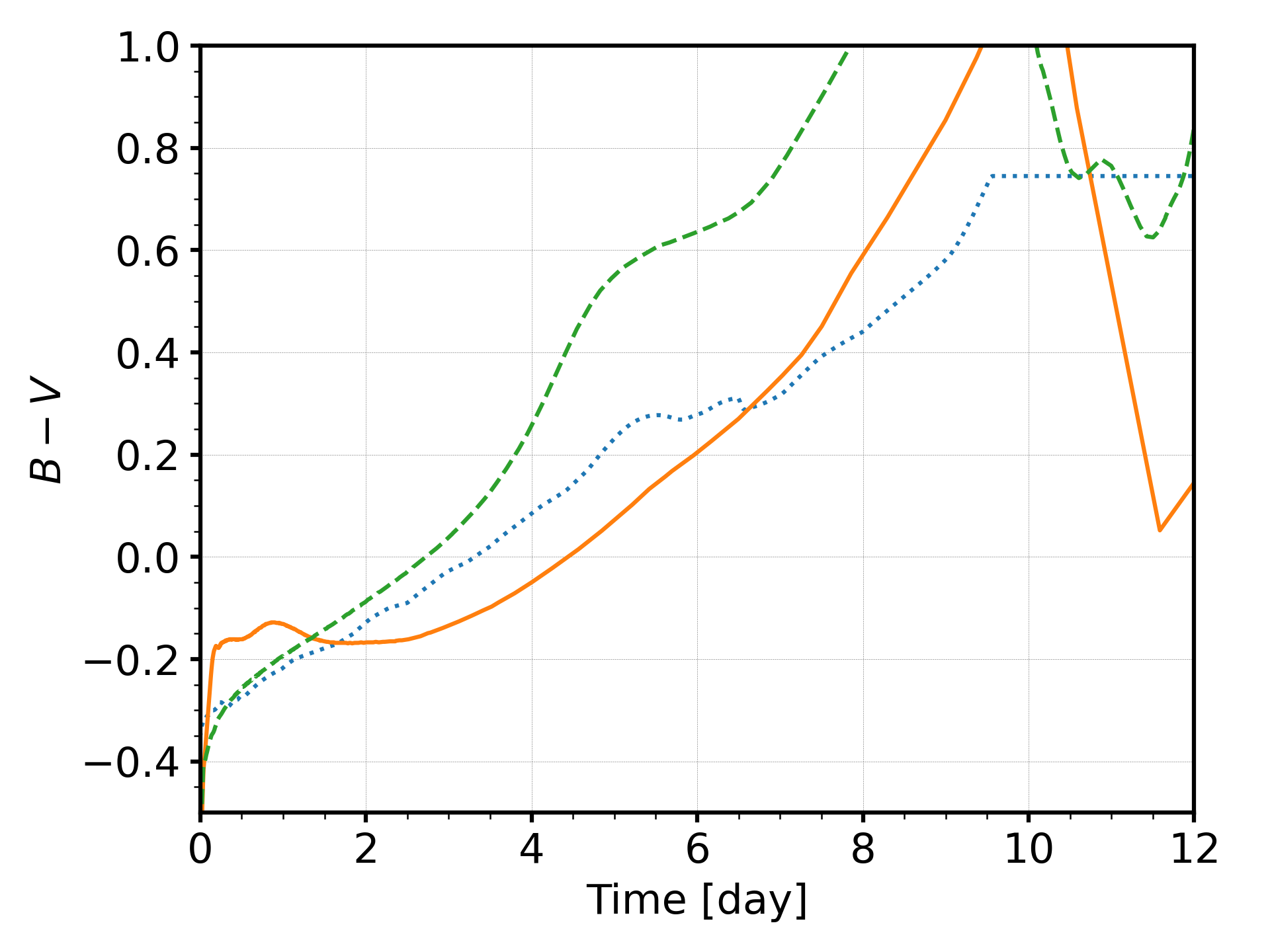}{0.47\textwidth}{}
                  }
        \caption{Same with Figure \ref{fig:p200snecresults}, but for the Sm11p400 progenitor model. The core mass cut is $1.48$\msun, and the ejecta kinetic energy is $2.32$\bethe.
        \label{fig:p400snecresults}}
        \end{figure*}

\begin{figure*}
        \gridline{\fig{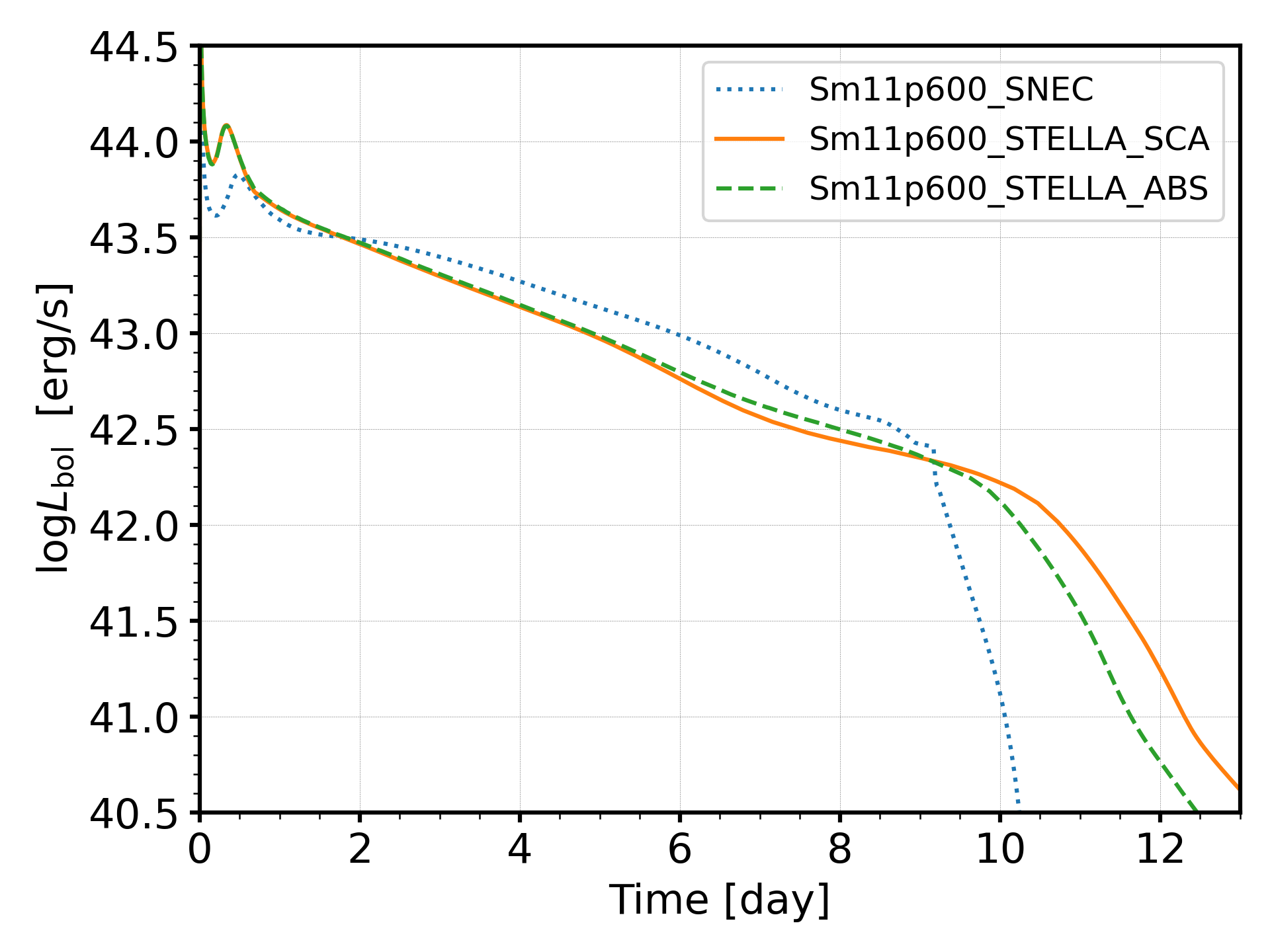}{0.47\textwidth}{}
                  \fig{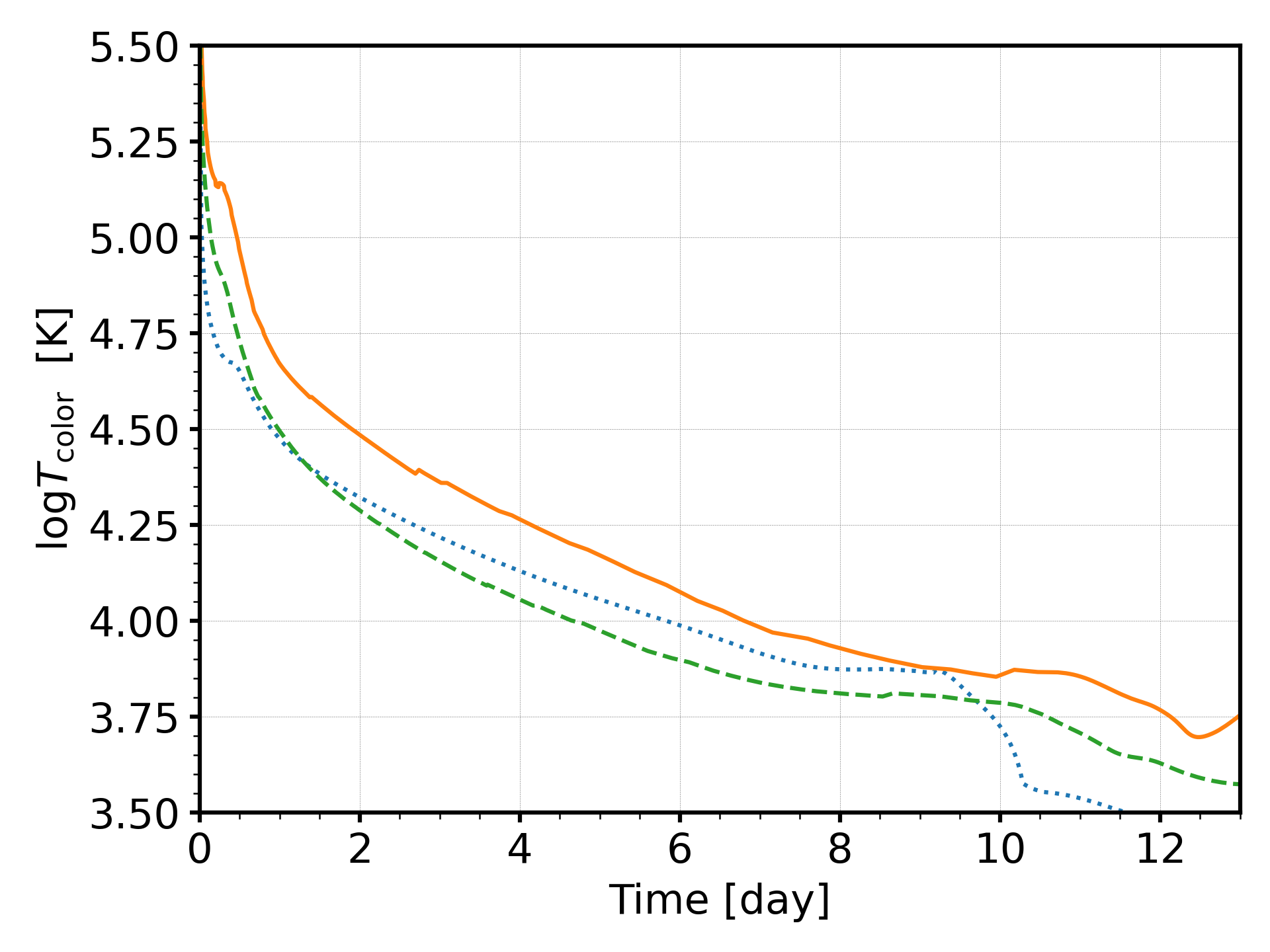}{0.47\textwidth}{}
                  }
        \gridline{\fig{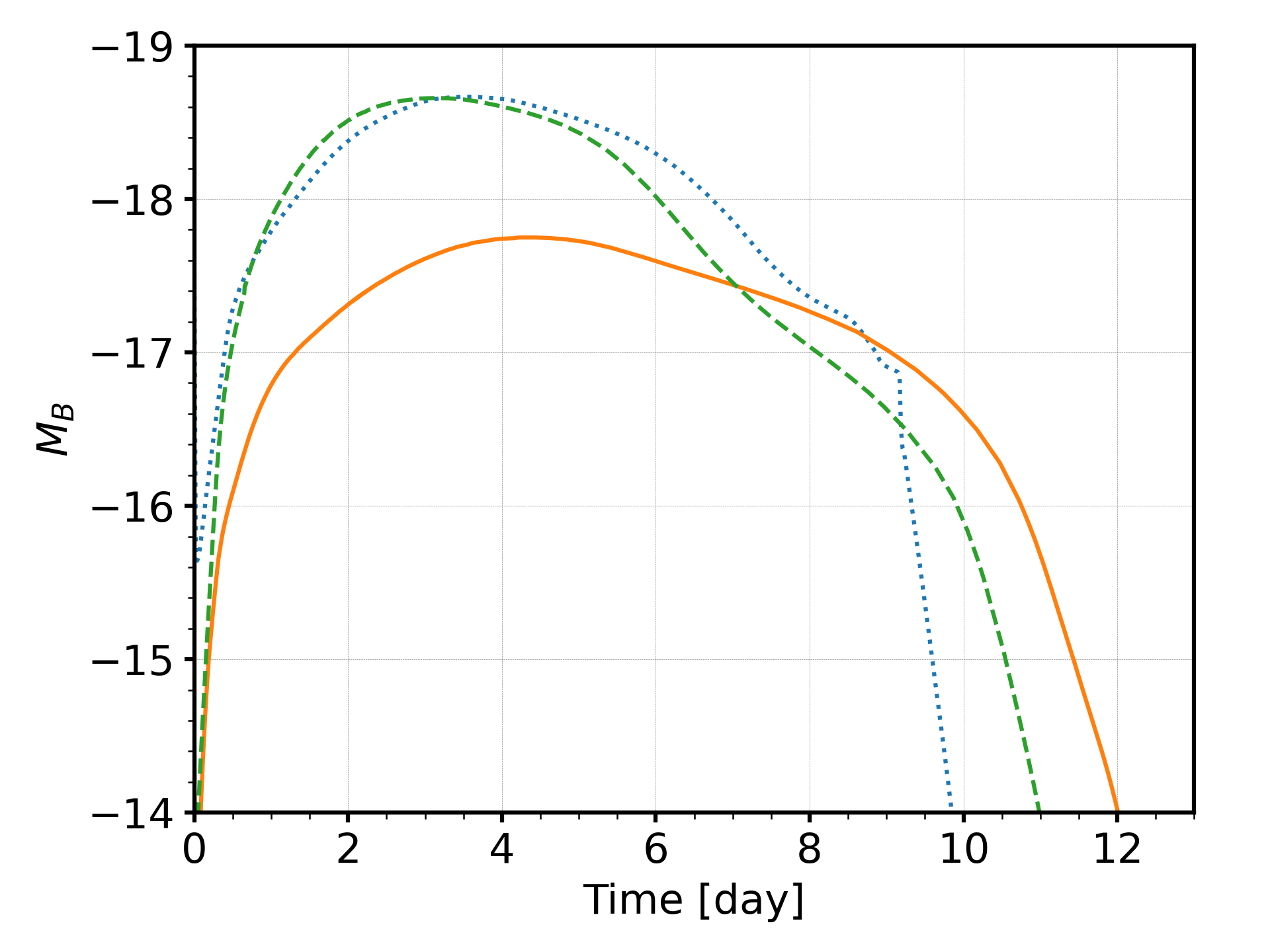}{0.47\textwidth}{}
                  \fig{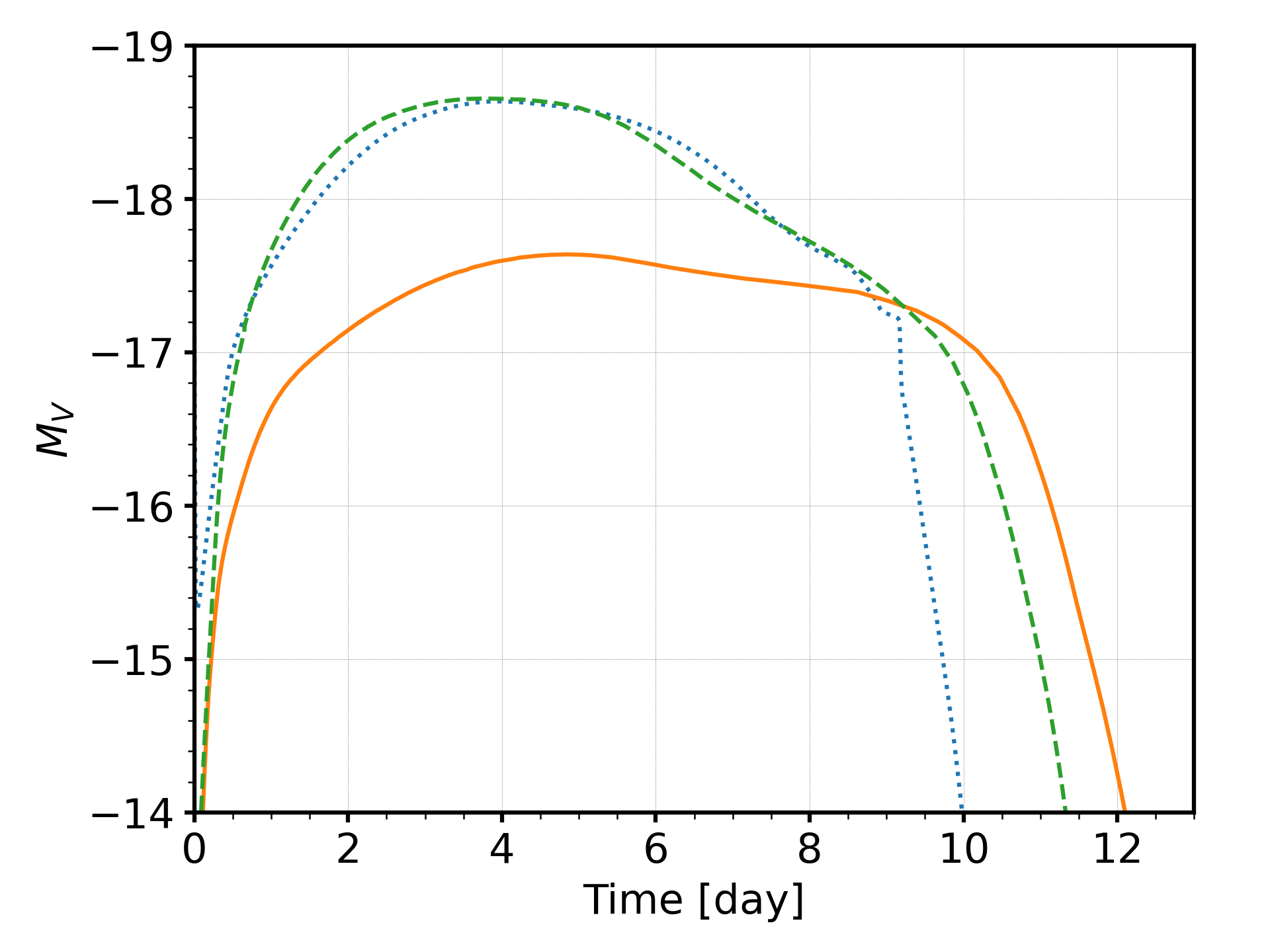}{0.47\textwidth}{}
                  }
        \gridline{\fig{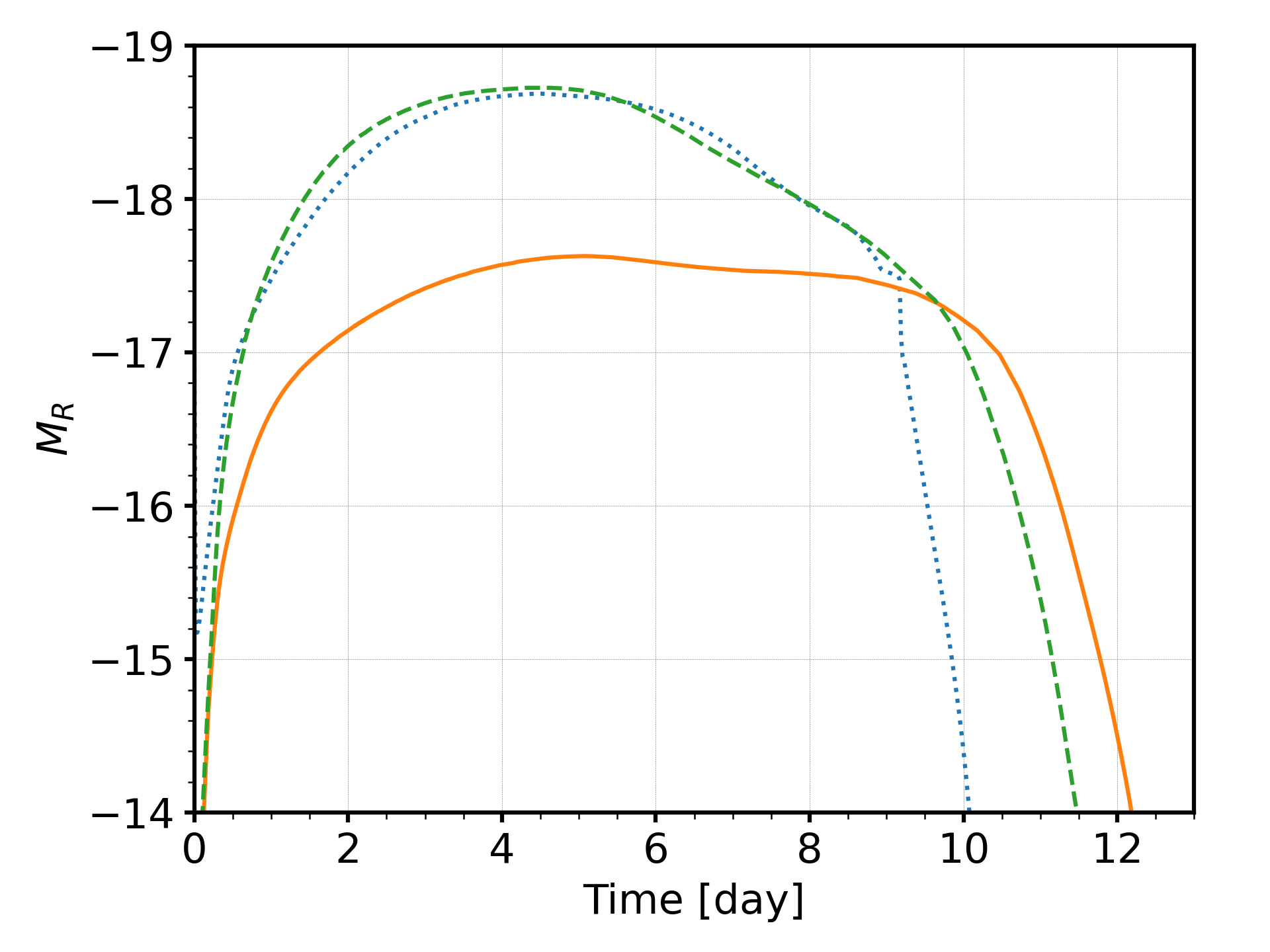}{0.47\textwidth}{}
                  \fig{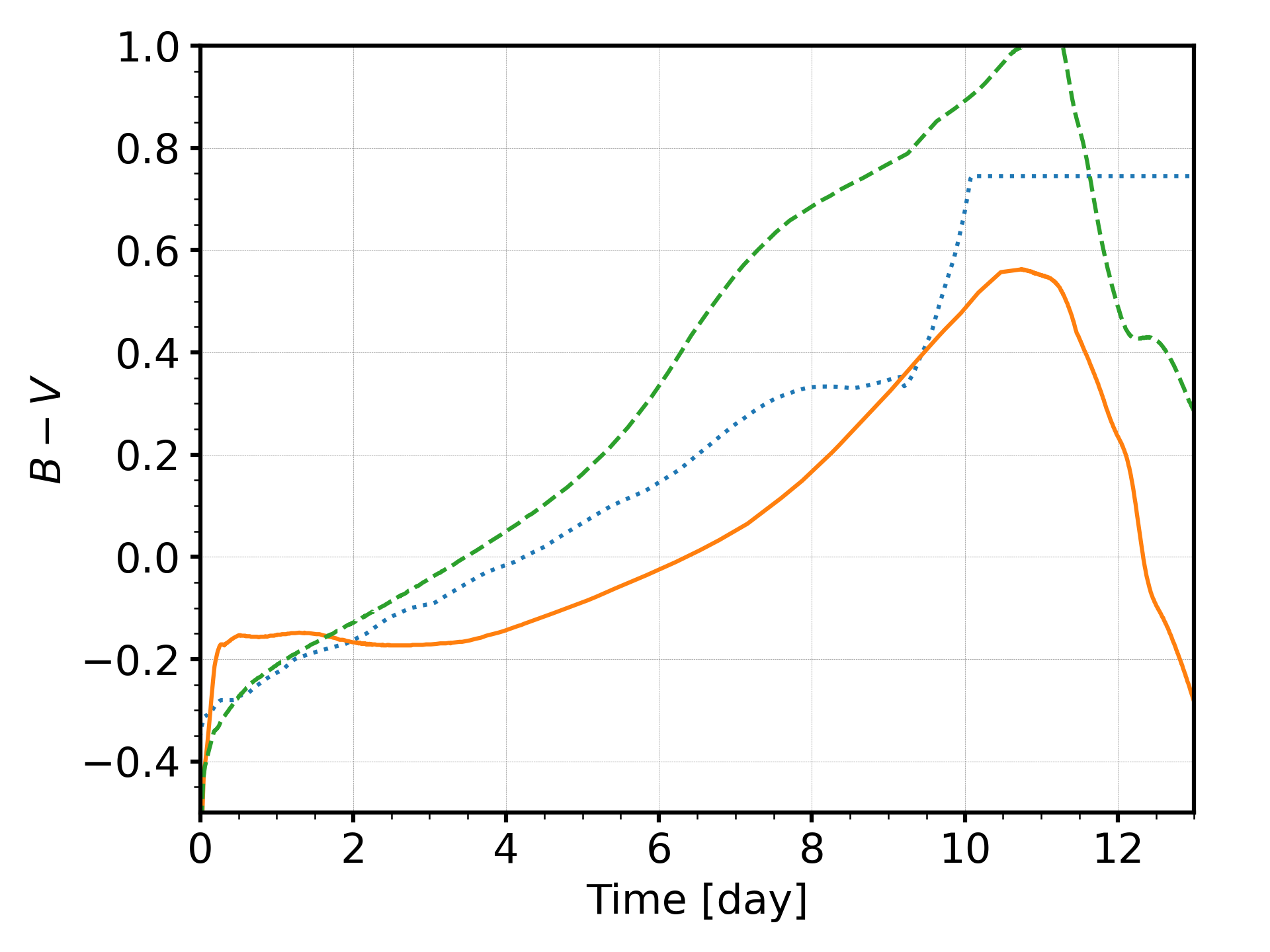}{0.47\textwidth}{}
                  }
        \caption{Same with Figure \ref{fig:p200snecresults}, but for the Sm11p600 progenitor model. The core mass cut is $1.41$\msun, and the ejecta kinetic energy is $2.14$\bethe.
        \label{fig:p600snecresults}}
        \end{figure*}

        In Figures \ref{fig:p200snecresults}, \ref{fig:p400snecresults}, and \ref{fig:p600snecresults}, we compare the STELLA\_SCA models with the corresponding STELLA\_ABS and SNEC models. For a given progenitor model and a SN energy, the evolution of the bolometric luminosities of the STELLA\_ABS models is similar to their STELLA\_SCA counterparts. 

        The plateau-like feature in the \snec{} models, on the other hand, appears significantly weaker when compared to the \stella models. This would be due to different treatments of opacity and radiative transfer. While the \snec{} code employs the opacity floor to account for line opacities in a rapidly expanding material, \stella{} uses an approximation method described in \citet{Friend83} and \citet{Eastman93} \citep[See also][]{Castor04}. For the radiative transfer, \stella{} solves multi-group radiative transfer equations while \snec{} solves Lagrangian hydrodynamics equations with the luminosity defined by radiative diffusion using the Rosseland-mean opacity. 

        In the optical light curves, we find more significant differences in the results from different methods compared to the bolometric luminosities. The $B$-, $V$-, and $R$-band peaks of STELLA\_SCA models are dimmer by more than a factor of 3 than those of the STELLA\_ABS and SNEC models. During the initial decline phase, the difference in their bolometric luminosities is small, so the $V$ magnitude difference should be attributed to different SEDs. The STELLA\_ABS models share the existence of the semi-plateau with the STELLA\_SCA models as can be seen in the top left panels of Figures \ref{fig:p200snecresults}-\ref{fig:p600snecresults}. Optical light curves of the SNEC models have a semi-plateau weaker and shorter than that of the STELLA models, analogous to their bolometric counterparts.
   
    The color temperatures of the STELLA\_SCA and STELLA\_ABS models are determined by the temperature that gives the best black-body fit to the SED. On the other hand, \snec{} defines the color temperature by the effective temperature at the Rosseland-mean photosphere.  As can be seen in Figures \ref{fig:p200snecresults}-\ref{fig:p600snecresults}, the STELLA\_SCA models give systematically higher $T_\mathrm{color}$ than of STELLA\_ABS and SNEC models during the first phase of $L_{\mathrm{bol}}$ where the $V$-band peak is reached. While the numerical solutions of radiative transfer equations of \stella{} do not give perfect black-body spectra at early times, the radiation during early times is dominated by continuum emission with little line features affecting the optical brightness because the ejecta temperature is well above 10000 K, and the black-body spectra of $T_{\mathrm{color}}$ can give a relatively good approximation for the SN spectra in the optical \citep{Dessart18}. Therefore, the different optical peaks in Figures \ref{fig:p200snecresults}-\ref{fig:p600snecresults} originate from different color temperatures, which is critically affected by Thomson scattering. 

    If the  matter and radiation at the photosphere were fully thermalized, the color temperature would be identical to the gas and effective temperatures at the photosphere. However, the radiation diluted from the hot and dense inner thermalized region may not be fully thermalized with the matter in the sparse outermost layer if free electrons are abundant. This is the case even if the outermost layer is optically thick due to Thomson scattering,  given that Thomson scattering does not lead to energy exchange between radiation and matter \citep[e.g.][]{Mihalas84,Baschek91,Ensman92,Blinnikov00}. In this case the color temperature of the spectrum would be significantly higher than the effective temperature defined by the Rosseland-mean photosphere. \citet{Blinnikov00} demonstrated this effect for SN 1987A-like supernovae (e.g., see Figures 7 and 9 of their paper). In our models, the STELLA\_SCA models are considerably fainter in the optical than STELLA\_ABS and SNEC models in the early time, which is analogous to  the results presented in Figure 13 of \citet{Blinnikov00}.\footnote{
    \citet{Blinnikov11} show that 1-temperature fits are not sufficient and at least 2-temperature fits are needed for describing emerging spectra near shock breakout. But even in the case of the 2-temperature-fit, both fitting temperatures are found to be higher when scattering is important in comparison with the pure absorption case.} 
    
	\citet{Bersten12} and \citet{Bersten18} also note that Thomson scattering plays an important role in radiative transfer in H-rich SN ejecta at early times. Their numerical method is based on RHD equations of \citet{Bersten11} which is similar to that of \snec, but instead of employing the effective temperature as the black-body temperature to obtain optical light curves, they follow the method of \citet{Ensman92} to approximately determine the thermalization temperature $T_{\mathrm{thm}}$. On a side note, the definition of the thermalization depth by \citet{Bersten12} is the location where $3\tau_{\mathrm{abs}}\tau_{\mathrm{sct}}\approx1$, whereas \citet{Ensman92} define it to be the location where $\sqrt{3\tau_{\mathrm{tot}}\tau_{\mathrm{sct}}}=2/3$. Here, $\tau_{\mathrm{tot}}$, $\tau_{\mathrm{abs}}$, and $\tau_{\mathrm{sct}}$ are respectively optical depths from the total opacity, pure absorption, and Thomson scattering. Another way of approximating the thermalization depth is presented in \citet{Baschek91} as $\tau_{th} = \int_R^{R_{\mathrm{th}}} \! \sqrt{3\kappa_{\mathrm{abs}}\kappa_{\mathrm{tot}}}\rho \, \mathrm{d}r = 2/3$, where $R_{\mathrm{th}}$ is the location of the thermalization depth, $\kappa_{\mathrm{abs}}$ is the absorptive opacity, and $\kappa_{\mathrm{tot}}$ is the total opacity. Here, $\kappa_{\mathrm{abs}}$ is defined by $\kappa_{\mathrm{abs}}=\kappa_{\mathrm{tot}}-\kappa_{\mathrm{sct}}$ where $\kappa_{\mathrm{sct}}$ is the Thomson scattering opacity. 

\begin{figure}
        \gridline{\fig{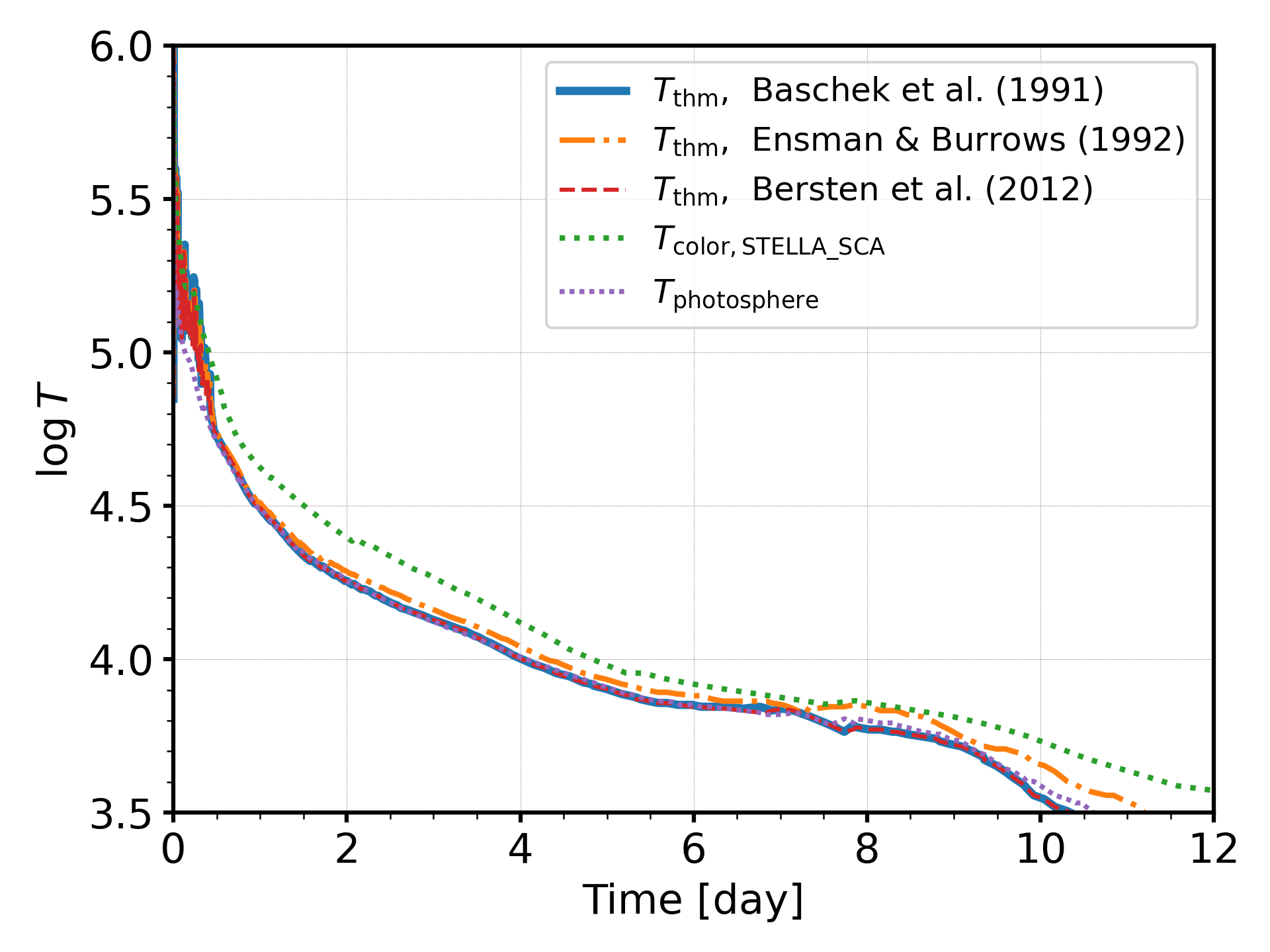}{0.47\textwidth}{}}
        \gridline{\fig{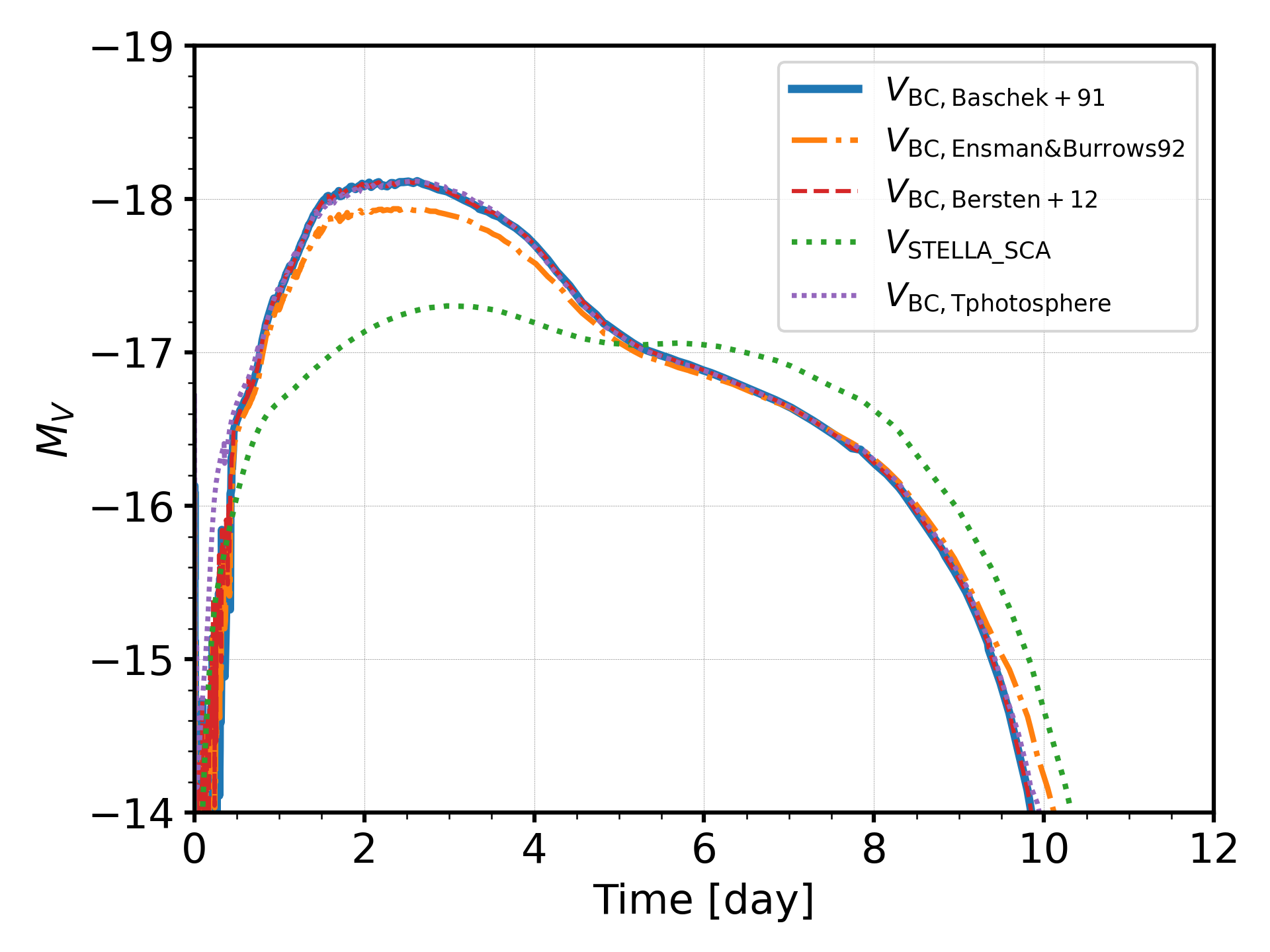}{0.47\textwidth}{}}
        \caption{Upper panel: Color temperature ($T_{\mathrm{color}}$) of the Sm11p400\_STELLA\_SCA model (dotted green) is compared with approximations of the thermalization temperature ($T_{\mathrm{thm}}$) obtained with the following methods: \citet[][Blue solid]{Baschek91}, \citet[][Orange dash-dotted]{Ensman92}, and \citet[][Red dashed]{Bersten12}. The purple dotted line denotes the temperatre at the Rosseland-mean photosphere. Lower panel: The $V$-band light curve of the Sm11p400\_STELLA\_SCA model is compared with the $V$-band light curves obtained by assuming the SEDs to be black-body spectra of $T_{\mathrm{thm}}$s and $T_\mathrm{photosphere}$. 
        \label{fig:ThermTest}}
        \end{figure}

        In Figure \ref{fig:ThermTest}, we present the approximated thermalization temperatures ($T_{\mathrm{thm}}$) obtained with these three different methods using the ejecta information of the Sm11p400\_STELLA\_SCA model and the corresponding $V$-band light curves generated from the black-body spectra of these approximated $T_{\mathrm{thm}}$ values. For comparison, we also show $T_{\mathrm{color}}$ and the $V$-band light curve of the Sm11p400\_STELLA\_SCA model. We find that the results from the analytical approximations and the full STELLA\_SCA model diverge from $t\simeq0.7$ d. The methods of \citet{Baschek91} and \citet{Bersten12} result in almost identical temperatures, while $T_{\mathrm{color}}$ from \citet{Ensman92} is slightly higher than the others. However, all of these analytical approximations give $T_{\mathrm{thm}}$ much lower than $T_{\mathrm{color}}$ of the STELLA\_SCA model, resulting in a significantly brighter optical peak.
        
\begin{figure}
            \includegraphics[width=\linewidth]{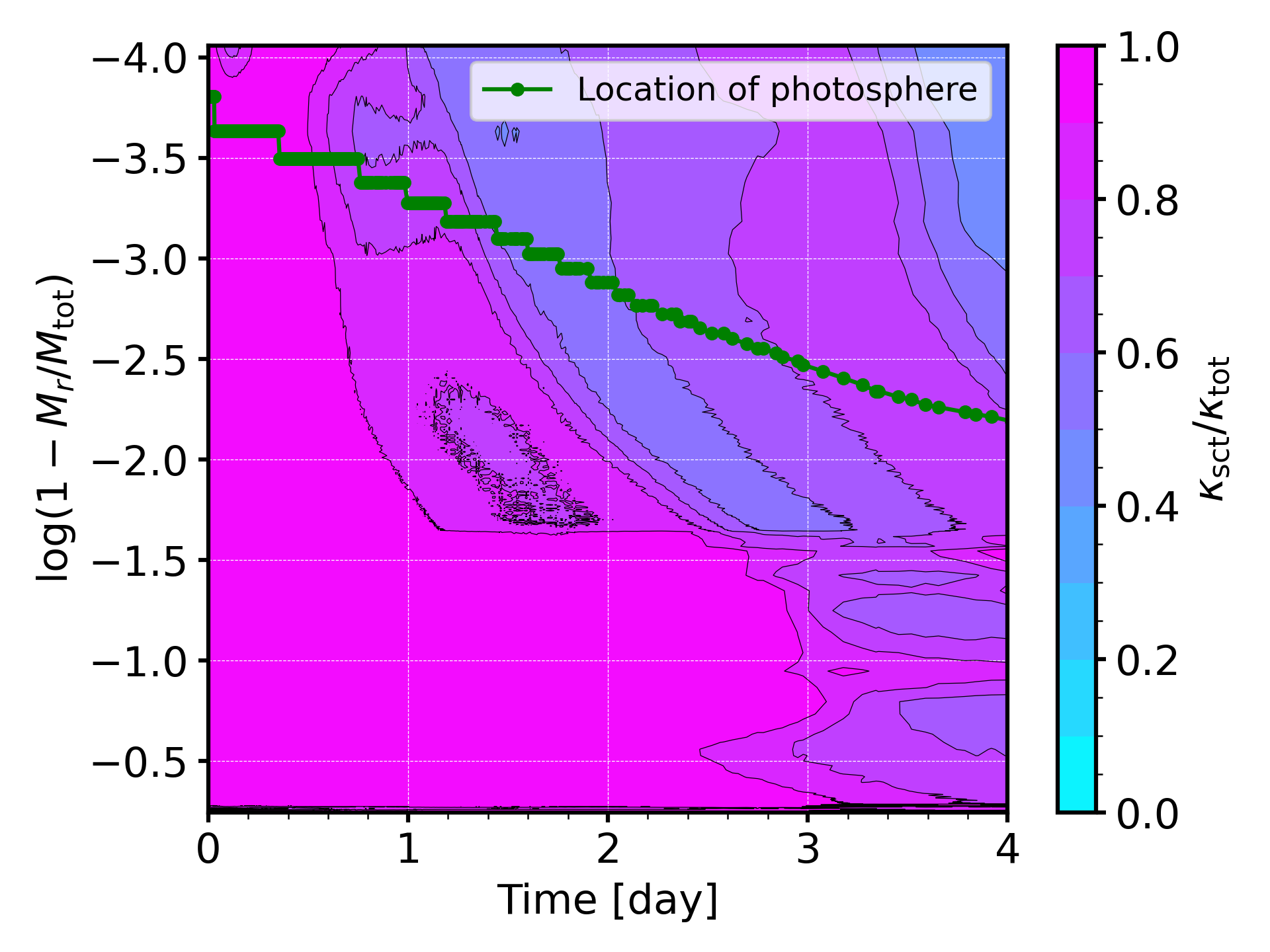}
            \caption{Internal profile of the ratio of the Thomson scattering opacity ($\kappa_{\mathrm{sct}}$) over the total opacity ($\kappa_{\mathrm{tot}}$), with the location of the Rosseland-mean photosphere displayed in red solid line.}
            \label{fig:p400_kskt}
        \end{figure}

        Note that both the methods of \citet{Baschek91} and \citet{Ensman92} are derived under the condition of $\kappa_{\mathrm{sct}}\approx\kappa_{\mathrm{tot}}\gg\kappa_{\mathrm{abs}}$. In Figure \ref{fig:ThermTest}, we indeed observe that $T_{\mathrm{thm}}$ and $T_{\mathrm{color}}$ from \stella{} match fairly well when $T>10^5$ K. However, the two begin to diverge as the ejecta cools down and the contribution of the absorption opacity to the total opacity becomes non-negligible. Figure \ref{fig:p400_kskt} shows the temporal evolution of the contribution of electron scattering to the total opacity ($\kappa_{\mathrm{sct}}/\kappa_{\mathrm{tot}}$) within the ejecta of the Sm11p400\_STELLA\_SCA model. The opacity is dominantly from Thomson scattering at $t\simeq0.7$ d, but the contribution of the absorption opacity is larger than 20\% at $t\approx3$ d when the $V$-band peak is reached. 

        In summary, we find that numerical models of the optical first peak of SNe IIb are crucially affected by the inclusion of the effects of Thomson scattering. Black-body spectra constructed from the effective temperature as assumed in the \snec{} code \citep{Morozova15,Piro17,Eldridge18} is an incomplete representation of radiative transfer inside ejecta where the non-thermalizing nature of Thomson scattering is relevant with its high contribution to the total opacity making the radiation temperature and the gas temperature significantly deviate from each other at the photosphere. We also conclude that approximating the thermalization depth as in \citet{Bersten12} and \citet{Bersten18} could still lead to a substantial overprediction (i.e., by more than 1 mag) of the optical brightness of the first peak of SNe IIb.

	\begin{deluxetable*}{lcccccc}
        \centering
        \tablecaption{Numerical explosion setups for the models for the study on the effects of chemical mixing  \label{tab:chemmodeltable}}
        \tablehead{
        \colhead{} & \multicolumn{3}{c}{Numerical Codes} & \colhead{} & \colhead{} & \colhead{} \\ \cmidrule{2-4}
        \colhead{Model Name} & \colhead{ZAMS $\rightarrow$ CC} & \colhead{CC $\rightarrow$ SBO} & \colhead{SBO $\rightarrow$ End} & \colhead{$M_{\mathrm{cut}}$} & \colhead{$E_{\mathrm{kin}}$} & 	\colhead{  Mixing  }
        } 
        \startdata
        Tm11p200\_STELLA\_SCA & \mesa & \stella & \stella & 1.40 & 1.20 & No \\
        Tm11p200\_noRTI & \mesa & \mesa & \stella & 1.54 & 1.19 & No \\
        Tm11p200\_mix1 & \mesa & \mesa & \stella & 1.57 & 1.19 & RTI-induced only \\
        Tm11p200\_mix2 & \mesa & \mesa & \stella & 1.57 & 1.19 & RTI + weak mixing \\
        Tm11p200\_mix3 & \mesa & \mesa & \stella & 1.57 & 1.19 & RTI + strong mixing   \\
        \hline
        Sm11p400\_STELLA\_SCA & \mesa & \stella & \stella & 1.48 & 1.83 & No \\
        Sm11p400\_noRTI & \mesa & \mesa & \stella & 1.44 & 1.83 & No \\
        Sm11p400\_mix1 & \mesa & \mesa & \stella & 1.44 & 1.83 & RTI-induced only \\
        Sm11p400\_mix2 & \mesa & \mesa & \stella & 1.44 & 1.83 & RTI + weak mixing \\
        Sm11p400\_mix3 & \mesa & \mesa & \stella & 1.44 & 1.83 & RTI + strong mixing   \\
        \enddata
        \tablecomments{ZAMS: Zero-age Main Sequence, CC: Core-collapse, SBO: Shock Breakout, $M_{\mathrm{cut}}$: excised core mass in units of \msun, $E_{\mathrm{kin}}$: final kinetic energy of the ejecta in units of $10^{51}$ ergs, Mixing: Information on the prescriptions of chemical mixing in the models.}
        \end{deluxetable*} 

\section{The effects of chemical mixing} \label{sec:chem}

    One of the remarkable characteristics observed in our SN IIb model predictions using \stella{} is the presence of a semi-plateau phase at early times (Figure \ref{fig:Henv_analysis}, \ref{fig:p400_energy_lcs}, \ref{fig:p200snecresults}, \ref{fig:p400snecresults} \& \ref{fig:p600snecresults}). However, such a short-duration plateau is not clearly found in the observations (Figure \ref{fig:observed}), suggesting that there may still be a crucial component missing in our simulations. In this regard, we address another potentially important factor that can determine the early-time light curves of SNe IIb: chemical mixing in the SN ejecta.

\begin{figure}
        \includegraphics[width=\linewidth]{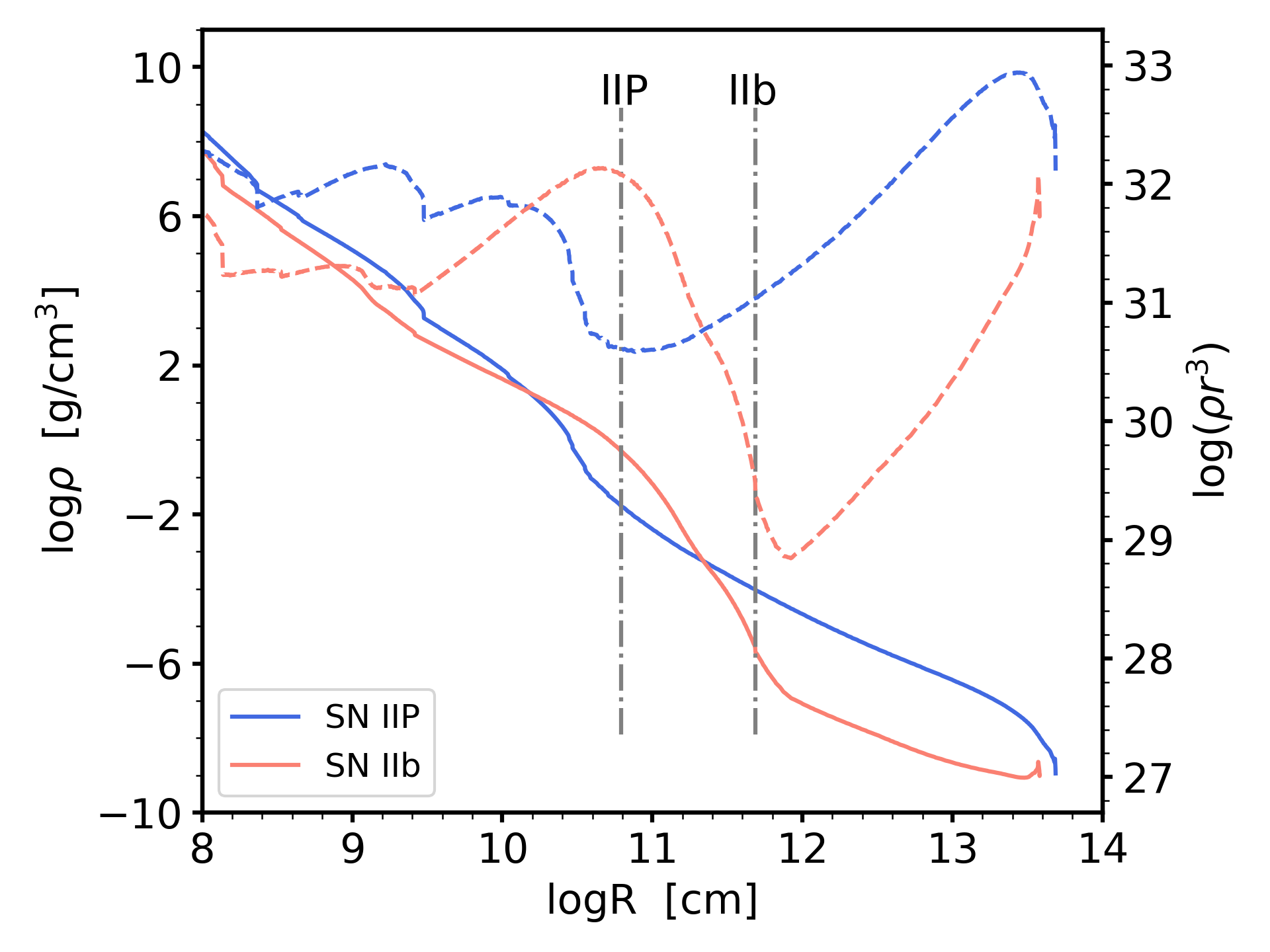}
        \caption{Density profiles of typical SN IIP (blue) and SN IIb (red) progenitors are shown in solid lines. Dashed lines are $\rho r^{3}$ profiles that determine whether the shock would accelerate or decelerate. Vertical dash-dotted lines indicate the boundary between the He core and the H-rich envelope ($X_{\mathrm{H}}=0.5$). Details on the models are provided in the text.}
        \label{fig:rhor3}
    \end{figure}

    It is commonly found in SN simulations with a cool supergiant progenitor having a H-rich envelope that the shock front accelerates when it traverses the He core where the density decreases rapidly, and decelerates when it moves into the H-rich envelope. A reverse shock is created as a result to form a dense shell, making this layer susceptible to the RTI \citep[e.g.,][]{Woosley95,Kifonidis03,Utrobin21}. In Figure \ref{fig:rhor3}, we show density profiles of typical SN IIP and SN IIb progenitors in a similar manner to Figure 8 of \citet{Kifonidis03}. The SN IIb progenitor is the Sm11p600 model, whose physical properties are presented in Table \ref{tab:presnparams}. The SN IIP progenitor model was evolved until the pre-SN stage from a 16\msun ZAMS mass single star, with its final mass, final radius, pre-SN H-rich envelope mass being 12.2\msun, 699.0\rsun, and 8.00\msun, respectively. Both progenitor models are RSGs.  


    The SN IIb progenitor has a less massive H-rich envelope and the density of the H-rich envelope is lower than that of the SN IIP progenitor. This leads to a more pronounced $\rho r^{3}$ gradient change near the He/H interface, shown in Figure \ref{fig:rhor3} with the dashed lines, in the SN IIb model than the SN IIP model. This implies the possibility of a fairly strong reverse shock, which may result in the RTI and a consequent chemical mixing. It is also possible that a chemical mixing between the He core and the H-rich envelope can be induced by asymmetric explosion. Therefore, we explore how the shock-cooling emission from the SN IIb progenitor envelope depends on various degrees of chemical mixing in the H-rich envelope.

    \subsection{Setup} \label{sec:chem-setup}
        
In order to incorporate the effects of the RTI within the 1D framework following \citet{Paxton18}, we use the \mesa{} code to make explosion models with our SN IIb progenitor models. We use the Tm11p200 and Sm11p400 progenitor models and produce five SNe models for each progenitor with the following suffixes: `\_STELLA\_SCA', `\_noRTI', `\_mix1', `\_mix2', and `\_mix3'. The STELLA\_SCA model is previously introduced in Section \ref{sec:snec}. For the noRTI model, the shock propagation is computed with \mesa{} until the forward shock closely reaches the surface and then handed over to \stella{}. But the RTI effects are not considered during the \mesa{} calculations. The SCA and noRTI models are essentially the same, except that the explosion and the resulting shock propagation until shock breakout is calculated with \stella{} and \mesa{}, respectively. The chemical profiles of STELLA\_SCA models are weakly mixed as mentioned in Section~\ref{sec:snec}, but the difference between the SCA and noRTI models are small and the stratified chemical structure is essentially preserved.

The mix1 models are produced following the method of \citet{Paxton18}, simulating the shock propagation including the RTI effects with \mesa{}.  The calculations with \mesa{} are stopped just before the forward shock reaches the surface of the ejecta and the model was transferred to \stella{} to calculate light curves. A limitation of this approach is that internal chemical mixing due to the RTI is only accounted for until the shock breakout because it is beyond the capacity of the current version of \stella{} to incorporate the effects of the RTI on the hydrodynamical and chemical structures of the ejecta.  For SNe IIP progenitors with more massive and dense H-rich envelopes, e.g.  $M_{\mathrm{env}}\simeq10$\msun, the forward shock takes about more than a day to reach the surface, during which the reverse shock generated at the core-envelope boundary can propagate into a sufficiently deep layer in the ejecta until the shock breakout such that further propagation of the reverse shock after the shock breakout would not significantly affect the resulting light curve. However, as shown in Figure 46 of \citet{Paxton18}, the forward shock passes through the heavily stripped H-rich envelopes of SNe IIb progenitors much faster than the case of SN IIP and the on-going reverse shock propagation into the deeper layers after the shock breakout is likely to have an impact on the light curve, which cannot be properly described with \stella.  For example, the forward shock takes $t\simeq0.2$ d to reach the surface in our models with $M_{\mathrm{env}}\simeq0.2$\msun. Therefore, the inward propagation of the reverse shock and the accompanying RTI effects may only partially be included. As a way of compensating for this limitation, we increase the value of the dimensionless constant $B$ in equations (26) and (35) of \citet{Duffell16} from 2.5 to 3.0 while following the default values of \citet{Paxton18} for the others, making the density smoothing due to the RTI more effective.  In this way, the chemical structure of the ejecta is influenced by a RTI-induced mixing, albeit partially.

        \begin{figure}
          \includegraphics[width=\linewidth]{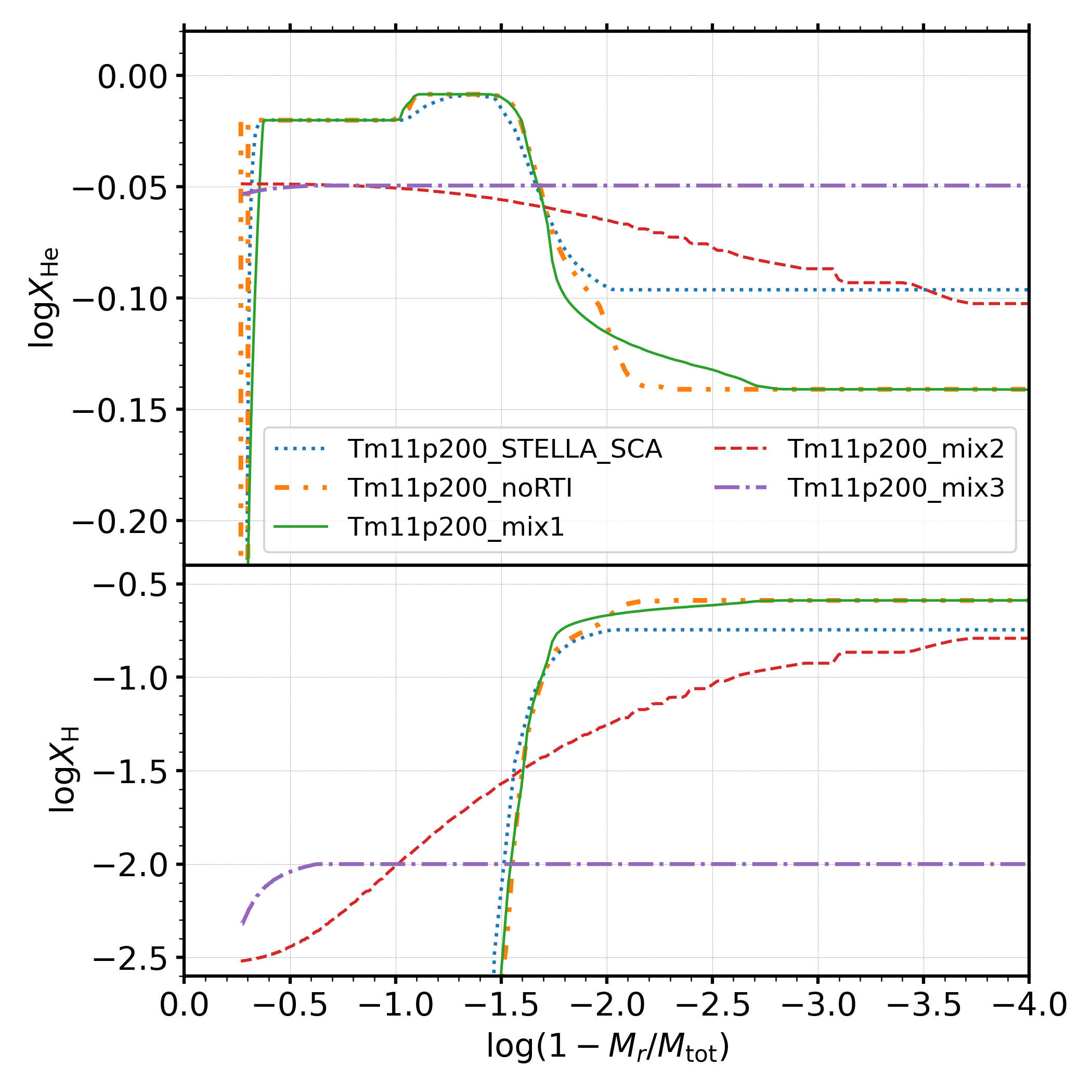}
          \caption{Initial chemical profiles of SNe models from the same progenitor Tm11p200. The STELLA\_SCA model is the same as Section \ref{sec:snec}, and the other models are computed with \mesa{} until the shock breakout then are handed over to \stella{} for the rest. The noRTI model is computed with \mesa{} without the RTI prescription. The mix1, mix2, and mix3 models are computed including the RTI prescription and different prescriptions of the chemical mixing. See the text for details.}
          \label{fig:p200_chem_mix}
        \end{figure}

        \begin{figure}
          \includegraphics[width=\linewidth]{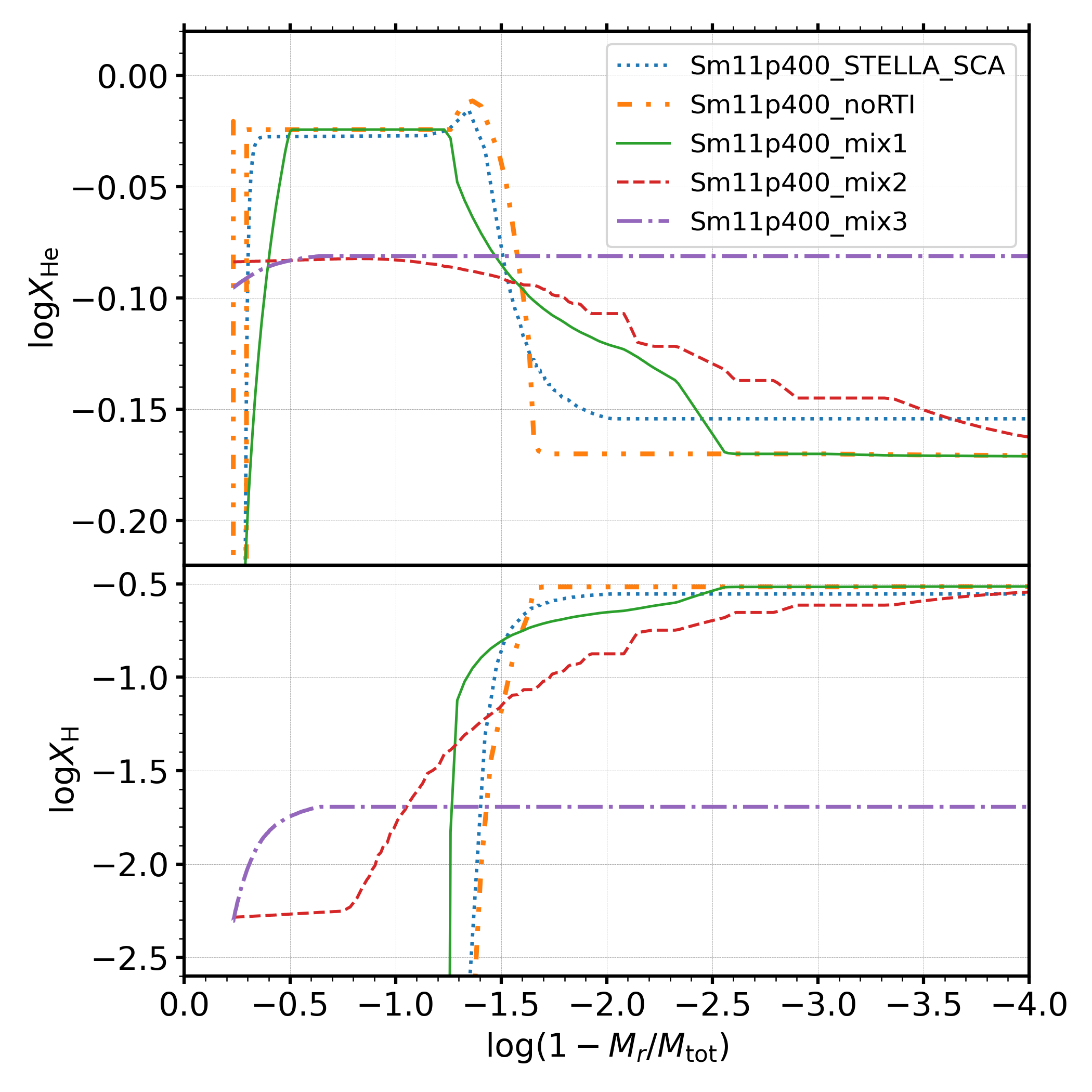}
          \caption{Same with Figure \ref{fig:p200_chem_mix}, but for the Sm11p400 progenitor model.}
          \label{fig:p400_chem_mix}
        \end{figure}              

We also construct models with stronger degrees of chemical mixing (mix2 and mix3 models) in order to account for the possible RTI-induced chemical mixing that might still occur after the shock breakout and take into account other possible causes of chemical mixing such as asymmetric explosions. For this purpose, we apply the boxcar smoothing method in the mix2 and mix3 models to impose additional chemical mixing at the onset of \stella calculations,  in addition to the mixing according to the RTI prescription applied to the mix1 models.  The degree of mixing of the mix2 model is relatively weaker than that of mix3. In Figures \ref{fig:p200_chem_mix} and \ref{fig:p400_chem_mix}, we show hydrogen and helium abundances inside the ejecta of the models having different degrees of chemical mixing at the time of the shock breakout. In Table \ref{tab:chemmodeltable}, we present numerical setups for our models including the final kinetic energy and excised core mass. See also Table \ref{tab:presnparams} for the pre-SN progenitor properties. 
        
        \begin{figure*}
        \gridline{\fig{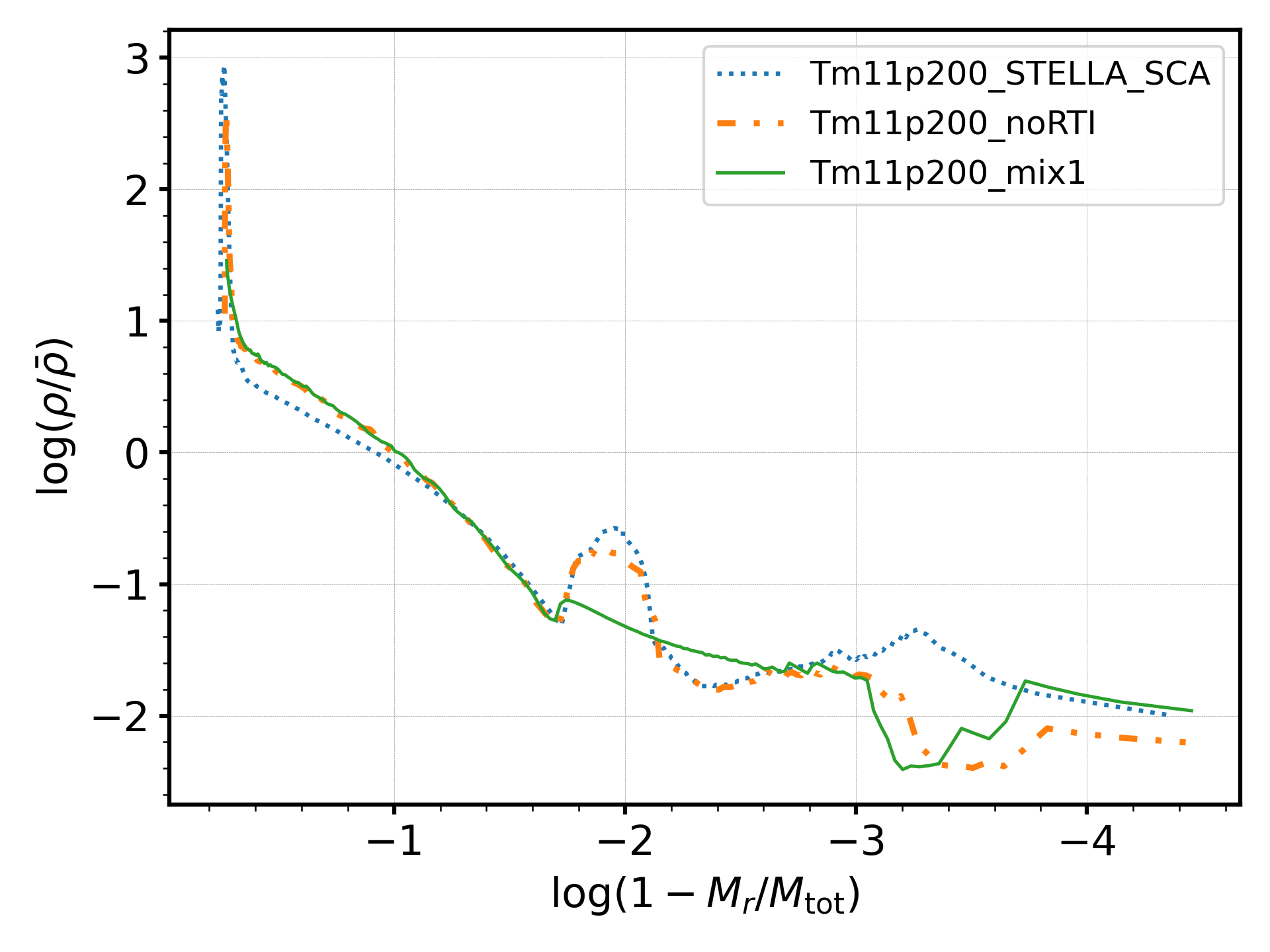}{0.47\textwidth}{}
                  \fig{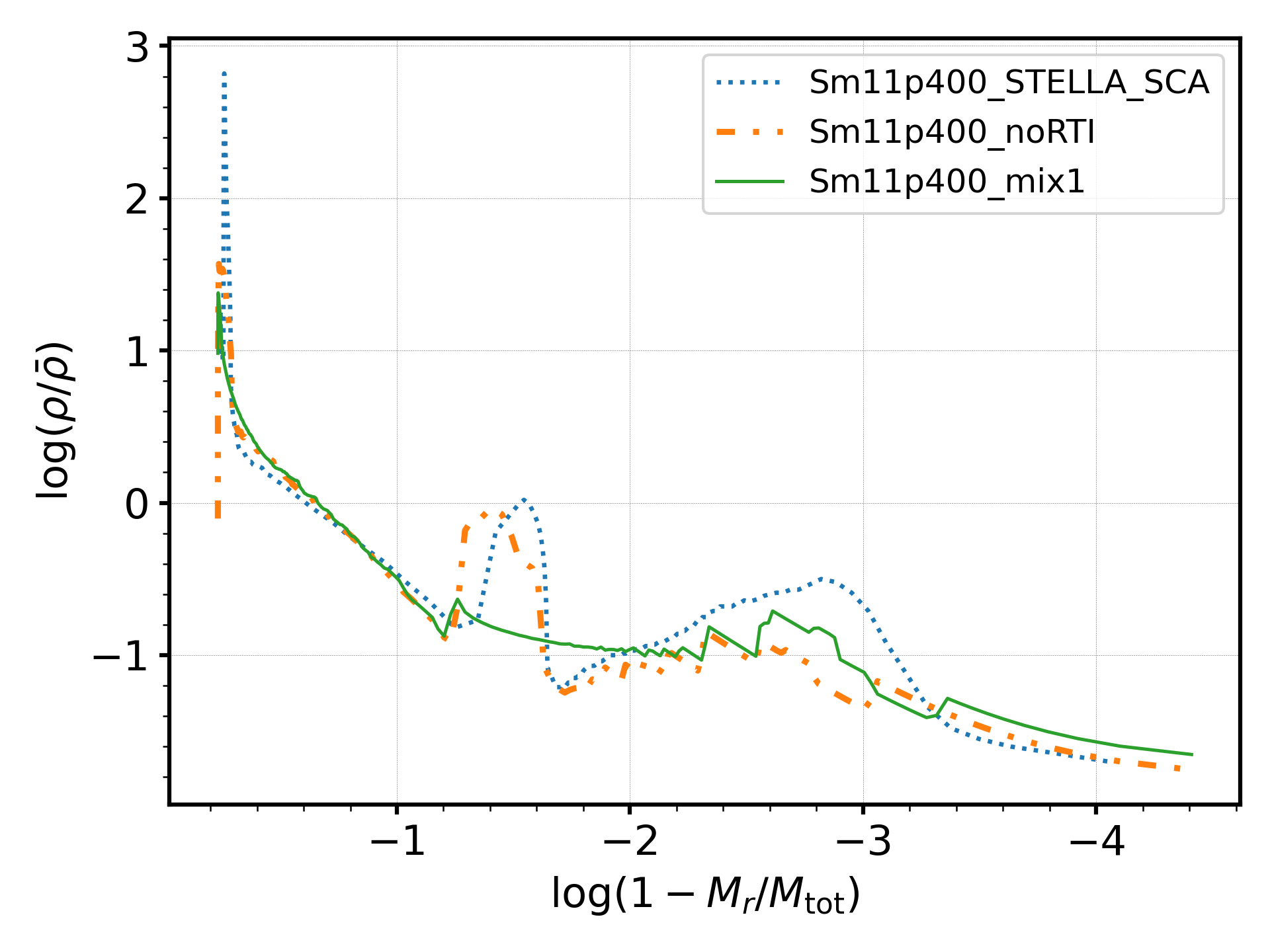}{0.47\textwidth}{}}
        \gridline{\fig{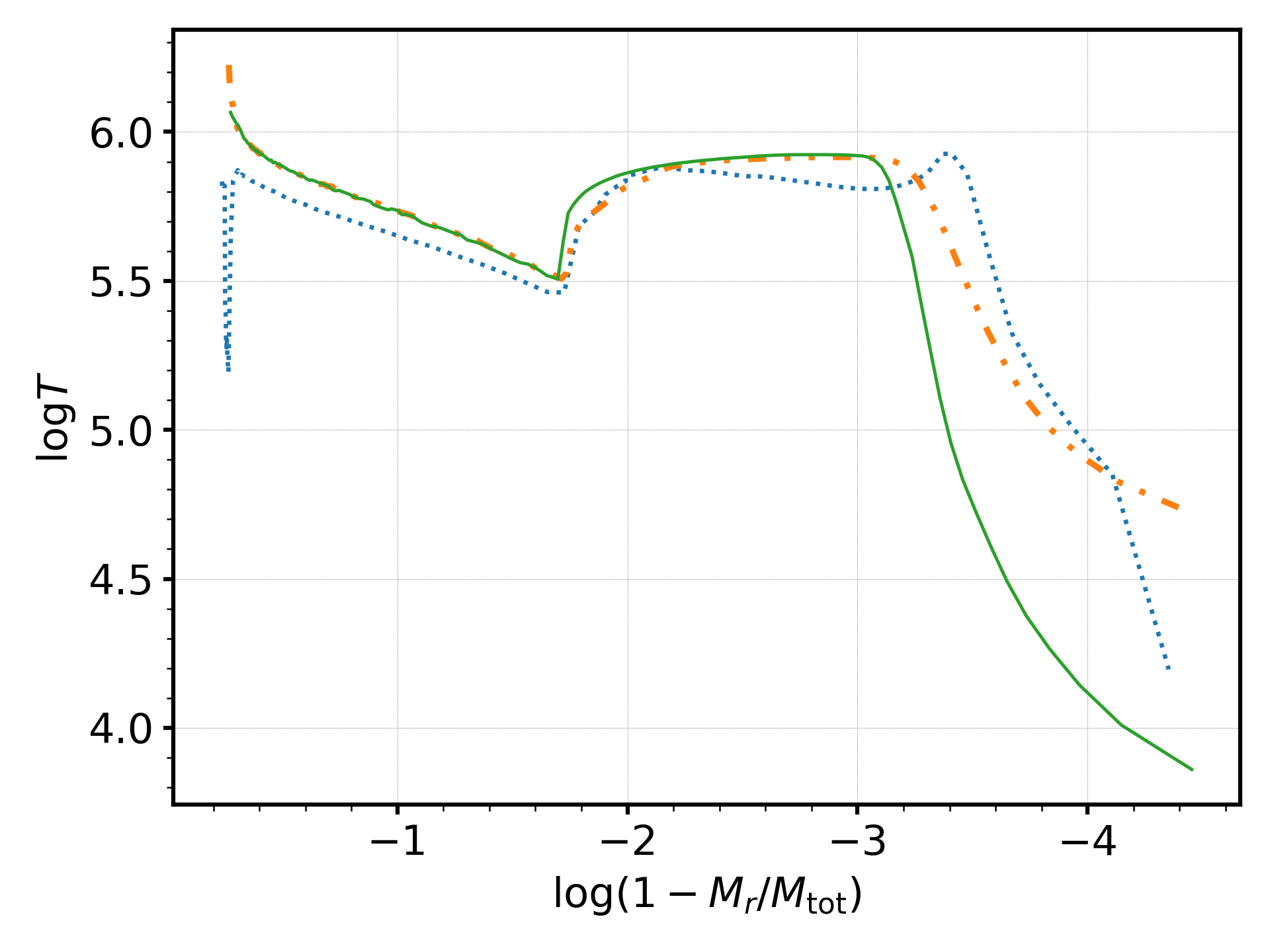}{0.47\textwidth}{}
                  \fig{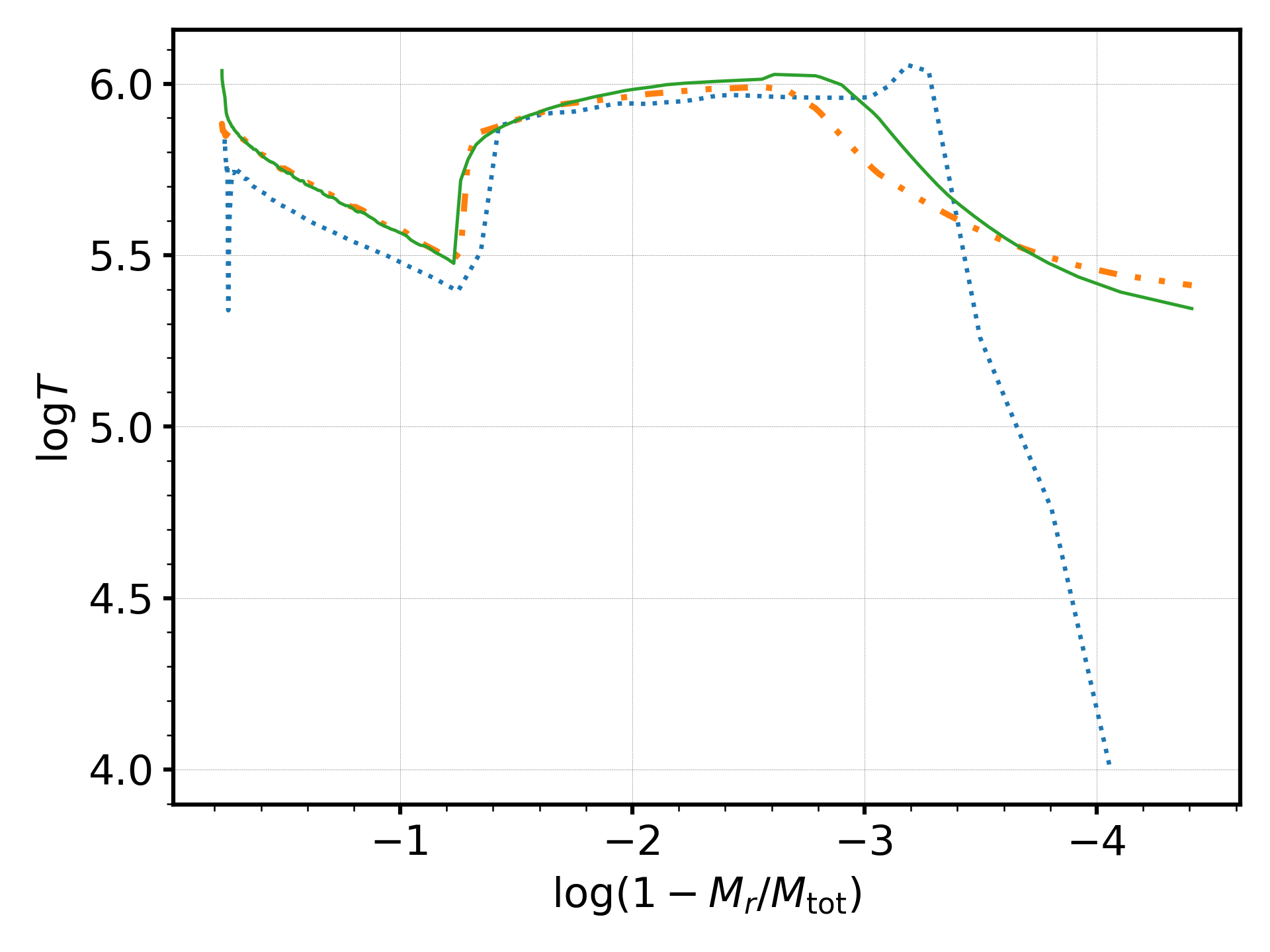}{0.47\textwidth}{}}
        \caption{Upper left: the density profiles of Tm11p200\_STELLA\_SCA (blue dotted), Tm11p200\_noRTI (orange dash-dotted), and Tm11p200\_mix1 (green solid) at the time of shock breakout. Y-axis is the density normalized by the average density of the ejecta ($\bar{\rho}=3M_{\mathrm{ej}}/4\pi R^{3}$), in a logarithmic scale. Upper right: same as the upper left panel but for their Sm11p400 counterparts instead of Tm11p200. Lower left: the temperature profiles of the Tm11p200 models. Lower right: the temperature profiles of the Sm11p400 models. 
        \label{fig:mixing_SBO_profile}}
        \end{figure*}    
        
        Note that the RTI also affects internal density and temperature profiles inside the ejecta as well as the chemical profiles, smoothing out the dense shell formed by the reverse shock. For example, Figure~29 of \citet{Paxton18} shows the density smoothing effect of the RTI when applied to a SN IIP. The upper left panel of Figure \ref{fig:mixing_SBO_profile} shows the density profiles of Tm11p200\_STELLA\_SCA, Tm11p200\_noRTI, and Tm11p200\_mix1 models just before the shock breakout. The upper right is the same but for their Sm11p400 counterparts. We find two dense shells respectively corresponding to the density jumps by the forward shock and the reverse shock in the STELLA\_SCA and noRTI models. On the other hand, these structures are smoothed out due to RTI-induced mixing in the mix1 model. The lower panels of Figure \ref{fig:mixing_SBO_profile} and \ref{fig:mixing_SBO_profile} show the temperature profiles at the same time. The influence of the RTI on the temperature structure is not so distinctive as on the density structure and the temperature structure can be more significantly affected by the adopted numerical codes (i.e., \mesa and \stella) that calculate the shock propagation.

        \begin{figure*}
        \gridline{\fig{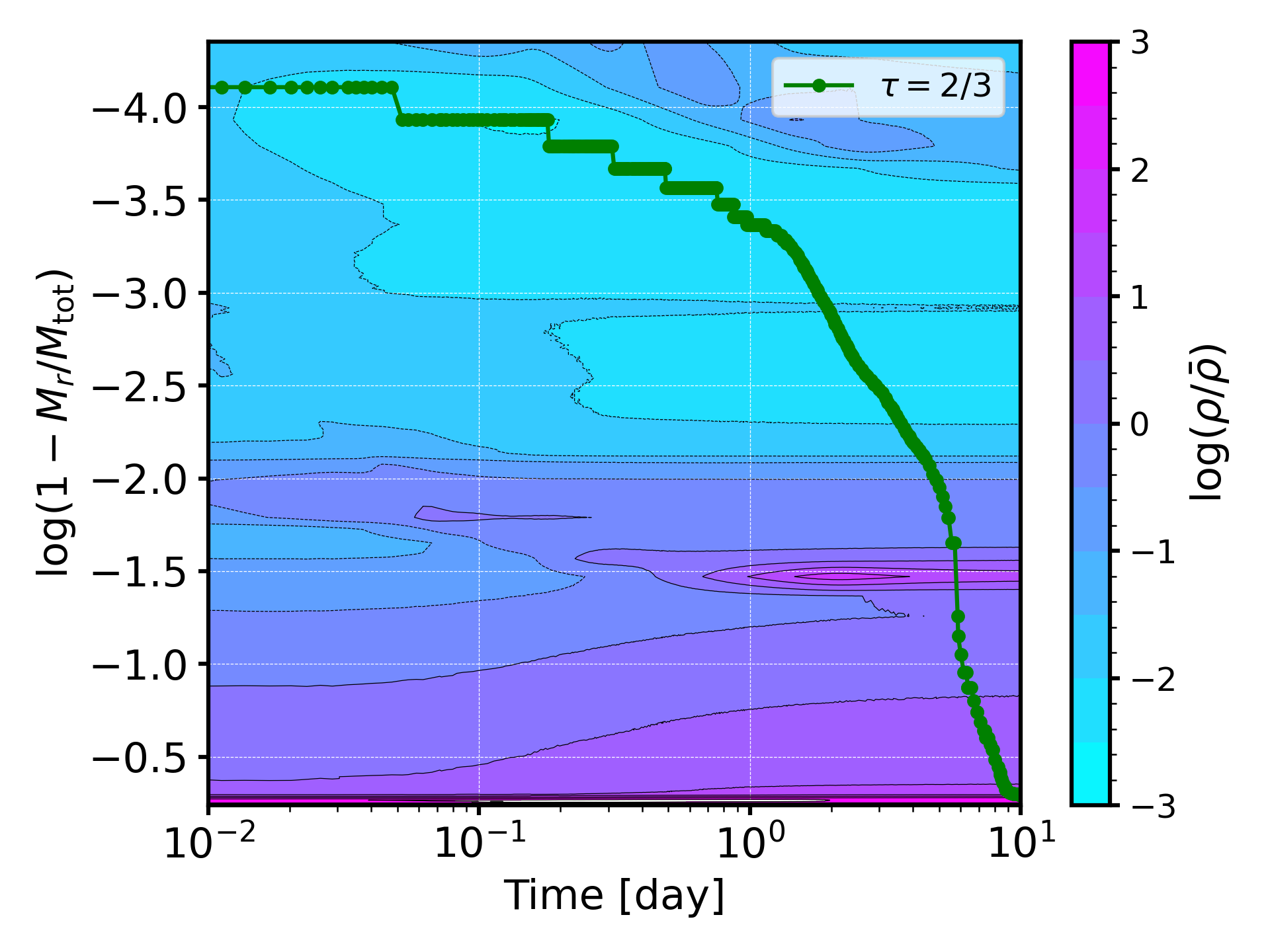}{0.47\textwidth}{Tm11p200\_STELLA\_SCA}
                  \fig{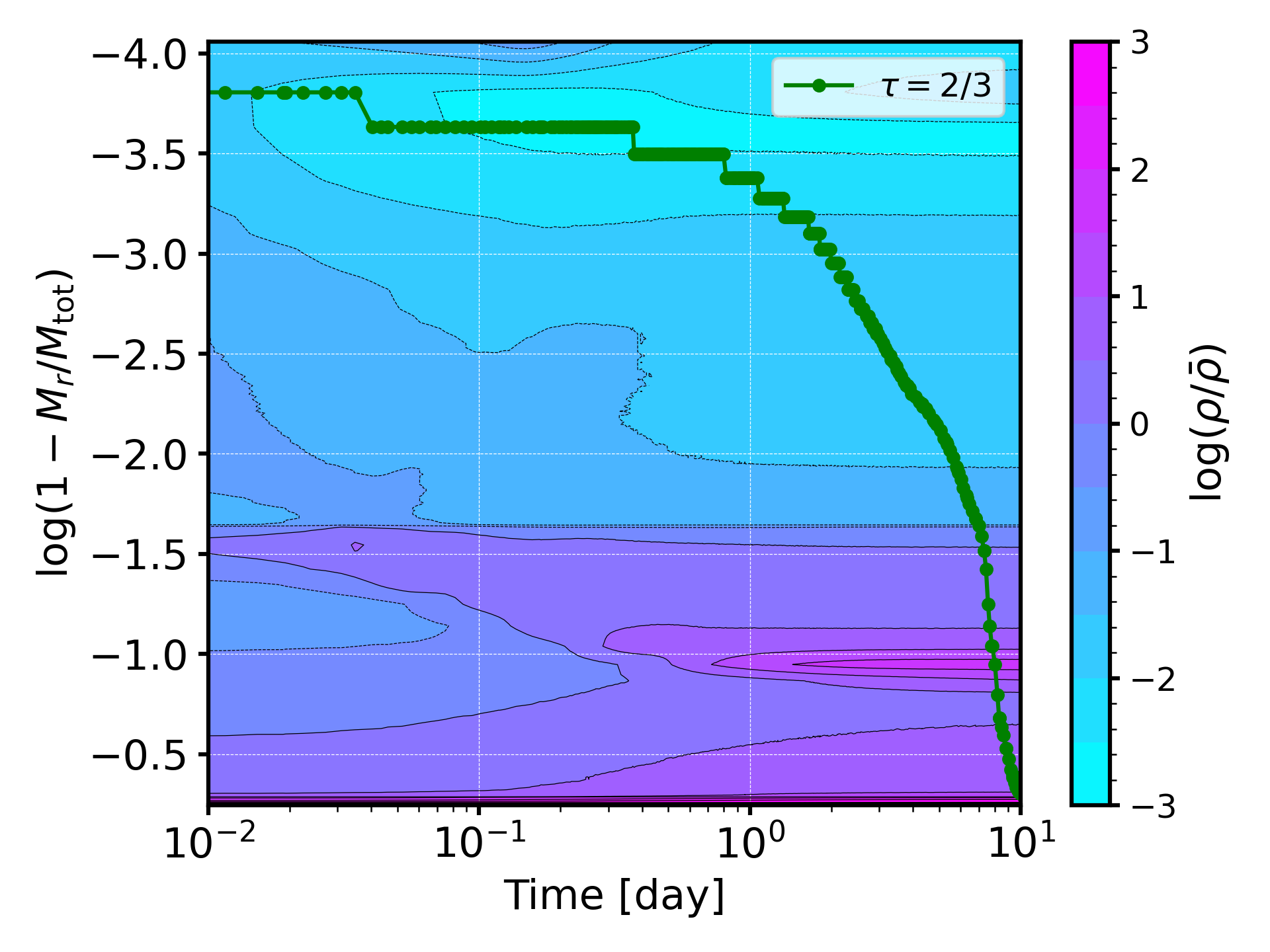}{0.47\textwidth}{Sm11p400\_STELLA\_SCA}}
        \gridline{\fig{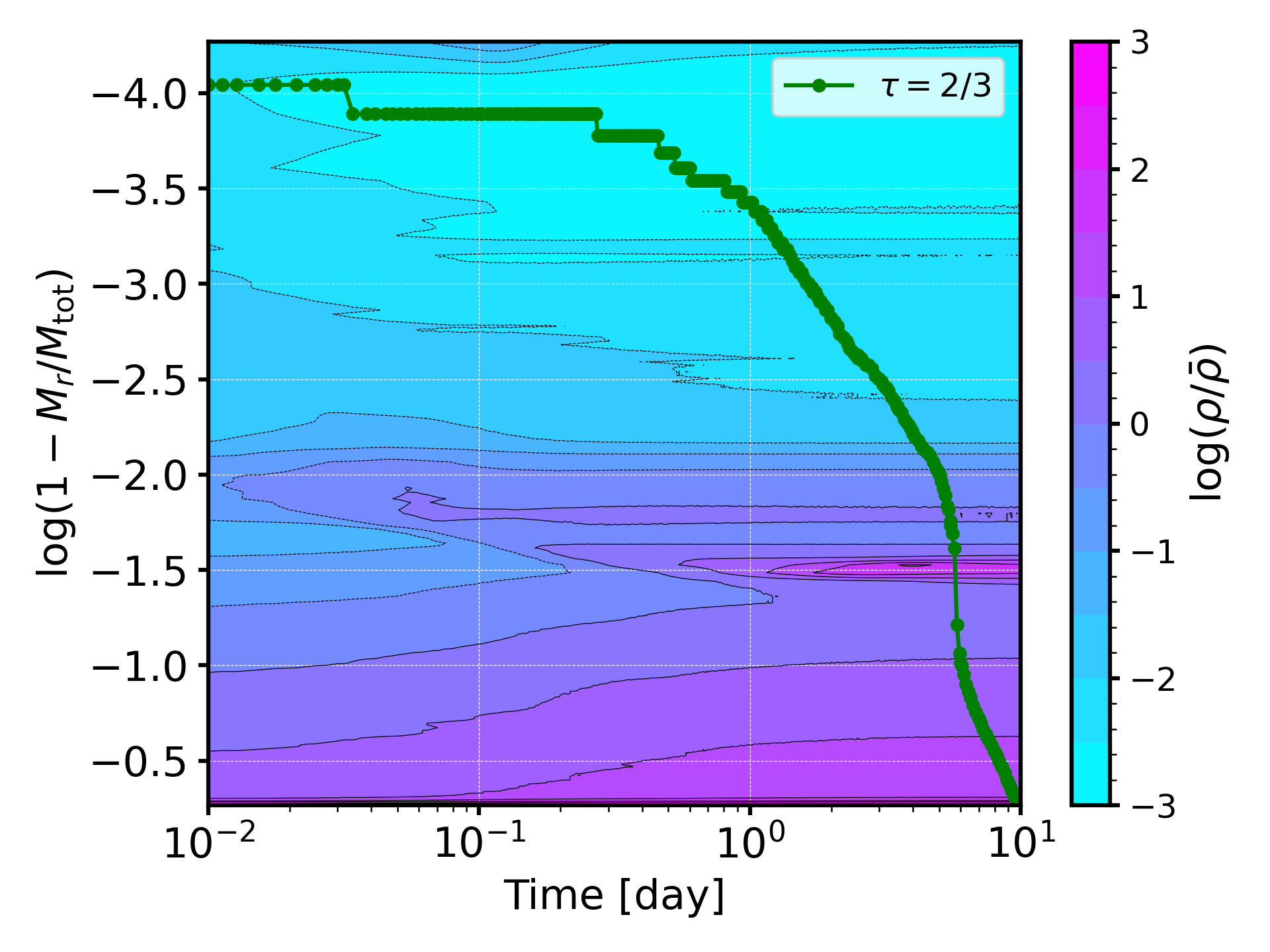}{0.47\textwidth}{Tm11p200\_noRTI}
                  \fig{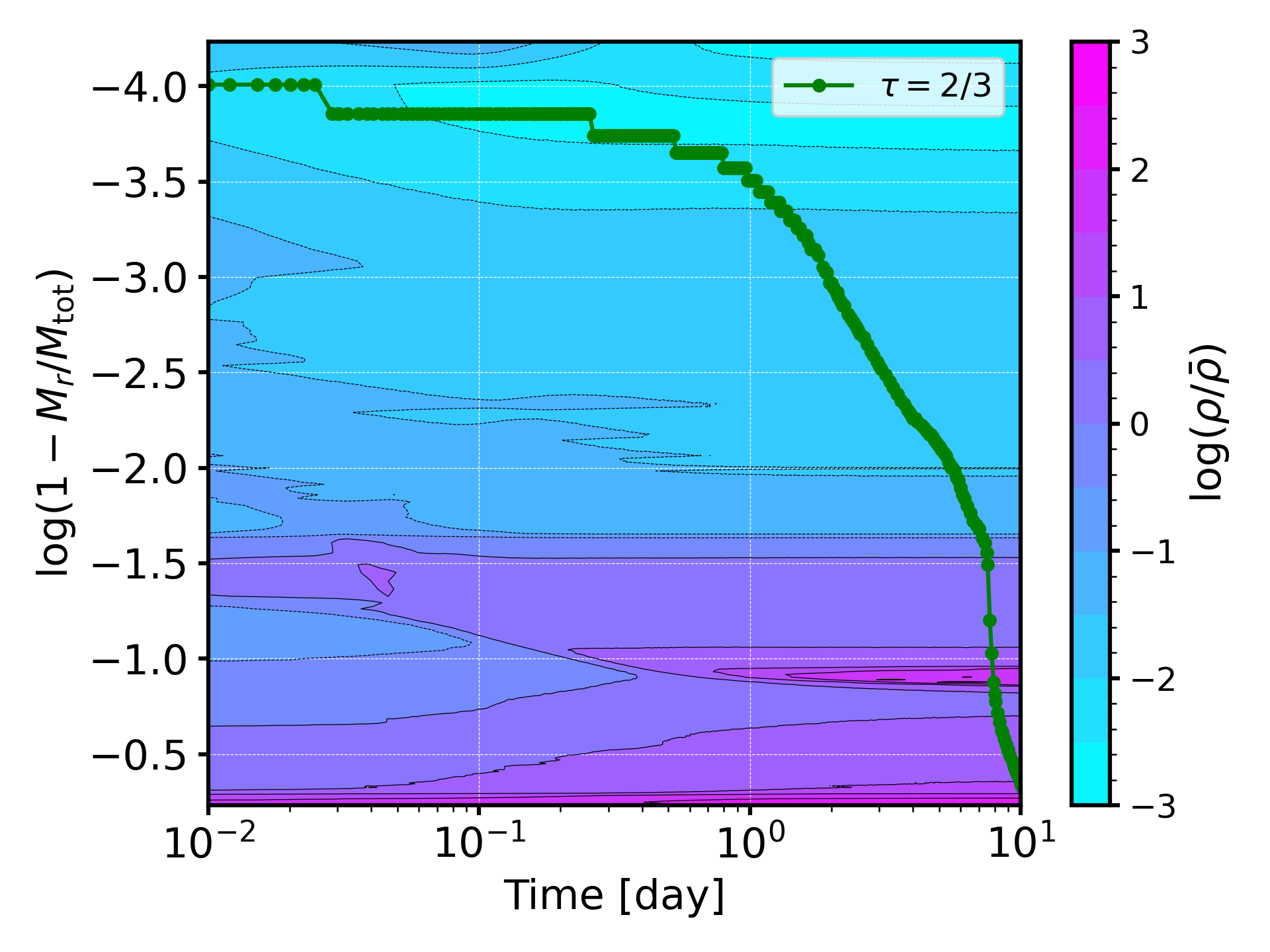}{0.47\textwidth}{Sm11p400\_noRTI}}
        \gridline{\fig{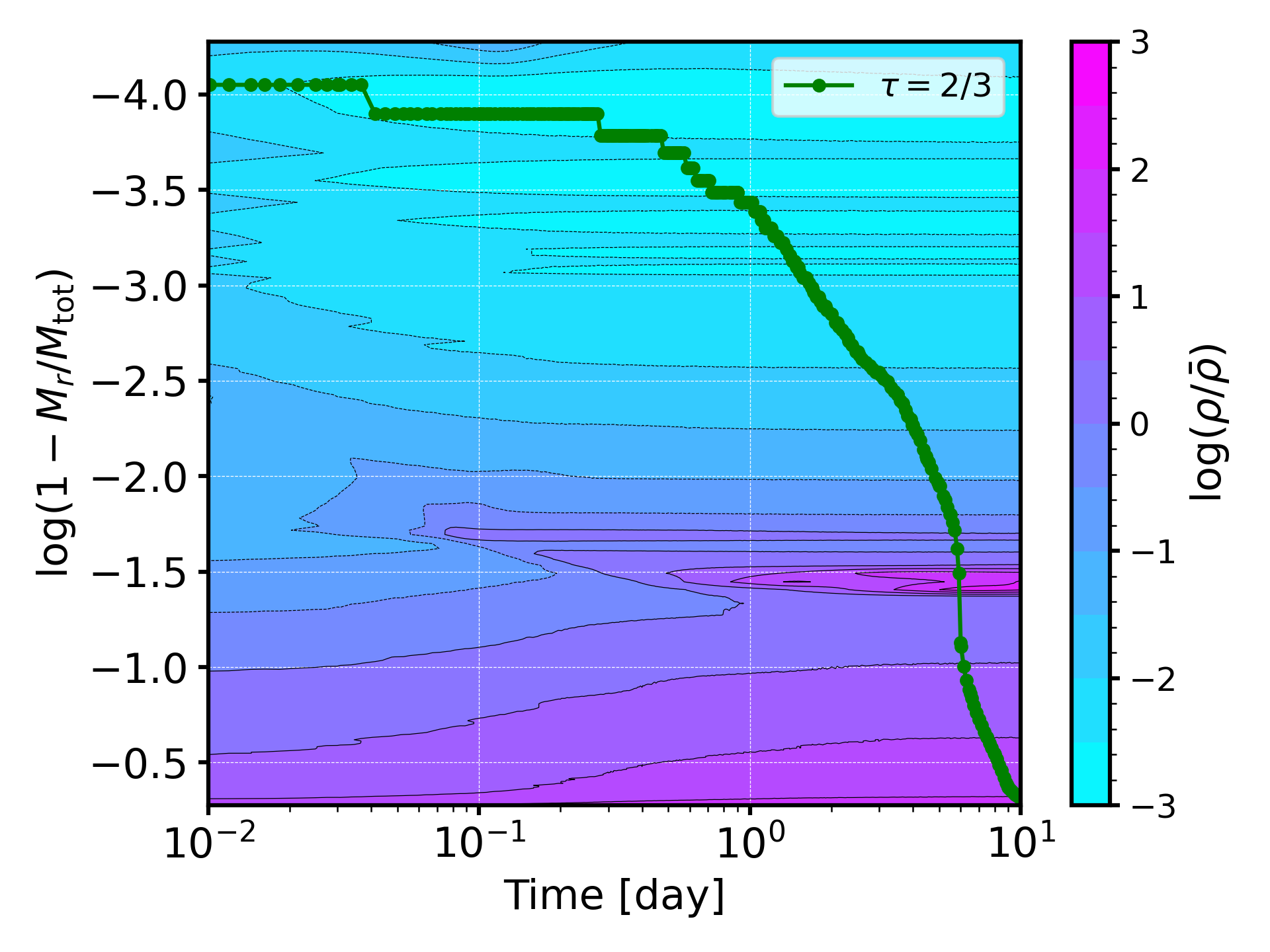}{0.47\textwidth}{Tm11p200\_mix1}
                  \fig{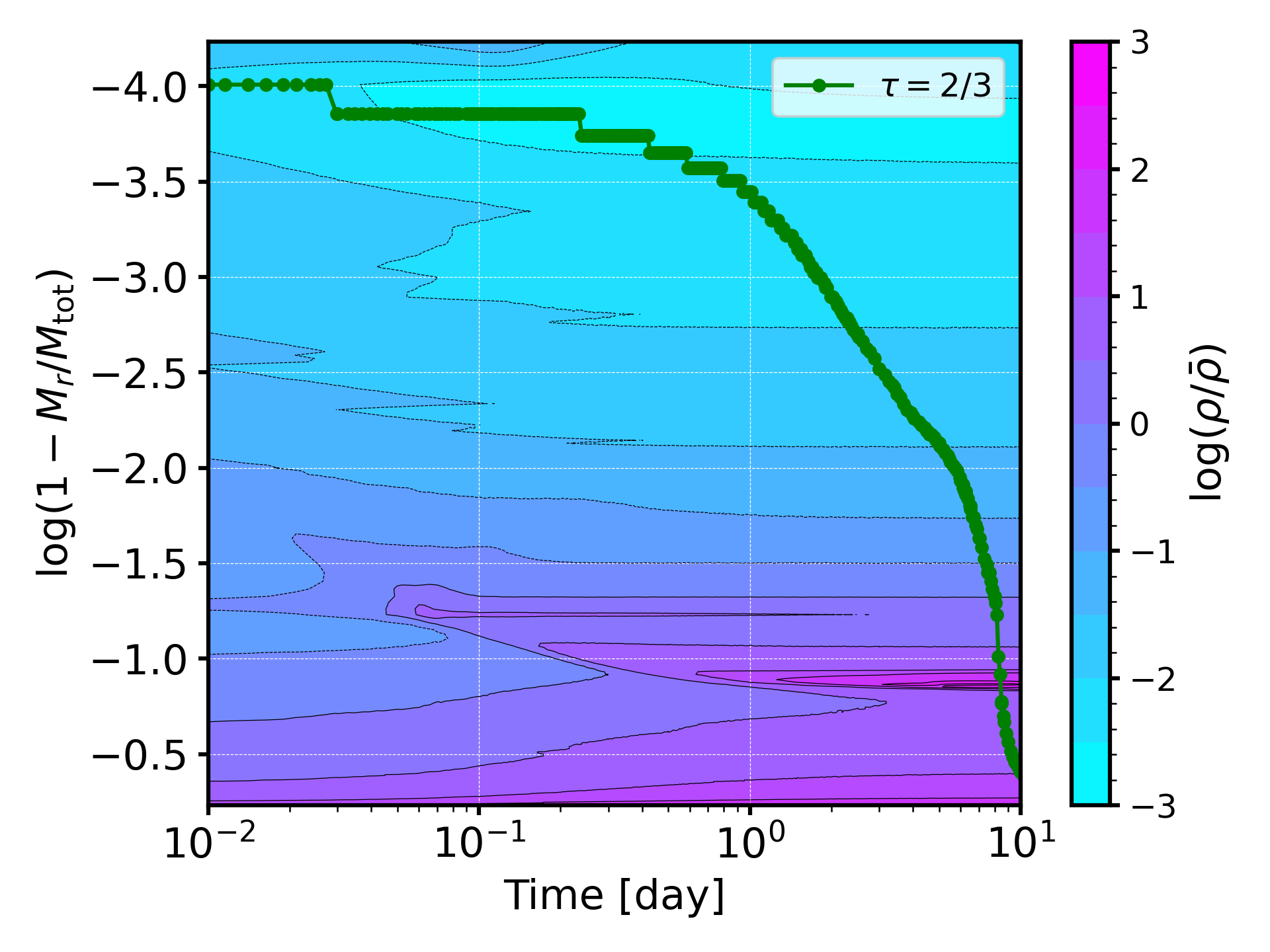}{0.47\textwidth}{Sm11p400\_mix1}}
        \caption{Temporal evolution of internal normalized density structures from $t=0.01$ d to $t=10$ d are presented for the Tm11p200 and Sm11p400 models, with the location of the Rosseland-mean photospheres shown in green lines.
        \label{fig:mixing_RTI_rhoevolution}}
        \end{figure*} 
        
        Figure \ref{fig:mixing_RTI_rhoevolution} shows how the normalized density structures of the models evolve after shock breakout. The on-going reverse shock propagating inwards forms a dense shell near $\log{(M_{r}/M_{\mathrm{tot}})}\approx-1.0\sim-1.5$ in all models, effectively nullifying the density smoothing due to the RTI that is applied just before the shock breakout. As discussed below, the photosphere propagates through the region affected by the reverse shock near the end of the semi-plateau phase. For future studies, either rigorous multi-D simulations or 1D RTI approximation extended to the post-shock breakout evolution of SNe should be considered to fully account for the density smoothing by the RTI and examine the effects of the resulting density structures on SNe IIb light curves. In this study, we focus our discussion on the effects of chemical mixing. 

    \subsection{Results} \label{sec:chem-results}
        
\begin{figure*}
        \gridline{\fig{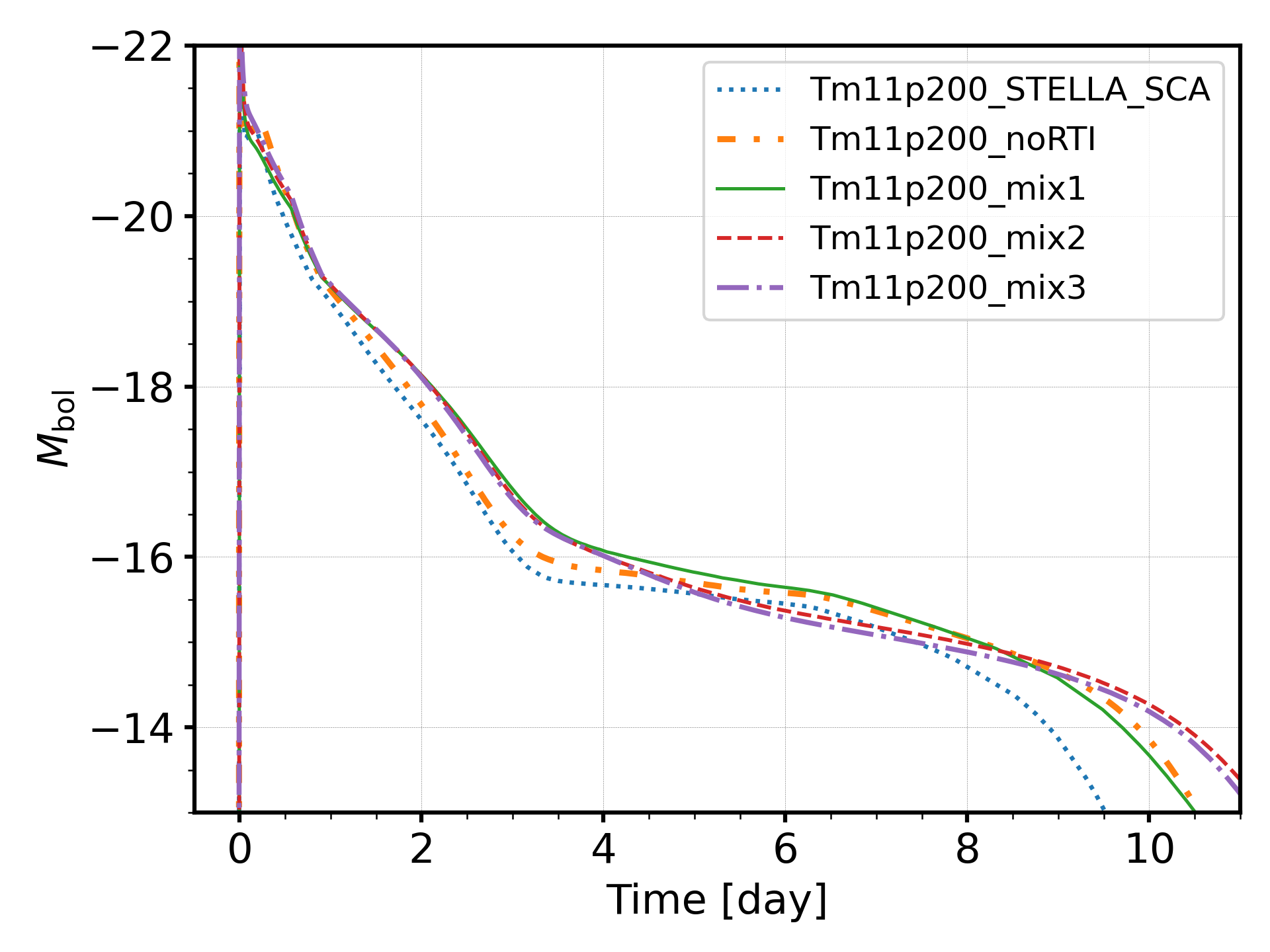}{0.47\textwidth}{}
                  \fig{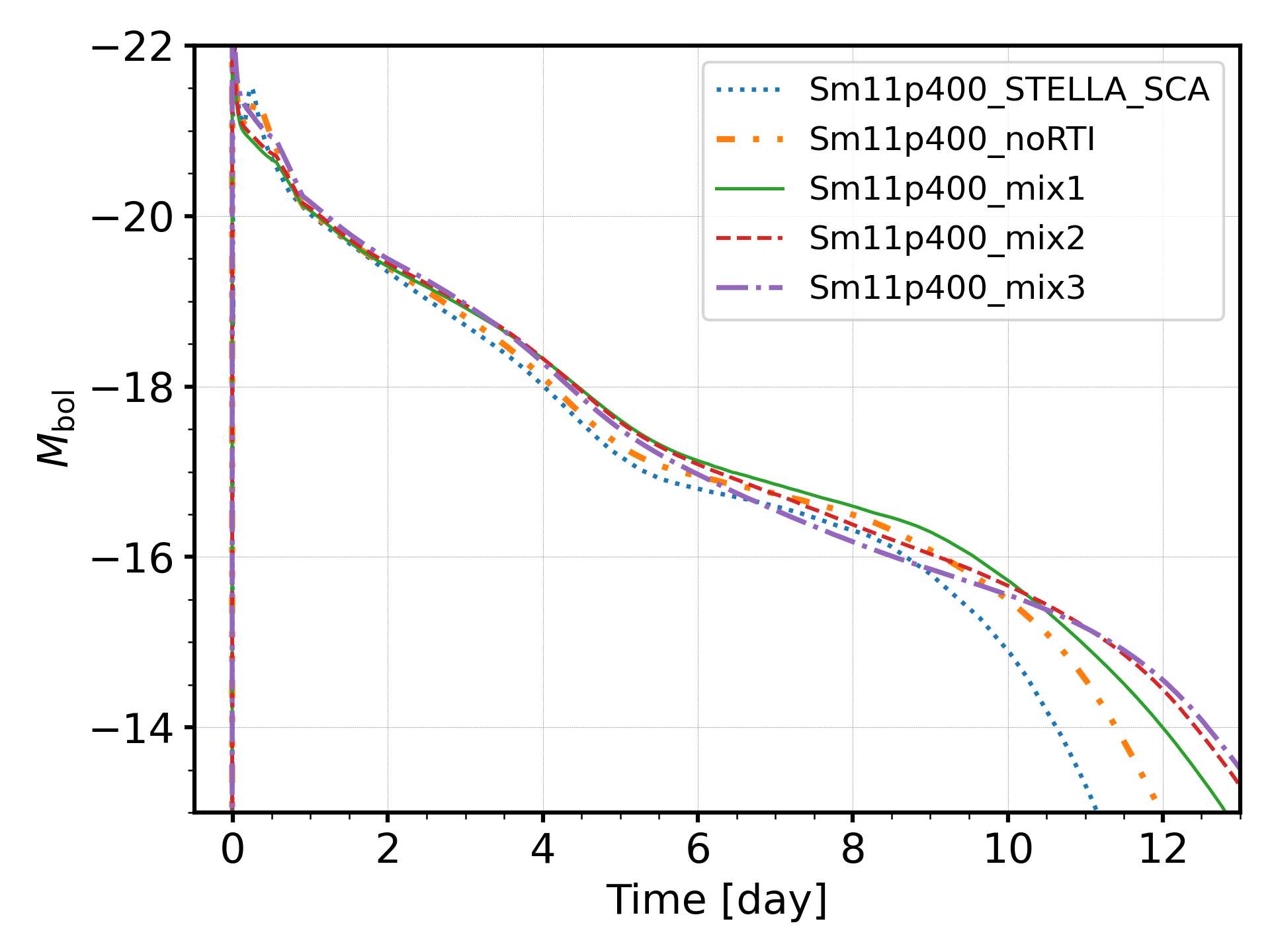}{0.47\textwidth}{}}
        \gridline{\fig{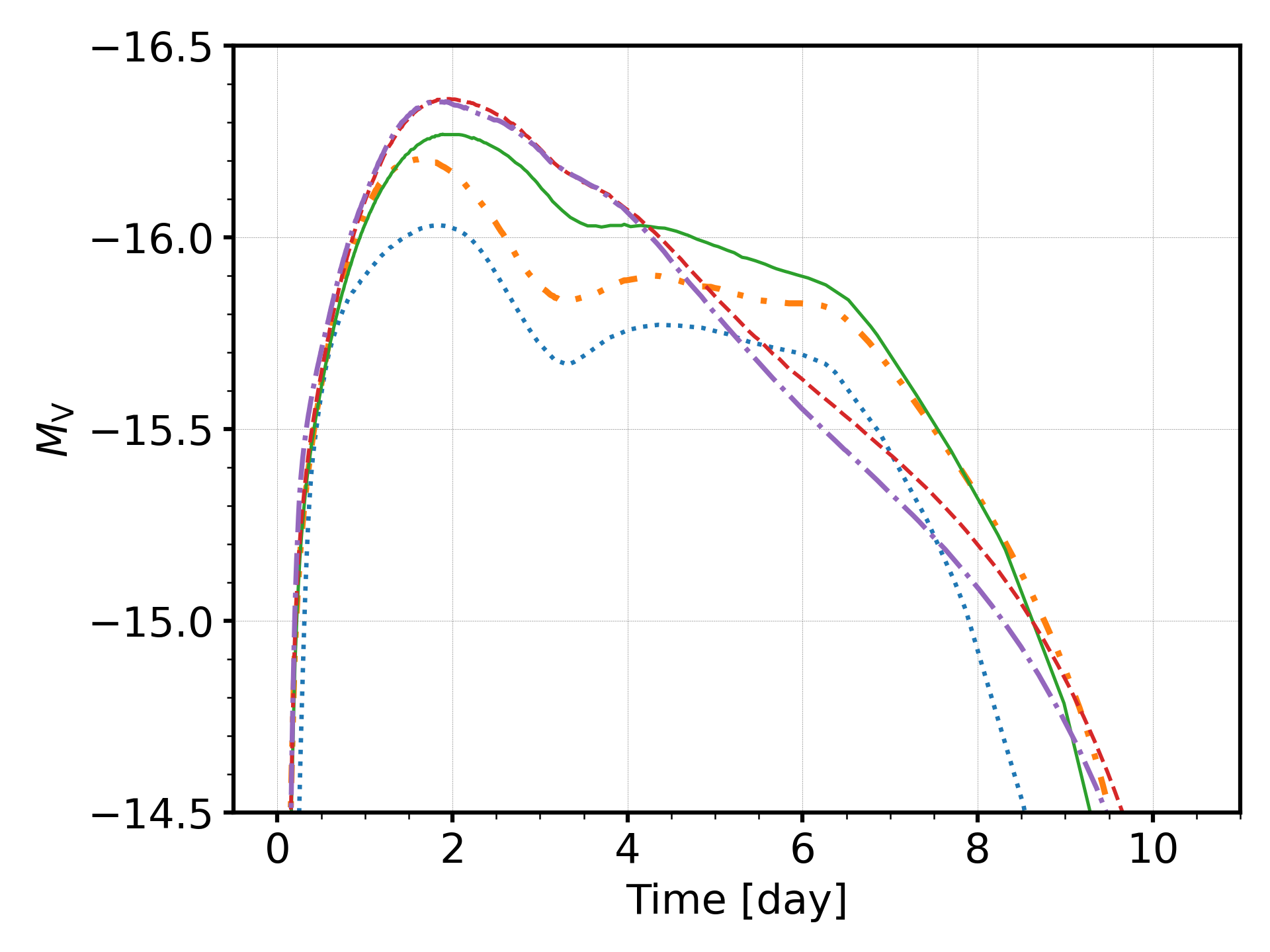}{0.47\textwidth}{}
                  \fig{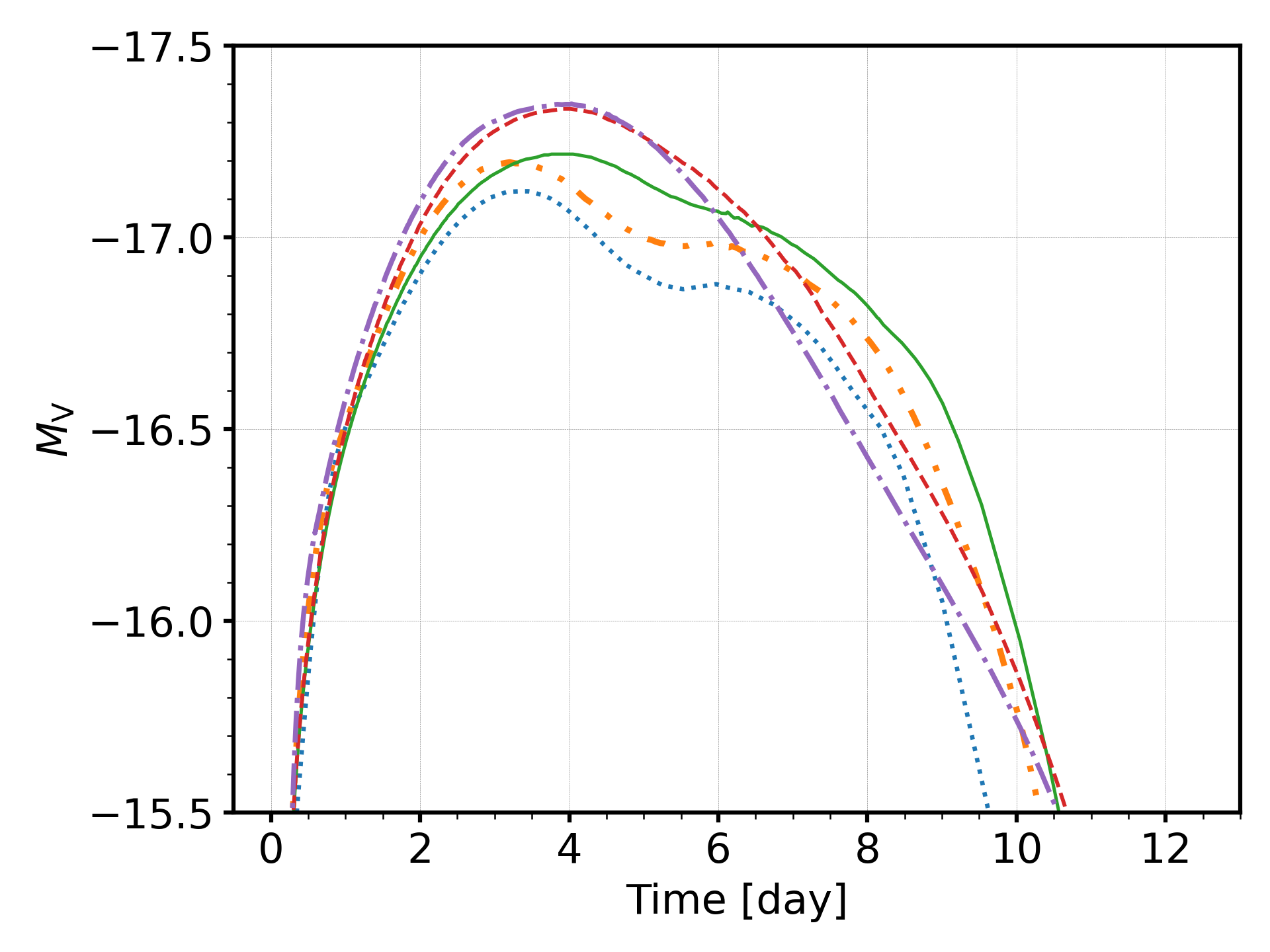}{0.47\textwidth}{}}
        \caption{Upper left: bolometric luminosities of Tm11p200\_STELLA\_SCA (blue), Tm11p200\_noRTI (orange), Tm11p200\_mix1 (green), Tm11p200\_mix2 (red), Tm11p200\_mix3 (purple) models. Upper right: same as the upper left panel but for the Sm11p400 models. Lower left: $V$-band light curves of the Tm11p200 models. Lower right: $V$-band light curves of the Sm11p400 models. 
        \label{fig:mixing_results}}
        \end{figure*}

        In Figure \ref{fig:mixing_results}, we show the bolometric and optical light curves produced by the SNe models presented in Table \ref{tab:chemmodeltable}. 

The $V$-band light curves of the Tm11p200 models show initial rise of $t\simeq2$~d. The $V$-band peak is brightest in the order of mix2 and mix3, mix1, noRTI, and STELLA\_SCA. The last two models show dimmer $V$-band peaks due to lower bolometric luminosity that may be originated from the hydrodynamical effects of the RTI. The mix1 model is dimmer by $\Delta M\approx0.1$ mags than mix2 and mix3 despite having more or less the same bolometric luminosity. From the $V$-band peak,  the light curves of Tm11p200\_STELLA\_SCA and Tm11p200\_noRTI decline until a local minimum at $t\simeq3.5$~d. Then, we see a bump or plateau-like feature at $t=3.5\sim6.5$~d after the explosion followed by a steep decline of the luminosity, a phenomenon that is already discussed above.  This semi-plateau can also be seen in the mix1 model, though with a lesser contrast in the slope between the semi-plateau phase and the post-peak decline phase. 

Although discussions on the semi-plateaus in the first peak of SNe IIb have been largely absent in the literature, we find similar substructures in numerical models from previous studies (Figure 10 of \citet{Bersten12}; Figure 1 of \citet{Moriya16}; Figure 4 of \citet{Bersten18}; Figure 6 of \citet{Dessart18}) For example, \citet{Dessart18} performed non-LTE numerical calculations with the CMFGEN code using progenitor models with extended H-rich envelopes that lost most of their hydrogen during the evolution, which matches the scope of our work. We find their  H\_R601 model with an extended H-rich envelope having $R = 601~R_\odot$ and $M_\mathrm{env} = 0.1 M_\odot$ shows plateau-like behavior until $t\simeq7$~d after the explosion. \citet{Moriya16} investigated rapidly fading SNe II models. Their model with $M_{\mathrm{env}}\simeq1$\msun shows a double-peaked optical light curve with the first peak showing a knee-like feature during its decline, analogous to our results. 

On the other hand, the mix2 and mix3 models do not exhibit a distinct semi-plateau, except for a subtle indication at $t\simeq3.5$~d. The bolometric light curve slopes change before and after $t\simeq3.5$~d for all models but with different degrees. At $t>3.5$ d, the mix3 model shows the steepest decline and the decline rate becomes lower with a lower degree of mixing, while the STELLA\_SCA and noRTI models show more or less constant bolometric luminosity during the semi-plateau phase.

As for the Sm11p400 models, the optical light curves of STELLA\_SCA and noRTI exhibit semi-plateaus around $t=5\sim6$ days. The Sm11p400\_mix1 model only displays a slight indication of a semi-plateau, while the mix2 and mix3 models do not exhibit this feature as in the corresponding Tm11p200 models. 


        \begin{figure*}
        \gridline{\fig{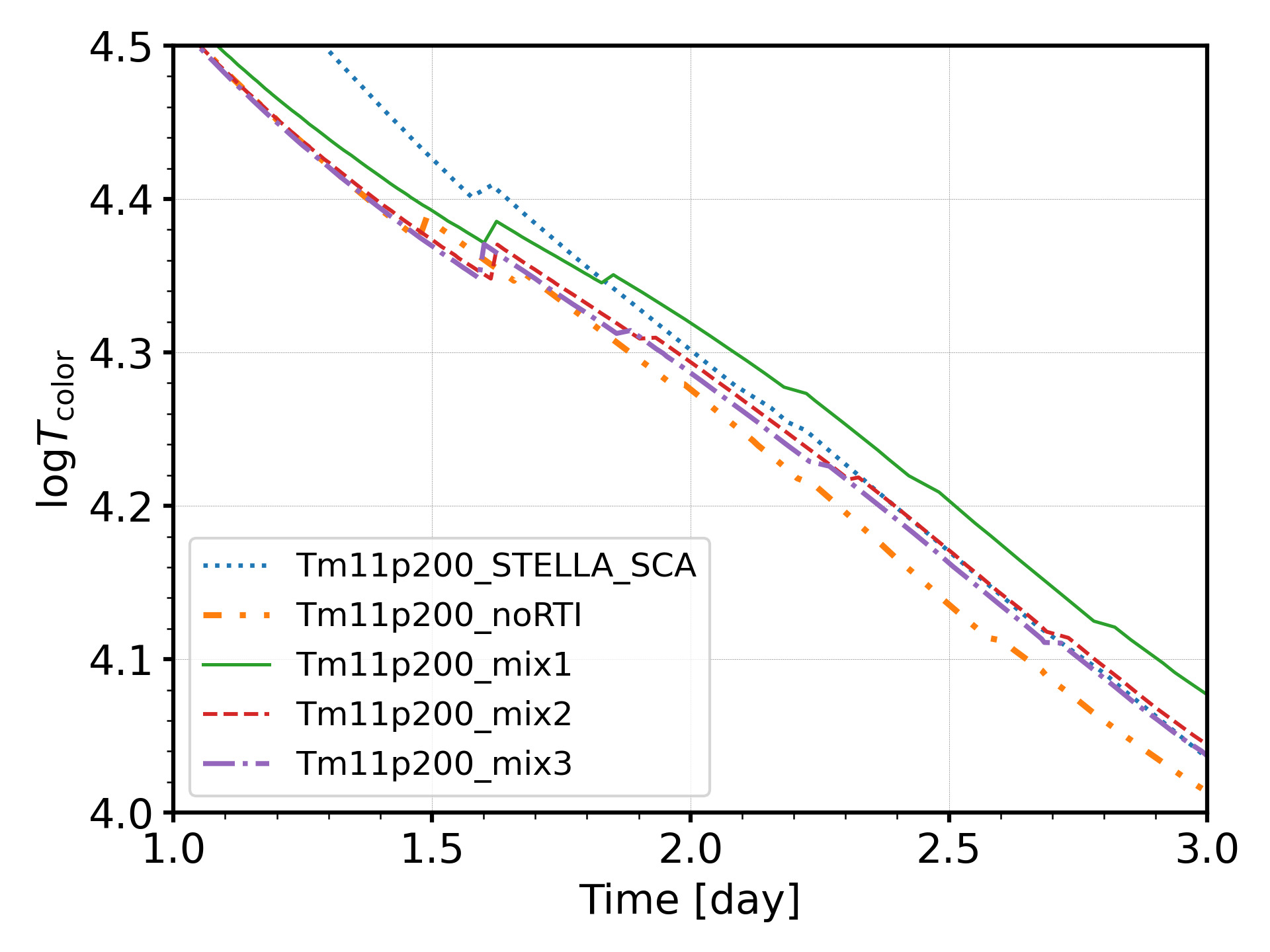}{0.47\textwidth}{}
                  \fig{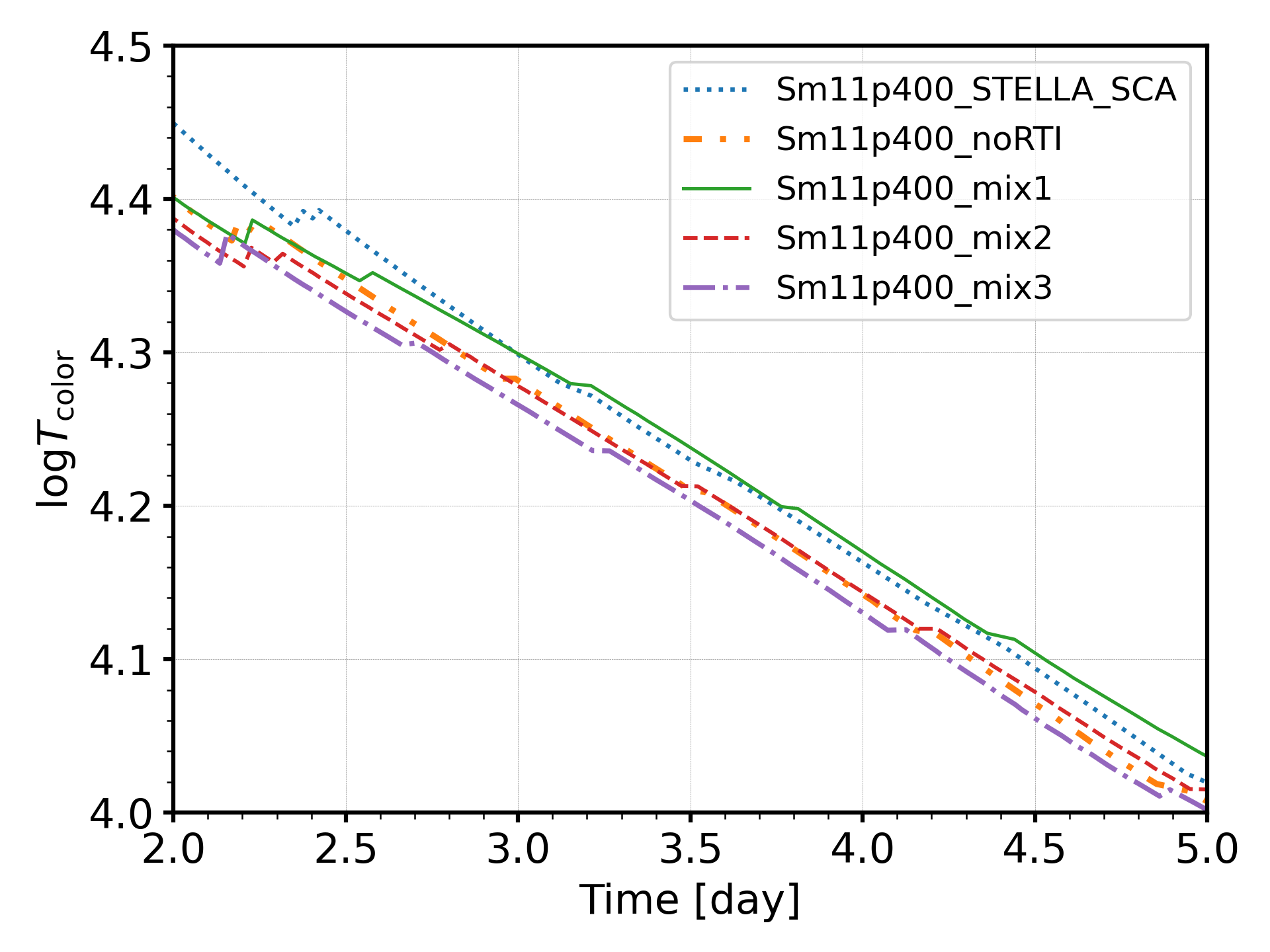}{0.47\textwidth}{}}
        \gridline{\fig{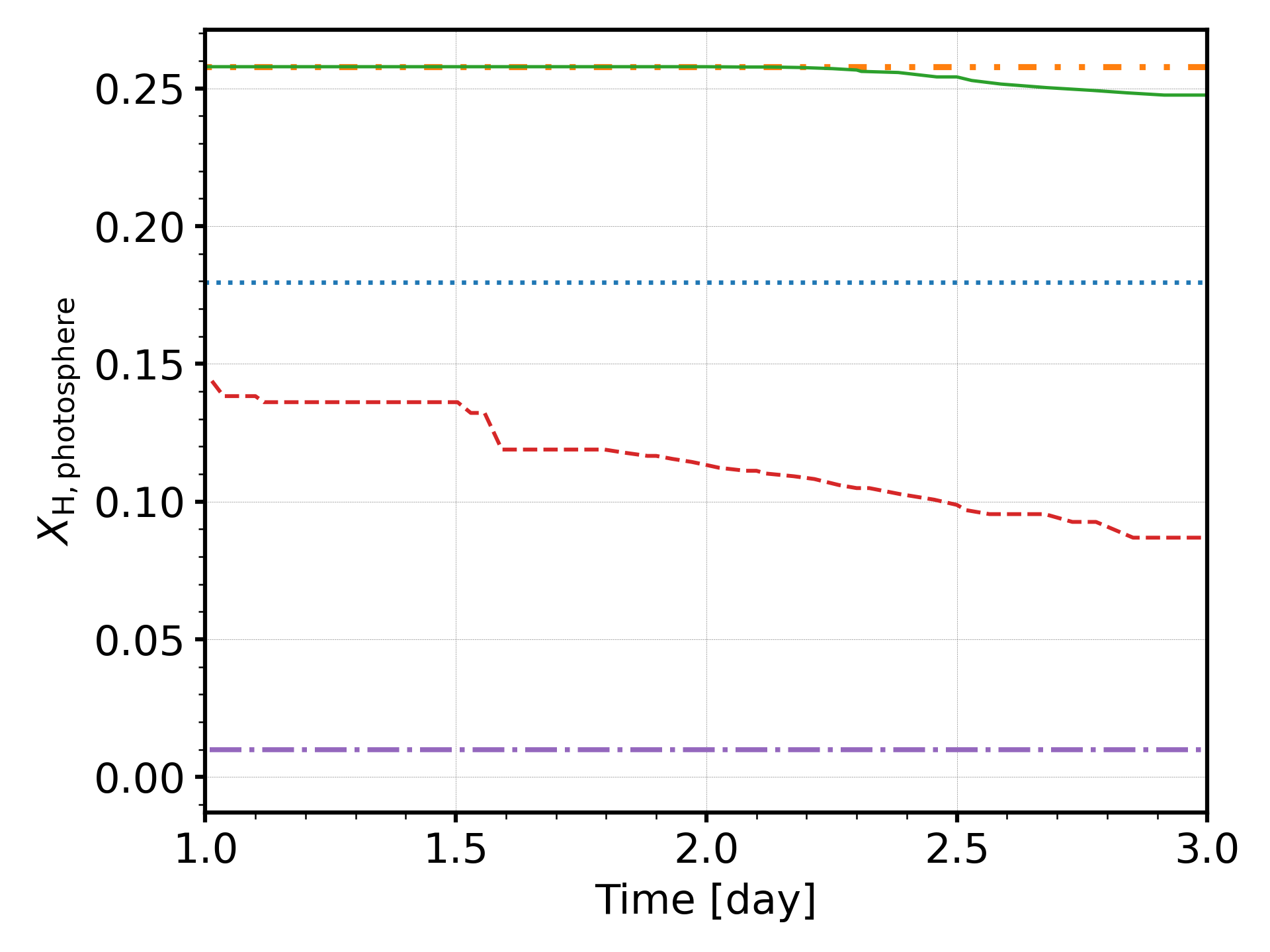}{0.47\textwidth}{}
                  \fig{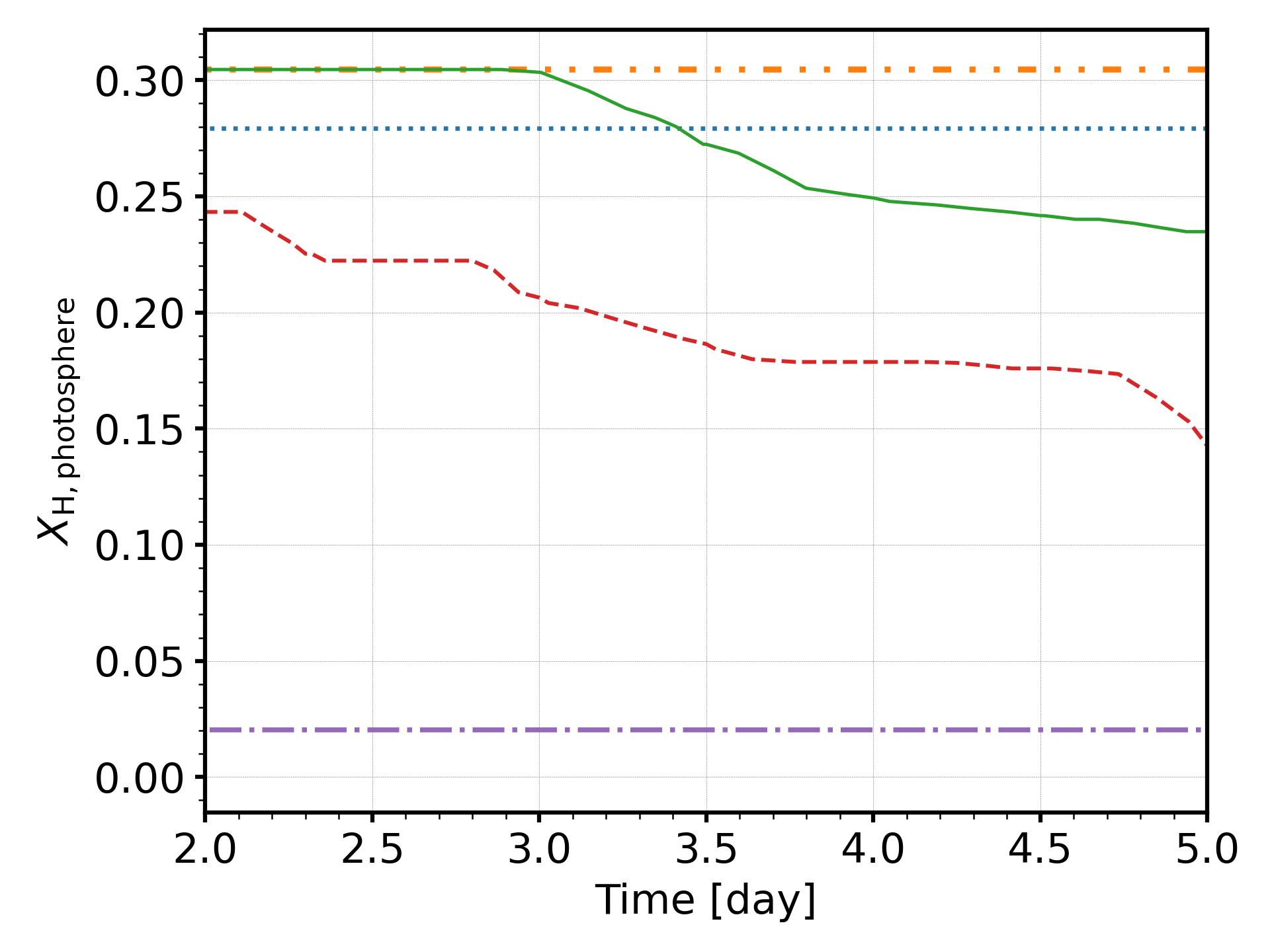}{0.47\textwidth}{}}
        \caption{Upper left: the color temperature of the Tm11p200 models. Upper right: same as the upper left but with the Sm11p400 models. Lower left: the abundance of hydrogen at the photosphere of the Tm11p200 models. Lower right: same as the lower left but with the Sm11p400 models. The temporal ranges of the plots are selected in order to highlight the models near the $V$-band peak.
        \label{fig:mixing_nearpeak}}
        \end{figure*} 
    
        In Figure \ref{fig:mixing_nearpeak}, the temporal evolution of the color temperatures of the Tm11p200(Sm11p400) models near the optical peak is presented. The observed variations in optical brightness among different models with varying degrees of mixing can be largely attributed to differences in color temperatures, similar to the case discussed in Section \ref{sec:snec}. For example, it is observed that the color temperature ($T_{\mathrm{color}}$) of the mix1 model is systematically higher than that of mix3. Due to the higher hydrogen abundance in mix1 compared to mix3 models, there is an increased number of free electrons given that the temperature near the photosphere is higher than 10000 K. This results in a deeper thermalization depth, a correspondingly higher color temperature, and a lower optical luminosity for the mix1 model despite the same bolometric luminosity with the mix3 model at the $V$-band peak. 
        

        \begin{figure*}
        \gridline{\leftfig{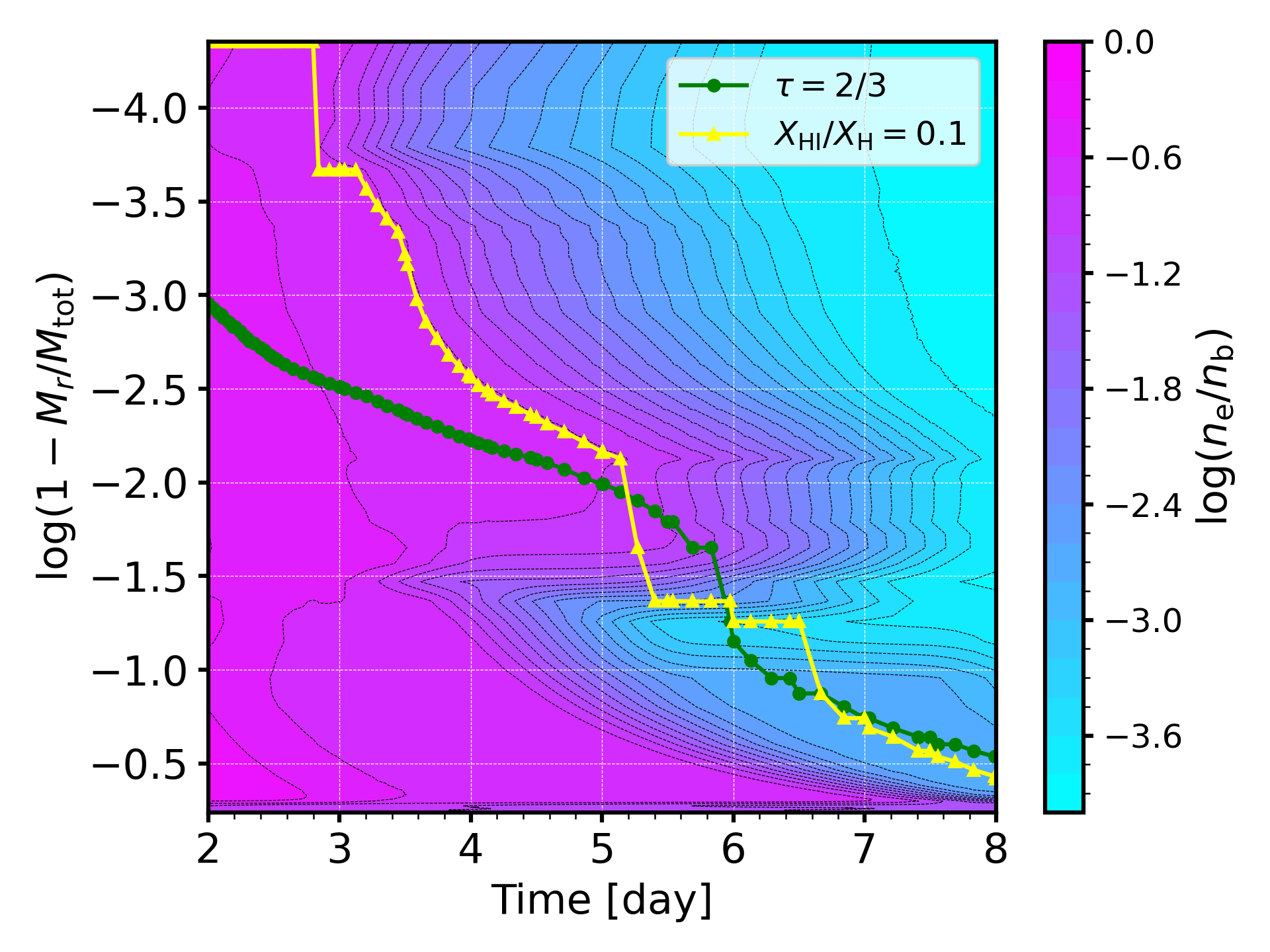}{0.47\textwidth}{Tm11p200\_STELLA\_SCA}
                  \rightfig{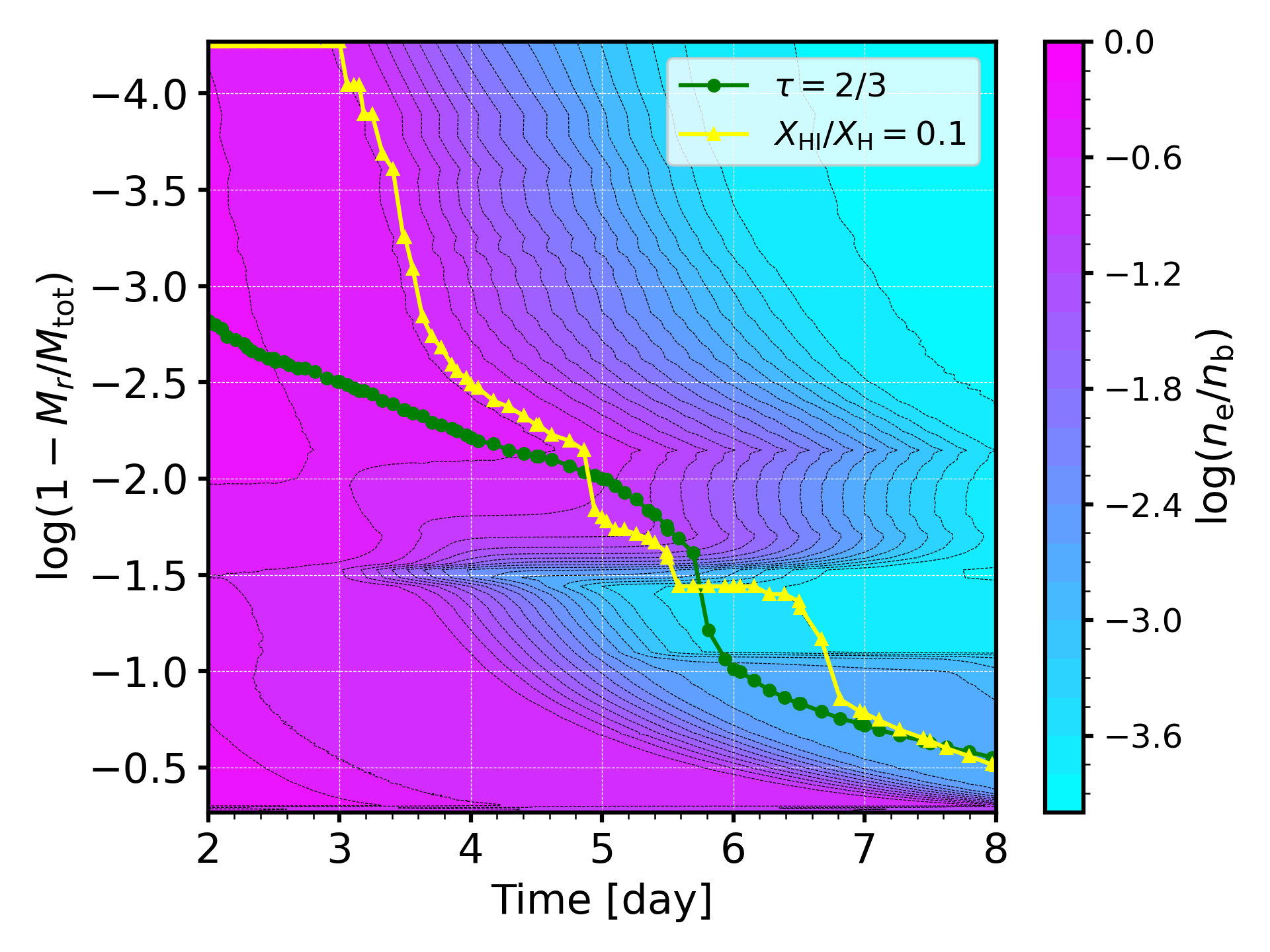}{0.47\textwidth}{Tm11p200\_noRTI}}
        \gridline{\leftfig{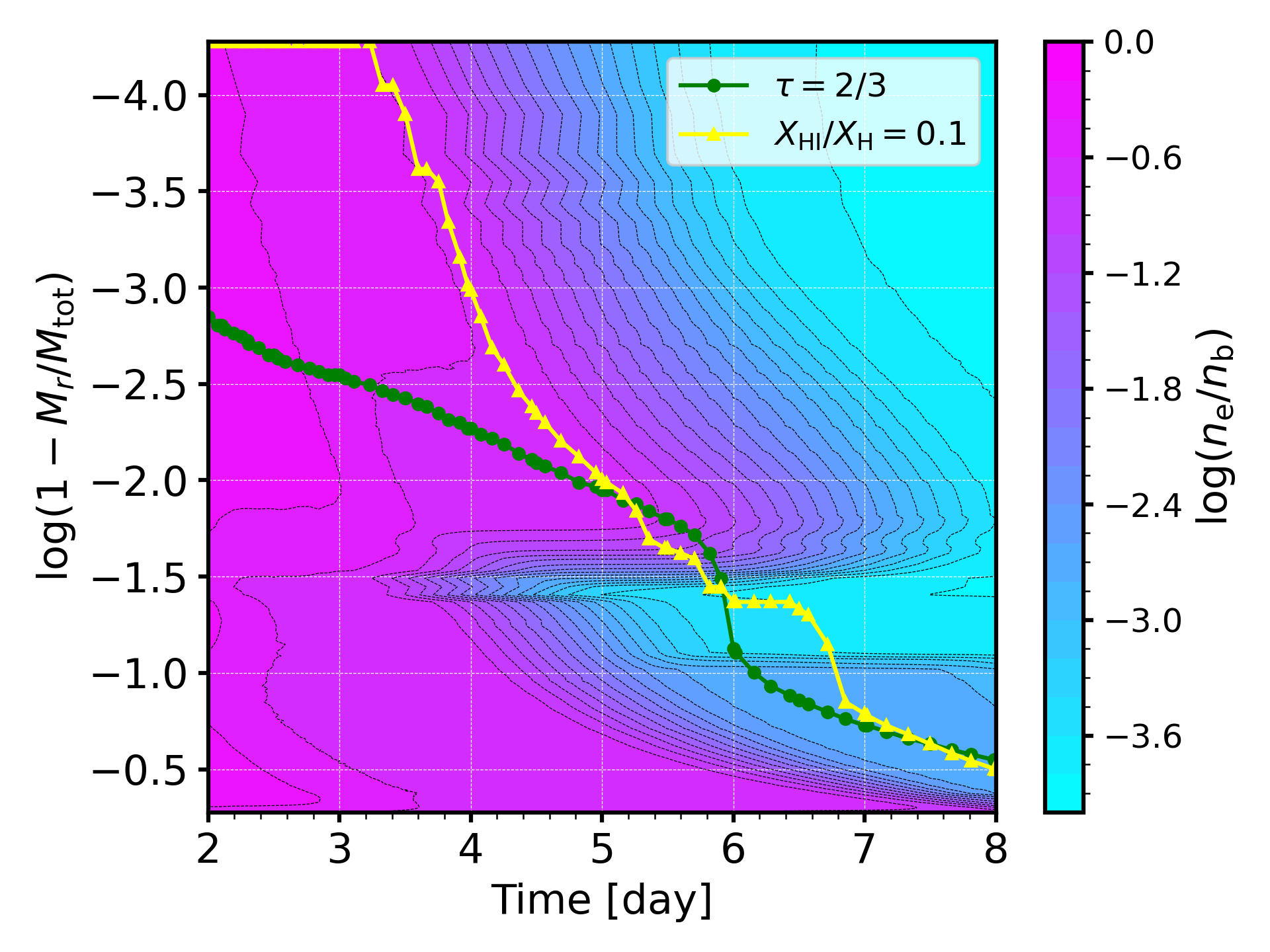}{0.47\textwidth}{Tm11p200\_mix1}
                  \rightfig{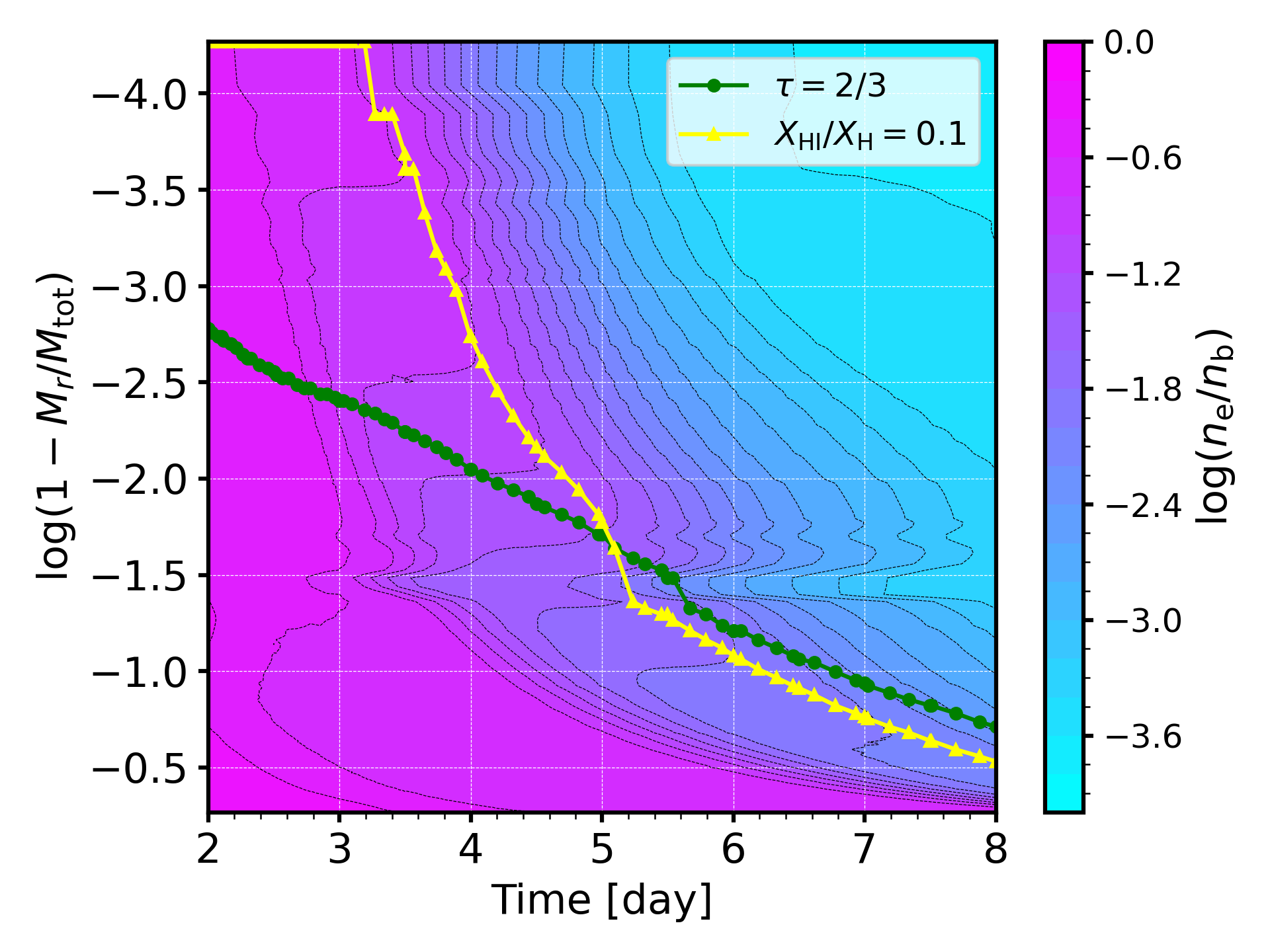}{0.47\textwidth}{Tm11p200\_mix2}}
        \gridline{\leftfig{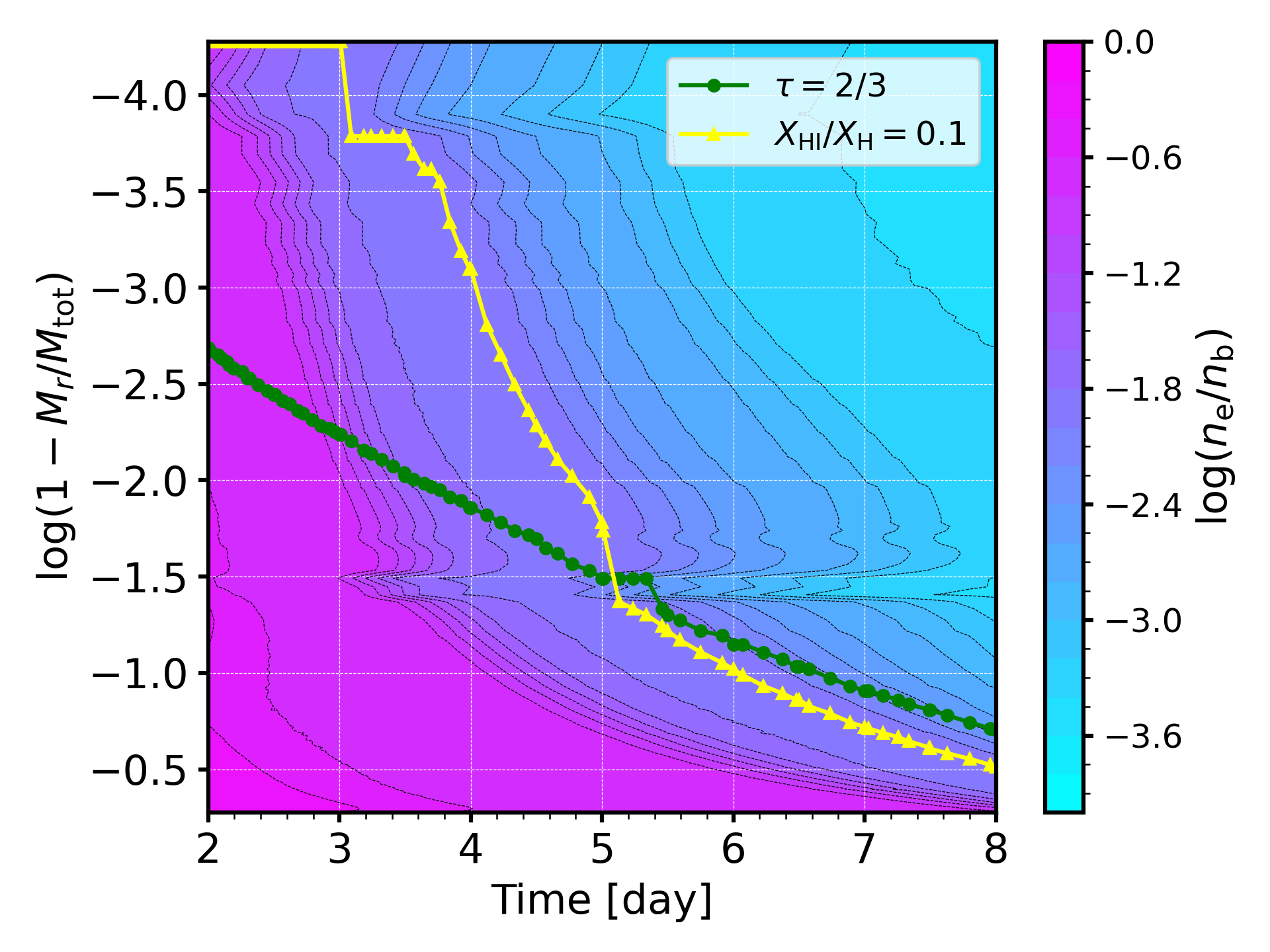}{0.47\textwidth}{Tm11p200\_mix3}}
        \caption{Contour plots of internal structures of free electron number density to baryon density ratio for the (a) Tm11p200\_STELLA\_SCA, (b) Tm11p200\_noRTI, (c) Tm11p200\_mix1, (d) Tm11p200\_mix2, and (e) Tm11p200\_mix3 models. In each panel, green lines with circles represent the location of the photosphere, and yellow lines with triangles roughly represent the H recombination front.
        \label{fig:p200_nenbKPH}}
        \end{figure*}  
        
        Unlike the peak brightness difference that is related to the SED, the decline rate from the optical peak is related to the inward propagation of the photosphere and the release of the thermal energy contained within the ejecta. The location of a photosphere is determined by density and opacity. The density decreases as the ejecta expands and the opacity changes as the ejecta cools down. Although absorptive opacities are also dependent on the temperature, the decrease of the free electron number density due to hydrogen recombination is the most significant factor when the photosphere propagates through a H-rich envelope. 
        
        
        Figure \ref{fig:p200_nenbKPH} shows the ratio of free electron to baryon number density ($n_{\mathrm{e}}/n_{\mathrm{b}}$), which serves as an indicator of ionization and the abundance of hydrogen. The figure also depicts the positions of the Rosseland-mean photosphere ($\tau=2/3$) and an approximate location of the H recombination front where 10\% of all hydrogen atoms exist in the neutral state ($X_{\mathrm{HI}}/X_{\mathrm{H}}=0.1$). In the contour plots, we see a rapid change of $n_{\mathrm{e}}/n_{\mathrm{b}}$ at $\log{(M_{r}/M_{\mathrm{tot}})}\approx-1.5$ for the STELLA\_SCA, noRTI, and mix1 models. This originates from the strong chemical stratification in the ejecta between the H-rich envelopes and the He cores. In the three models, the photosphere recedes rapidly into the inner layers once it reaches the bottom of the H-rich envelope since the opacity below this layer, having a much smaller electron number density, is significantly lower than the H-rich envelope. On the other hand, the mix2 and mix3 models exhibit a rather smooth change in $n_{\mathrm{e}}/n_{\mathrm{b}}$ and the recession of the photosphere is also fairly continuous as a result of rather a strong chemical mixing between the H-rich envelope and the He core. 

        The hydrogen recombination starts in the outermost layers at $t\simeq3$ d and the recombination front propagates inward as the inner layers cool down due to the expansion of the ejecta. The span of the semi-plateau phase in the SCA, noRTI, and mix1 models corresponds to the period between the onset of hydrogen recombination near the ejecta surface and the photosphere reaching the He/H chemical interface. 
        
        \citet{Kasen09} found that the different opacity prescriptions and different He abundances in the H-rich envelope drastically affect the light curves at the plateau phase of SNe IIP. For a higher He abundance, the early light curve brightness increases and the light curve during the plateau/linear decline phase becomes shorter \citep[cf.][]{Swartz91,Moriya16}. \citet{Blinnikov93} also finds that a chemical mixing between the H-rich envelope and the He core may affect the slope of the SN II light curve. The models of these studies have much more massive H-rich envelopes than our models, but our results are qualitatively in good agreement with the previous findings.  

        \begin{figure*}
        \gridline{\leftfig{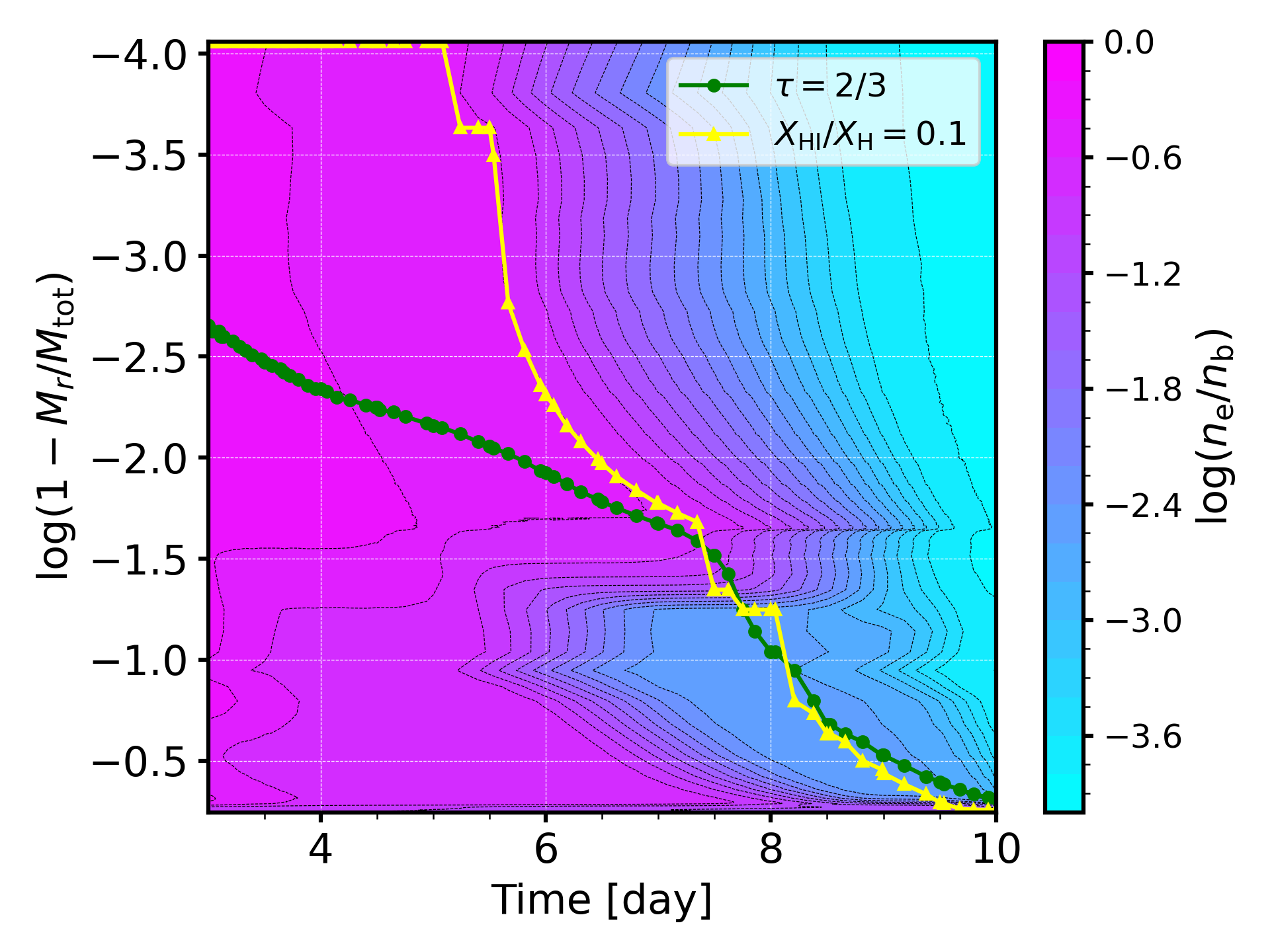}{0.47\textwidth}{Sm11p400\_STELLA\_SCA}
                  \rightfig{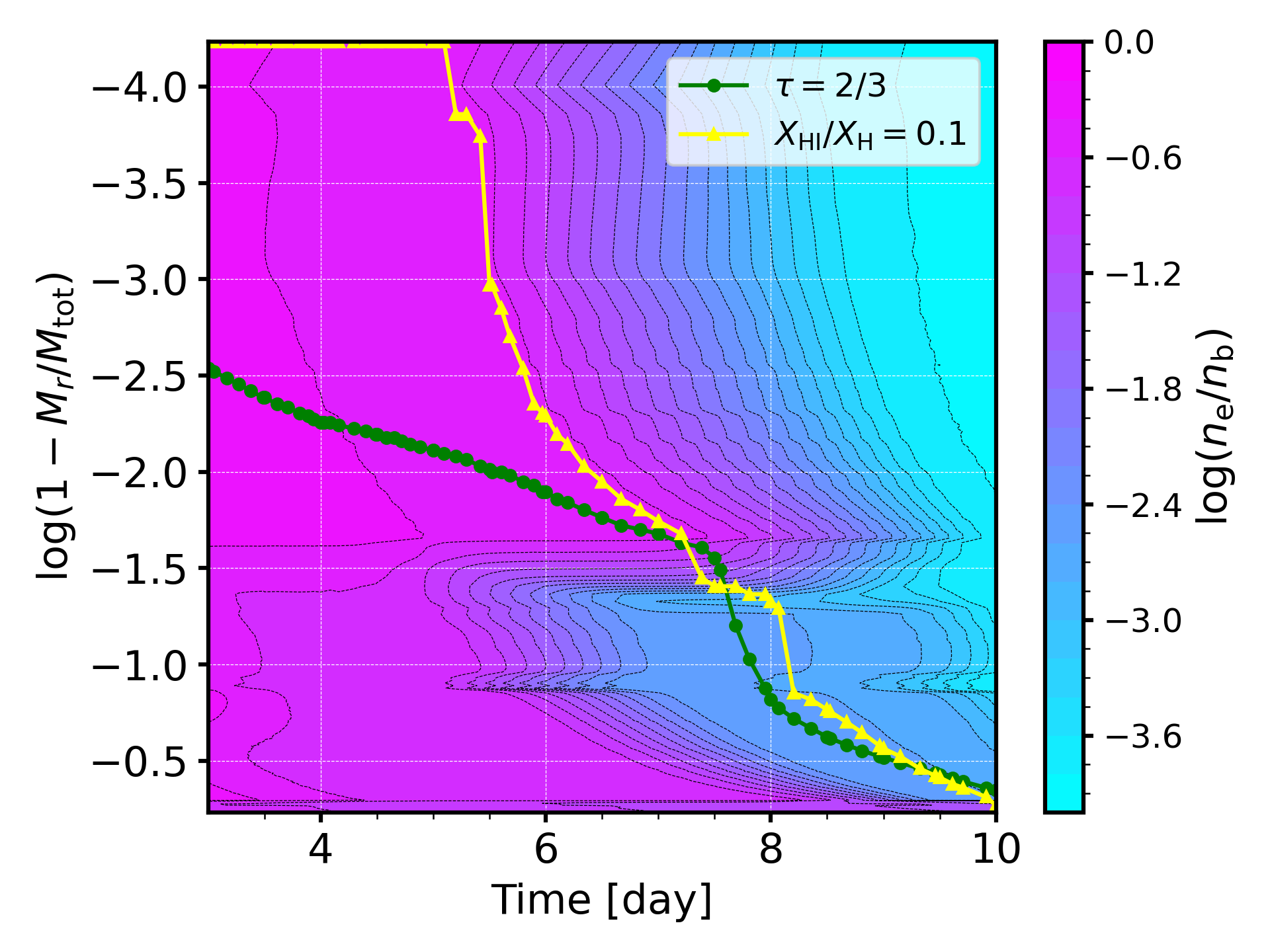}{0.47\textwidth}{Sm11p400\_noRTI}}
        \gridline{\leftfig{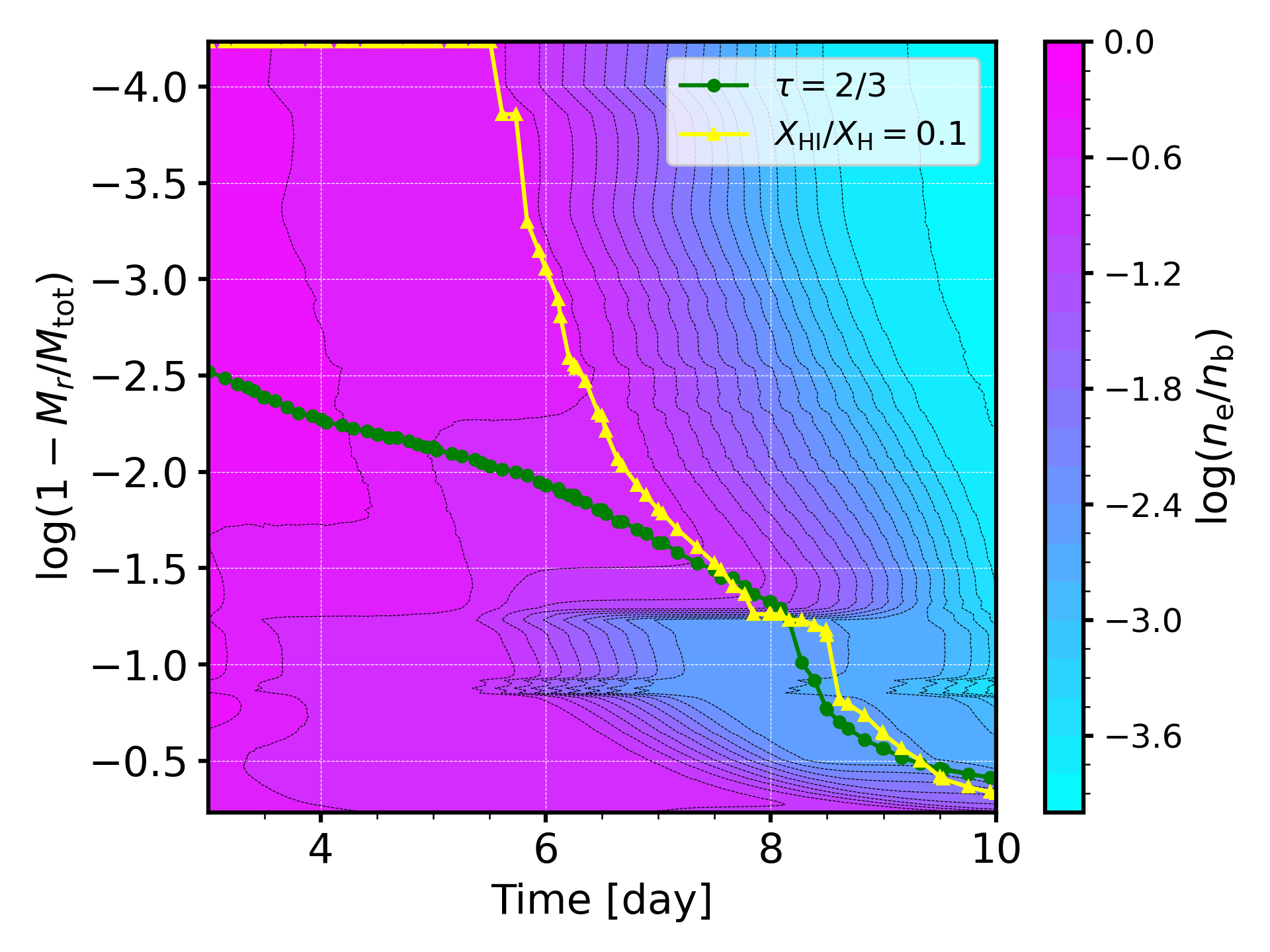}{0.47\textwidth}{Sm11p400\_mix1}
                  \rightfig{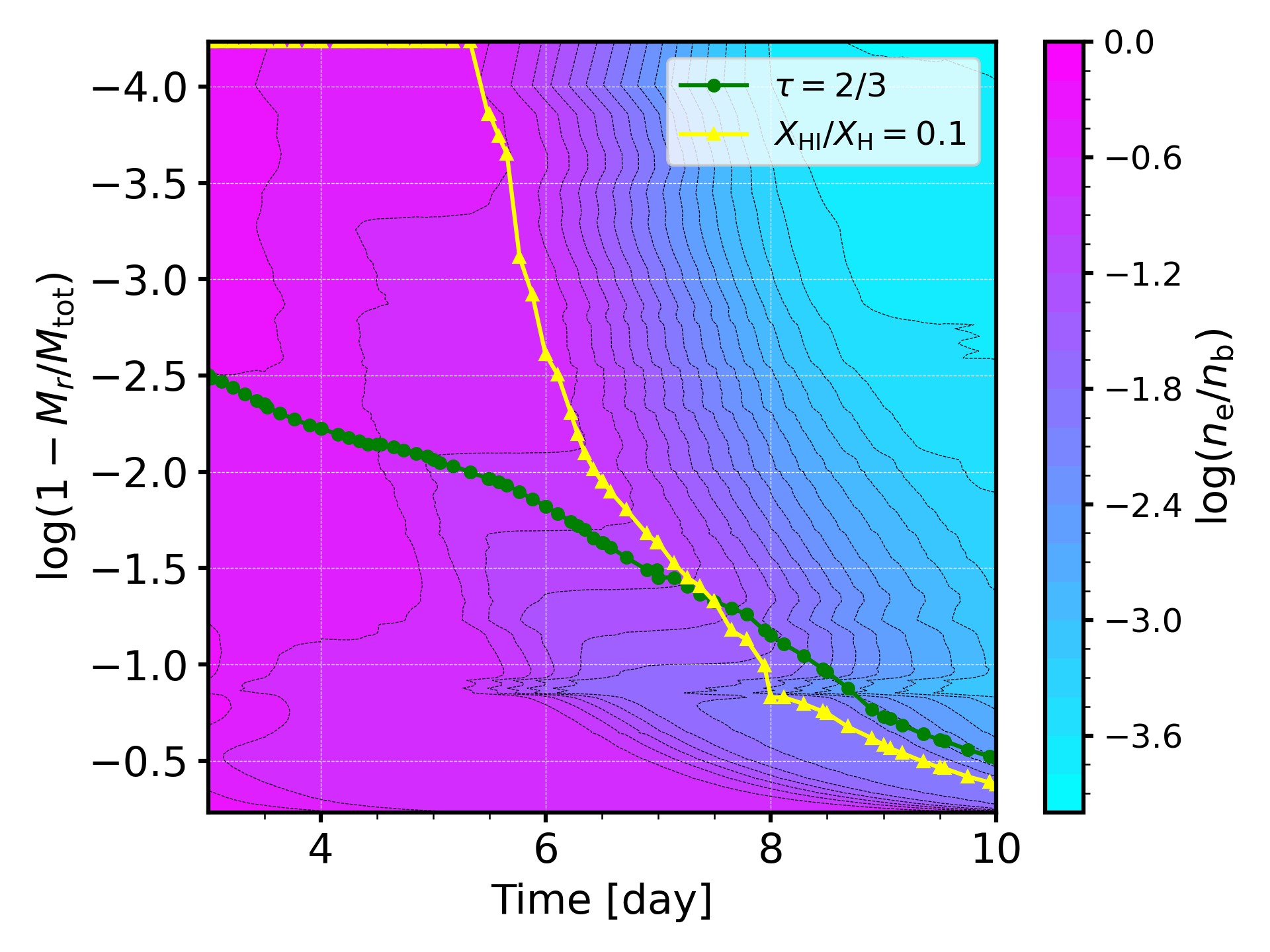}{0.47\textwidth}{Sm11p400\_mix2}}
        \gridline{\leftfig{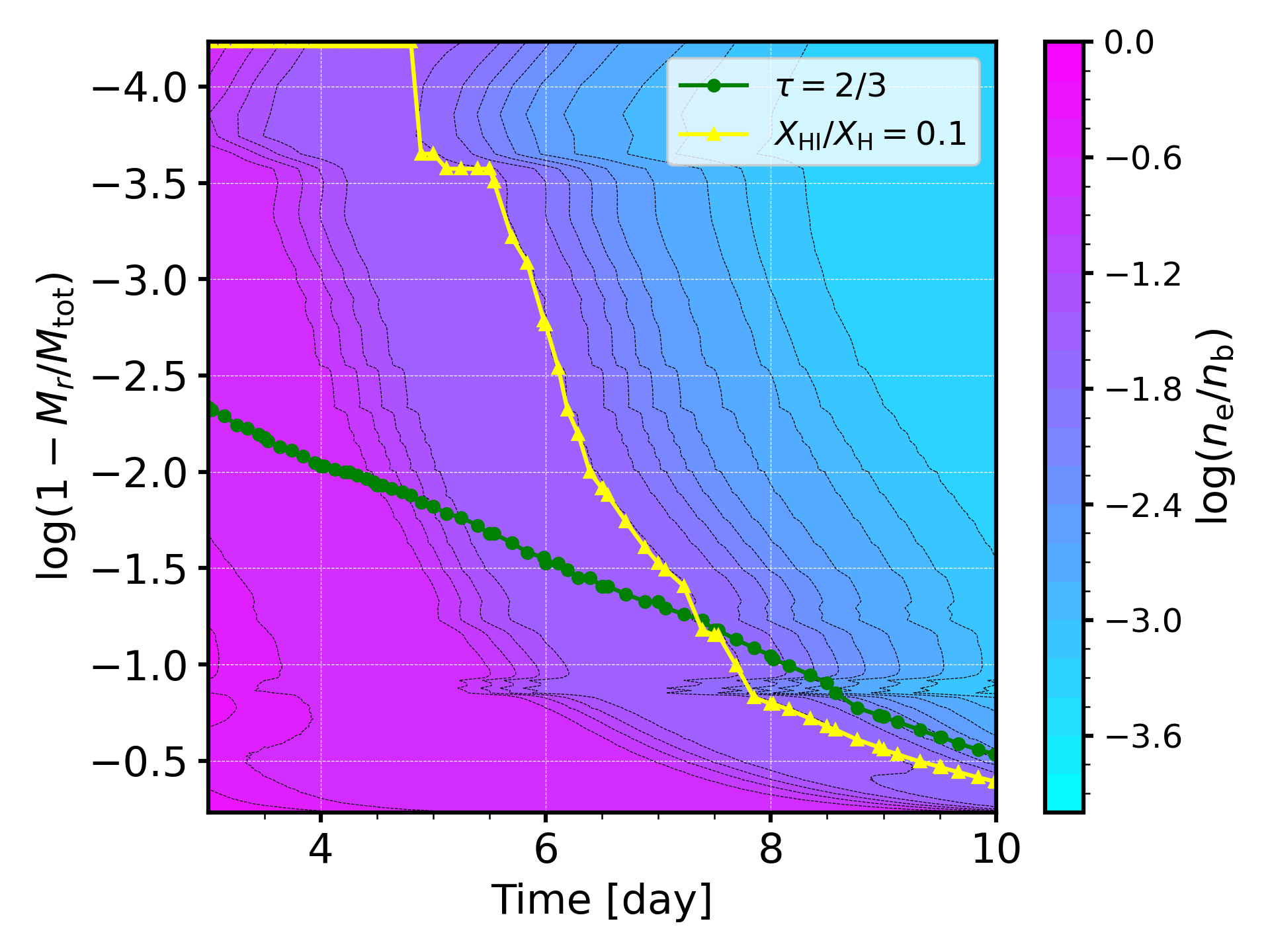}{0.47\textwidth}{Sm11p400\_mix3}}
        \caption{Same with Figure \ref{fig:p200_nenbKPH}, but for the Sm11p400 models. 
        \label{fig:p400_nenbKPH}}
        \end{figure*}       


\section{Comparison to observations} \label{sec:obs}

    \begin{figure}
    \gridline{\fig{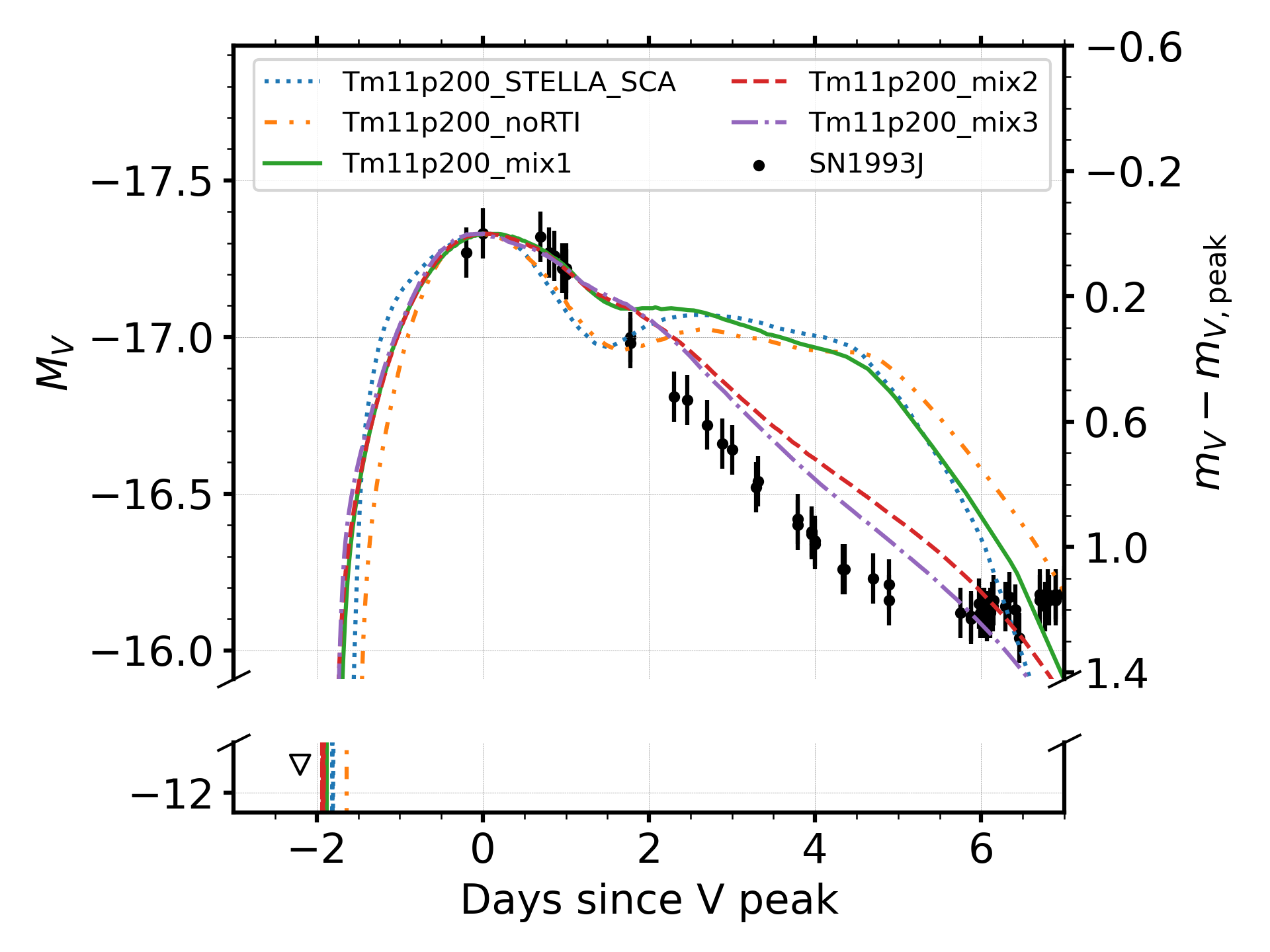}{0.47\textwidth}{}}
    \gridline{\fig{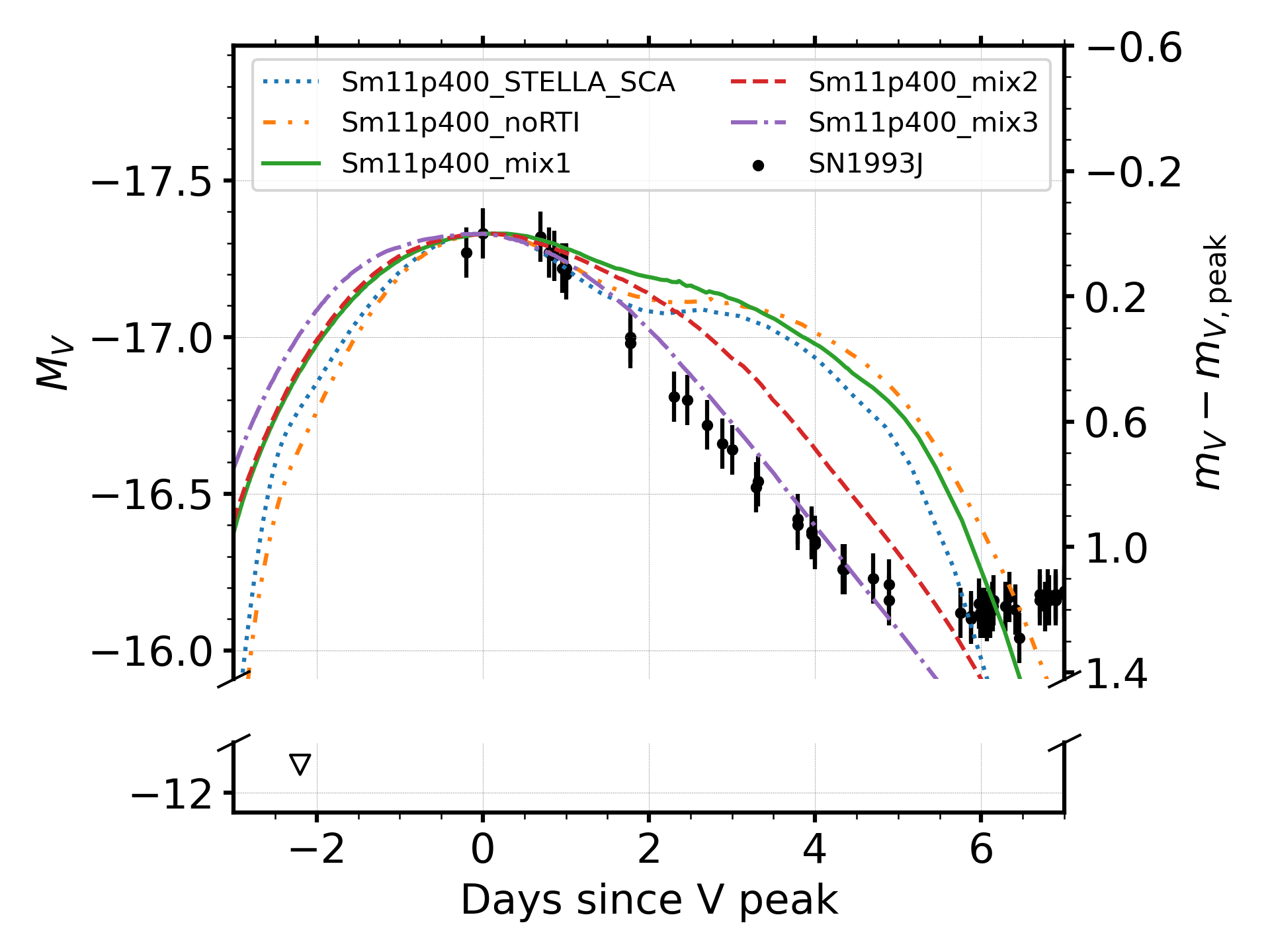}{0.47\textwidth}{}}
    \caption{$V$-band light curve of SN 1993J (black points, data from \citet{Guillochon17}) is compared to the Tm11p200 (upper panel) and Sm11p400 (lower panel) models. SN 1993J is shown in absolute magnitudes, adopting the distance of $d=3.63\pm0.14$ Mpc to M81 by \citet{Durrell10} and Galactic Extinction of 0.220 mags by \citet{Schlafly11}. The time and magnitudes of the models are shifted with respect to the $V$-band peak of each model. 
    \label{fig:SN1993J_models}}
    \end{figure}

In Figure \ref{fig:SN1993J_models}, we compare the $V$-band first peak of SN 1993J to that of Tm11p200 and Sm11p400 models. The Tm11p200 models show more or less the same initial rise to the peak, but the post-peak evolution differs greatly for varying degrees of chemical mixing in the ejecta. The STELLA\_SCA, noRTI, and mix1 models show a plateau-like feature that lasts for four days after the peak while the mix2 and mix3 models show a quasi-linear decline with a slope of $\sim$0.3 mags/day. The post-peak decline of SN 1993J appears to be almost linear during the same time period. The rise times of the Tm11p200 models agree with the non-detection limit shown in a downward triangle. However, the $V$-band peak brightness of the models ($M_{V}>-16.5$; See the lower left panel of Figure \ref{fig:mixing_results}) is too low compared to $M_V=-17.3$ of SN 1993J. In panel (b), the Sm11p400 models are compared to SN 1993J. The $V$-band peak magnitude $M_{V,\mathrm{peak}}=-17.3$ of the Sm11p400\_STELLA\_SCA model is in agreement with the observation (see Table \ref{tab:snparams}), but the rise times of the Sm11p400 models are too long. 

    We are unable to find a model that can reproduce the fast rise time and the bright optical first peak of SN 1993J simultaneously in our model grid, and it would require a combination of different explosion parameters and new progenitor models having different H-rich envelope structures. For example, we would probably need a progenitor with a H-rich envelope as extended as Sm11p400 but with a reduced mass to explain SN 1993J. Note also that CSM interaction of the SN ejecta at early times might lead to the excessive brightness that cannot be easily explained by standard progenitor models \citep{Pellegrino23}.

    In Figure \ref{fig:observed}, we find an indication of a semi-plateau feature in the $V$-band light curve of SN 2013df around $t \simeq 6$~d, but the photometric uncertainty is large and such feature is absent in $B$- and $R$-bands \citep{MoralesGaroffolo14}. The $V$-band data of SN 2011dh only contain a single data point that can be identified as the first peak \citep{Arcavi11}. SN 1993J and SN 2016gkg display distinct initial peaks, which are subsequently followed by a nearly straight decline of over 1.0 magnitude until the rebrightening phase caused by the heating of \nifs{} \citep{Richmond94,Bersten18}. ZTF18aalrxas is another SN IIb that can be identified with a quasi-linear decline from the first peak \citep{Fremling19}. High-cadence photometry of SN 2017jgh performed with \textit{Kepler} shows a peculiar first peak, but the photometric uncertainty is too large \citep{Armstrong21}. In the case of ground-based photometry, an $i$-band data point near $t\simeq13$ d before the \nifs peak seems to imply a plateau-like feature, yet it is absent in $g$- and $r$-bands. 
    
    Therefore, we conclude that the majority of the observed double-peaked SNe IIb have a quasi-linear decline phase right after the first peak.\footnote{SN 2017czd is an exception, which shows a plateau in its light curves for 20 days after the discovery. \citet{Nakaoka19} estimates its explosion energy to be 0.5\bethe{} and \nifs{} mass to be 0.003\msun, both lower than most SNe IIb.  Also, the inferred H-rich envelope mass is 0.4\msun, much larger than those of other SNe IIb (See Table \ref{tab:presnparams}).} This provides evidence for a significant chemical mixing between the H-rich envelope and the He-core in the ejecta, implying that we have to consider the effects of chemical mixing when inferring the progenitor properties from the early-time light curves of SNe IIb.

    A few issues need to be addressed before firmly connecting the linear decline from the first peak of the SN IIb light curve and the chemical mixing. First, in this study, we do not consider the possible presence of CSM around the progenitor, which may affect the light curves at early times as in the case of SNe IIP \citep[e.g.][]{Moriya11,Moriya18,Dessart13,GonzoGaitan15,Morozova17,Forster18,Kozyreva22}. Second, our discussion in Section \ref{sec:chem} is based on a simple approximation of multi-D phenomena by a 1D framework that does not fully incorporate multi-D aspects of hydrodynamic instabilities such as density clumping and non-spherical matter and velocity structures of the ejecta. We treat the chemical mixing as a microscopic mixing in this study rather than a multi-dimensional macroscopic mixing. The difference between the two affects the nebular phase spectra strongly \citep{Fransson89,Ergon22}. Previous non-LTE calculations also show that density clumping tends to reduce the degree of ionization and the effective opacity, which might affect the light curves during the photospheric phase \citep{Dessart18clumping,Dessart21,Ergon22}. 

    Note also that the mixing between the H-rich envelope and the He core is found to be weak in the 2D simulations by \citet{Iwamoto97}, in contrast to the case of SN 1987A and SNe IIP, because of the relatively small H-rich envelope mass. However, this study was based on the assumption of spherical symmetric SN explosion. Observations of the Cas A SN IIb remnant reveal strong asymmetric structures of the ejecta \citep{Lawrence95,Reed95,DeLaney10,Rest11,Milisavljevic13CasA}. By performing a 3D simulation of a neutrino-driven explosion from a SN IIb progenitor model, \citet{Wongwathanarat17} could produce an ejecta model that resembles the asymmetric morphology of radioactive $^{44}$Ti in Cas A. However, the result of \citet{Wongwathanarat17} show He/H mixing in SNe IIb near the chemical interface to be much less than in the case of SNe IIP or what is assumed in our mix2 and mix3 model, in good agreement with \citet{Iwamoto97}. 
These previous multi-D simulations only follow the evolution of the ejecta until about 1 day after the explosion, and the possibility of further mixing from the reverse shock generated at the He/H interface on a longer timescale still needs to be explored.

    \begin{figure}
    \gridline{\fig{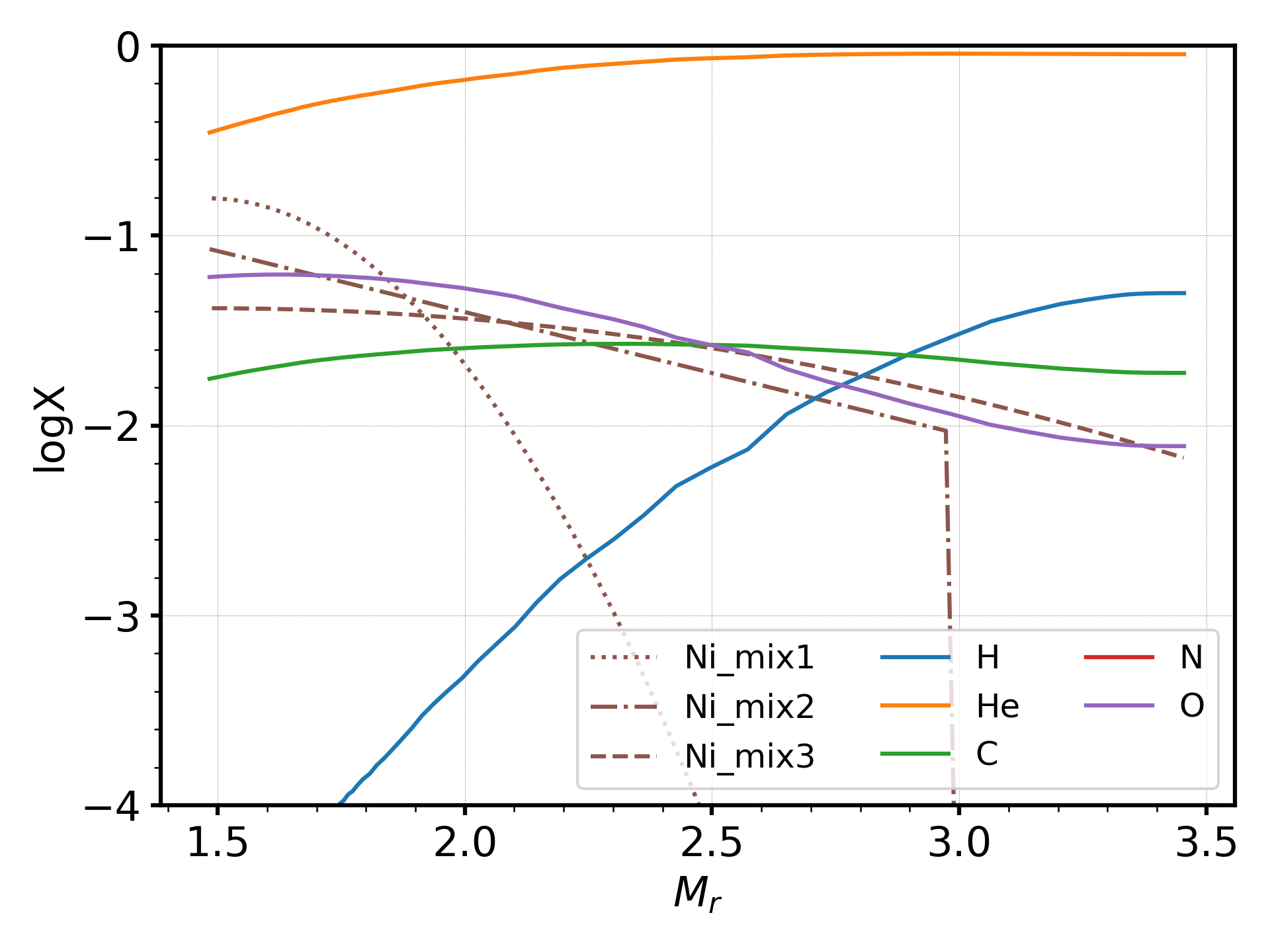}{0.47\textwidth}{}}
    \gridline{\fig{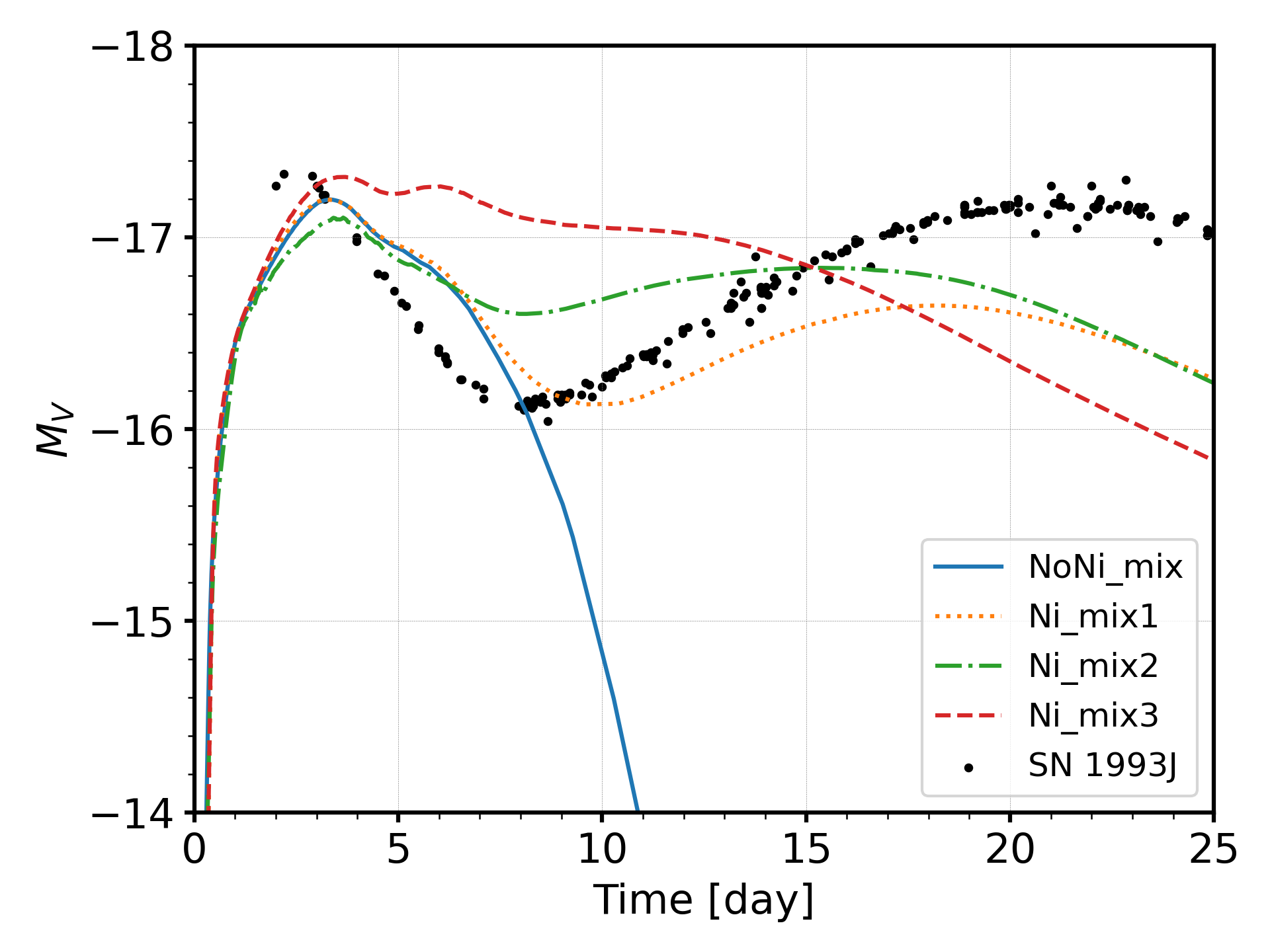}{0.47\textwidth}{}}
	    \caption{ 
\emph{Upper panel}: The initial chemical profile of the Sm11p400\_STELLA\_NoNi\_mix model. The brown lines denote \nifs{} distributions with varying degrees of mixing for Sm11p400\_STELLA\_Ni\_mix1, Sm11p400\_STELLA\_Ni\_mix2, and Sm11p400\_STELLA\_Ni\_mix3 models as indicated by the figure labels.  
\emph{Lower panel:} The corresponding $V$-band light curves. See the text for details}
    \label{fig:nickelmixing}
    \end{figure}      

    Lastly, our models are devoid of radioactive \nifs, since we focus on the first peak. If \nifs{} were also mixed into the H-rich envelope, it would have a significant impact on the early-time light curves due to the resulting radioactive heating \citep[cf.][]{Yoon19}. In order to examine the effects of \nifs mixing, we present three different SN models from the Sm11p400 progenitor in Figure \ref{fig:nickelmixing}: one without \nifs{} (Sm11p400\_STELLA\_NoNi\_mix) and three of different \nifs{} profile with varying degrees of \nifs{} mixing (Sm11p400\_STELLA\_Ni\_mix1, Ni\_mix2, and Ni\_mix3). The explosions of these models are calculated with \stella{}. The final kinetic energy of the ejecta in these models is 1.83\bethe. The level of chemical mixing in the Sm11p400\_STELLA\_NoNi\_mix model, which does not include \nifs{}, is comparable to that of the Sm11p400\_mix2 model. In the Sm11p400\_STELLA\_Ni\_mix1 and Sm11p400\_STELLA\_Ni\_mix3 models, \nifs{} is artificially introduced into the same ejecta utilized in the Sm11p400\_STELLA\_NoNi\_mix model, with a mass of $M_{\mathrm{Ni}}=0.05$\msun{}. In these two models, two different  \nifs{} mixing parameters (i.e., $f_{\mathrm{Ni}}=0.2$ and $0.8$) are adopted, where $f_\mathrm{Ni}$ is defined as the following:  
    \begin{equation} 
        X_{\mathrm{Ni}} \propto \exp\left[-\frac{(M_{\mathrm{r}}-M_{\mathrm{cut}})^2}{(f_{\mathrm{Ni}}(M_{\mathrm{tot}}-M_{\mathrm{cut}}))^2}\right]
        \label{eqn:nifsmixing}
    \end{equation}
    with $X_{\mathrm{Ni}}$, $M_{\mathrm{tot}}$, and $M_{\mathrm{cut}}$ respectively being the mass fraction of \nifs, the total mass, and the mass coordinate at the inner boundary. In this way, we assume \nifs{} follows a Gaussian distribution in each model as in \citet{Yoon19} and \citet{Jin23}.  Here $f_{\mathrm{Ni}}=0.2$ and $0.8$ correspond to the cases of relatively weak and strong mixing, respectively, as shown in the lower panel of the figure. We also construct the Ni\_mix2 model, with which we mimic the \nifs{} distribution of \citet{Wongwathanarat17} such that \nifs is only mixed in the inner 70\% of the total ejecta mass as shown in Figure~\ref{fig:nickelmixing}.

    In Figure \ref{fig:nickelmixing}, we observe that the Ni\_mix1 model with $f_\mathrm{Ni} = 0.2$ produces a double peaked optical light curve similar to those seen in SNe IIb. The time and magnitude differences between the first peak and the local brightness minimum (``valley'') are respectively $\Delta t \simeq 6.5$ days and $\Delta M \simeq 1.1$ mags in the model, which are comparable to $\Delta t \simeq 5$ days and $\Delta M \simeq 1.2$ mags of SN 1993J. However, in the case of strong \nifs{} mixing with $f_\mathrm{Ni} = 0.8$,  the heating from \nifs{} mixed into the H-rich envelope already plays an important role at early times and the main peak of the Ni\_mix3 model appears much earlier than that of the Ni\_mix1 model, almost merging with the first peak unlike any observed SNe IIb. In the case of Ni\_mix2, its optical light curve exhibits distinguished two peaks similar to Ni\_mix1, but the magnitude difference between the first peak and the valley is only $\Delta M \simeq 0.5$ mags, which is much smaller than what is observed in SN 1993J. This implies that while the first peak can be explained with a significant mixing of hydrogen and helium in the outer layers, \nifs{} should not be mixed into the outermost layers.

Note that previous studies on SNe IIb argue for a relatively strong \nifs{} mixing (i.e., up to the outer edge of the He core) in the ejecta based on the analysis of the light curve around the main peak \citep{Bersten12, Bufano14, Ergon15, Taddia18, Ergon23}. The SN IIb investigated in these studies either have a much narrower optical first peak than SN 1993J (i.e., SN 2011dh and SN 2011hs) or lack any notable early time first peak (i.e., SN 2020acat and most SNe II of the sample of \citet{Taddia18}), implying a much smaller H-rich envelope
in their progenitors than SN 1993J.  Therefore, their work cannot be directly compared to our result on the \nifs{} mixing. Note also that none of our models cannot properly reproduce the overall light curve shape of SN 1993J, and a precise constraint on the \nifs{} distribution in the ejecta still needs a comprehensive study that considers a wider parameter space in terms of the ejecta mass to energy ratio, \nifs{} mass, and SN energy. 
As discussed earlier, the morphology of the light curve might also be significantly affected by multi-dimensional effects of chemical mixing and density clumping. Future work should address these issues for the confirmation of our conclusion.

\section{Conclusion} \label{sec:con}

    Double-peaked optical light curves from SNe IIb provide crucial information for the structures of extended, H-rich envelopes of their supergiant progenitors before the explosion. The first peak produced by thermal radiation of the shock-heated envelope is affected by various factors such as the envelope mass, the envelope radius, and the explosion energy. Conversely, we can use the observed optical brightness, rise time, and width of the light curve around the first peak to infer the progenitor properties to constrain the current theories on the stellar evolution and mass-loss history of massive stars. Only a limited number of observations are available that encompass the entire duration, from the initial rise to the subsequent decline, of the first peak due to the short timescales of this feature, but with an increasing number of high-cadence transient surveys underway worldwide, coupled with immediate follow-up observations, improved statistics are anticipated. A comprehensive understanding of the nature of the first peak is therefore highly required. 

    In this study, we focus on the effects of Thomson scattering and chemical mixing on the bolometric and optical light curves. For this purpose, we calculate radiation hydrodynamics simulations of SNe IIb originating from cool supergiant progenitors that have extended H-rich envelopes with relatively low masses ($0.1\lesssim M/$\msun$\lesssim 0.2$).
    
    We compare SN light curve models computed with two different 1D-RHD codes \snec{} and \stella{} (see Section \ref{sec:snec}). In order to compute the optical light curves, \snec{} assumes blackbody radiation of the Rosseland-mean effective temperature, in contrast to \stella{} that solves multi-group radiative transfer equations while including the effects of electron scattering. The former is equivalent to assuming the opacity of the ejecta to be purely absorptive, which is not consistent with the hot ionized H-rich outer layers of the ejecta where Thomson electron scattering makes a significant contribution to the radiative transfer. We find that the first peak in the optical of the \stella{} models is fainter by more than a factor of 3 than the corresponding \snec model. This is because the effect of Thomson scattering considered in \stella{} calculations causes the thermalization depth to be located at a deeper and hotter layer compared to the corresponding \snec{} model, resulting in a higher color temperature.

    Some previous studies have utilized analytic approximations to estimate the thermalization temperature to calculate spectral energy distributions (SEDs) without the need to solve complex radiative transfer equations. However, we find that these approximations hold true only when the ejecta is considerably hotter ($T>10^5$ K) compared to its photospheric temperature when the first peak is formed.

    Additionally, we explore the influence of chemical mixing within the ejecta on the shape of the light curves around the optical peak by performing calculations of SN IIb models with varying degrees of chemical mixing (Section \ref{sec:chem}). Our results reveal that when there is no or limited chemical mixing between the H-rich envelope and the He core, a short semi-plateau phase lasting for a few days is observed. However, in the presence of strong chemical mixing, this semi-plateau phase disappears, and the light curve exhibits a quasi-linear decline from the first peak.

    

    
    By comparing our models with the well-observed SNe IIb including SN 1993J and SN 2016gkg, we find that the light curve morphology at around the first peak aligns more closely with the models with a significant chemical mixing than those with a no/weak chemical mixing (Section \ref{sec:obs}). This provides evidence for a significant mixing between the H-rich envelope and the He core in SNe IIb originating from cool supergiants. On the other hand, we also observe that a pronounced chemical mixing of \nifs{} into the outermost H-rich envelope leads to the disappearance of the double-peak feature, which contradicts the observation. To validate the existence of the implied chemical structure (significant mixing between the H-rich envelope and the He core, along with no significant mixing of \nifs{} into the H-rich outermost layers), it is essential to conduct future multi-dimensional numerical simulations and investigate the effects of density clumping and non-spherical symmetry on the light curve.
    

    In short, our findings highlight the importance of considering Thomson scattering and chemical mixing effects in double-peaked SN IIb models. Properly accounting for these factors is crucial for accurately predicting the optical brightness of the first peak and the overall shape of the early-time light curve. We plan to expand this investigation by encompassing a broader range of progenitor models and employing various numerical methods, including non-LTE radiative transfer codes like CMFGEN, which will provide a more comprehensive understanding of the properties of the SN IIb first peak.

    \begin{acknowledgments}
        This work has been supported by the National Research Foundation of Korea (NRF) grant (NRF-2019R1A2C2010885). We are grateful to Taebum Kim for creating the python package for the Kippenhahn diagram. S.B. is grateful to Marat Potashov for discussions. His work on the
development of numerical radiation hydrodynamics is supported by RSCF grant 21-11-00362.
    \end{acknowledgments}

\bibliography{references}{}

\begin{thebibliography}{}
\expandafter\ifx\csname natexlab\endcsname\relax\def\natexlab#1{#1}\fi
\providecommand{\url}[1]{\href{#1}{#1}}
\providecommand{\dodoi}[1]{doi:~\href{http://doi.org/#1}{\nolinkurl{#1}}}
\providecommand{\doeprint}[1]{\href{http://ascl.net/#1}{\nolinkurl{http://ascl.net/#1}}}
\providecommand{\doarXiv}[1]{\href{https://arxiv.org/abs/#1}{\nolinkurl{https://arxiv.org/abs/#1}}}

\bibitem[{{Aldering} {et~al.}(1994){Aldering}, {Humphreys}, \&
  {Richmond}}]{Aldering94}
{Aldering}, G., {Humphreys}, R.~M., \& {Richmond}, M. 1994, \aj, 107, 662,
  \dodoi{10.1086/116886}

\bibitem[{{Arcavi} {et~al.}(2011){Arcavi}, {Gal-Yam}, {Yaron}, {Sternberg},
  {Rabinak}, {Waxman}, {Kasliwal}, {Quimby}, {Ofek}, {Horesh}, {Kulkarni},
  {Filippenko}, {Silverman}, {Cenko}, {Li}, {Bloom}, {Sullivan}, {Nugent},
  {Poznanski}, {Gorbikov}, {Fulton}, {Howell}, {Bersier}, {Riou},
  {Lamotte-Bailey}, {Griga}, {Cohen}, {Hachinger}, {Polishook}, {Xu},
  {Ben-Ami}, {Manulis}, {Walker}, {Maguire}, {Pan}, {Matheson}, {Mazzali},
  {Pian}, {Fox}, {Gehrels}, {Law}, {James}, {Marchant}, {Smith}, {Mottram},
  {Barnsley}, {Kandrashoff}, \& {Clubb}}]{Arcavi11}
{Arcavi}, I., {Gal-Yam}, A., {Yaron}, O., {et~al.} 2011, \apjl, 742, L18,
  \dodoi{10.1088/2041-8205/742/2/L18}

\bibitem[{{Armstrong} {et~al.}(2021){Armstrong}, {Tucker}, {Rest},
  {Ridden-Harper}, {Zenati}, {Piro}, {Hinton}, {Lidman}, {Margheim}, {Narayan},
  {Shaya}, {Garnavich}, {Kasen}, {Villar}, {Zenteno}, {Arcavi}, {Drout},
  {Foley}, {Wheeler}, {Anais}, {Campillay}, {Coulter}, {Dimitriadis}, {Jones},
  {Kilpatrick}, {Mu{\~n}oz-Elgueta}, {Rojas-Bravo}, {Vargas-Gonz{\'a}lez},
  {Bulger}, {Chambers}, {Huber}, {Lowe}, {Magnier}, {Shappee}, {Smartt},
  {Smith}, {Barclay}, {Barentsen}, {Dotson}, {Gully-Santiago}, {Hedges},
  {Howell}, {Cody}, {Auchettl}, {B{\'o}di}, {Bogn{\'a}r}, {Brimacombe},
  {Brown}, {Cseh}, {Galbany}, {Hiramatsu}, {Holoien}, {Howell}, {Jha},
  {K{\"o}nyves-T{\'o}th}, {Kriskovics}, {McCully}, {Milne}, {Mu{\~n}oz}, {Pan},
  {P{\'a}l}, {Sai}, {S{\'a}rneczky}, {Smith}, {S{\'o}dor}, {Szab{\'o}},
  {Szak{\'a}ts}, {Valenti}, {Vink{\'o}}, {Wang}, {Zhang}, \&
  {Zsidi}}]{Armstrong21}
{Armstrong}, P., {Tucker}, B.~E., {Rest}, A., {et~al.} 2021, \mnras, 507, 3125,
  \dodoi{10.1093/mnras/stab2138}

\bibitem[{{Arnett}(1980)}]{Arnett80}
{Arnett}, W.~D. 1980, \apj, 237, 541, \dodoi{10.1086/157898}

\bibitem[{{Baklanov} {et~al.}(2013){Baklanov}, {Blinnikov}, {Potashov}, \&
  {Dolgov}}]{Baklanov13}
{Baklanov}, P.~V., {Blinnikov}, S.~I., {Potashov}, M.~S., \& {Dolgov}, A.~D.
  2013, Soviet Journal of Experimental and Theoretical Physics Letters, 98,
  432, \dodoi{10.1134/S0021364013200034}

\bibitem[{{Balakina} {et~al.}(2021){Balakina}, {Pruzhinskaya}, {Moskvitin},
  {Blinnikov}, {Wang}, {Xiang}, {Lin}, {Rui}, \& {Wang}}]{Balakina21}
{Balakina}, E.~A., {Pruzhinskaya}, M.~V., {Moskvitin}, A.~S., {et~al.} 2021,
  \mnras, 501, 5797, \dodoi{10.1093/mnras/staa3383}

\bibitem[{{Baschek} {et~al.}(1991){Baschek}, {Scholz}, \& {Wehrse}}]{Baschek91}
{Baschek}, B., {Scholz}, M., \& {Wehrse}, R. 1991, \aap, 246, 374

\bibitem[{{Bersten} {et~al.}(2011){Bersten}, {Benvenuto}, \&
  {Hamuy}}]{Bersten11}
{Bersten}, M.~C., {Benvenuto}, O., \& {Hamuy}, M. 2011, \apj, 729, 61,
  \dodoi{10.1088/0004-637X/729/1/61}

\bibitem[{{Bersten} {et~al.}(2012){Bersten}, {Benvenuto}, {Nomoto}, {Ergon},
  {Folatelli}, {Sollerman}, {Benetti}, {Botticella}, {Fraser}, {Kotak},
  {Maeda}, {Ochner}, \& {Tomasella}}]{Bersten12}
{Bersten}, M.~C., {Benvenuto}, O.~G., {Nomoto}, K., {et~al.} 2012, \apj, 757,
  31, \dodoi{10.1088/0004-637X/757/1/31}

\bibitem[{{Bersten} {et~al.}(2018){Bersten}, {Folatelli}, {Garc{\'\i}a}, {van
  Dyk}, {Benvenuto}, {Orellana}, {Buso}, {S{\'a}nchez}, {Tanaka}, {Maeda},
  {Filippenko}, {Zheng}, {Brink}, {Cenko}, {de Jaeger}, {Kumar}, {Moriya},
  {Nomoto}, {Perley}, {Shivvers}, \& {Smith}}]{Bersten18}
{Bersten}, M.~C., {Folatelli}, G., {Garc{\'\i}a}, F., {et~al.} 2018, \nat, 554,
  497, \dodoi{10.1038/nature25151}

\bibitem[{{Blinnikov} {et~al.}(2000){Blinnikov}, {Lundqvist}, {Bartunov},
  {Nomoto}, \& {Iwamoto}}]{Blinnikov00}
{Blinnikov}, S., {Lundqvist}, P., {Bartunov}, O., {Nomoto}, K., \& {Iwamoto},
  K. 2000, \apj, 532, 1132, \dodoi{10.1086/308588}

\bibitem[{{Blinnikov} \& {Bartunov}(1993)}]{Blinnikov93}
{Blinnikov}, S.~I., \& {Bartunov}, O.~S. 1993, \aap, 273, 106

\bibitem[{{Blinnikov} {et~al.}(1998){Blinnikov}, {Eastman}, {Bartunov},
  {Popolitov}, \& {Woosley}}]{Blinnikov98}
{Blinnikov}, S.~I., {Eastman}, R., {Bartunov}, O.~S., {Popolitov}, V.~A., \&
  {Woosley}, S.~E. 1998, \apj, 496, 454, \dodoi{10.1086/305375}

\bibitem[{{Blinnikov} {et~al.}(2006){Blinnikov}, {R{\"o}pke}, {Sorokina},
  {Gieseler}, {Reinecke}, {Travaglio}, {Hillebrandt}, \&
  {Stritzinger}}]{Blinnikov06}
{Blinnikov}, S.~I., {R{\"o}pke}, F.~K., {Sorokina}, E.~I., {et~al.} 2006, \aap,
  453, 229, \dodoi{10.1051/0004-6361:20054594}

\bibitem[{{Blinnikov} \& {Tolstov}(2011)}]{Blinnikov11}
{Blinnikov}, S.~I., \& {Tolstov}, A.~G. 2011, Astronomy Letters, 37, 194,
  \dodoi{10.1134/S1063773711010051}

\bibitem[{{Bufano} {et~al.}(2014){Bufano}, {Pignata}, {Bersten}, {Mazzali},
  {Ryder}, {Margutti}, {Milisavljevic}, {Morelli}, {Benetti}, {Cappellaro},
  {Gonzalez-Gaitan}, {Romero-Ca{\~n}izales}, {Stritzinger}, {Walker},
  {Anderson}, {Contreras}, {de Jaeger}, {F{\"o}rster}, {Gutierrez}, {Hamuy},
  {Hsiao}, {Morrell}, {Olivares E.}, {Paillas}, {Parker}, {Pian}, {Pickering},
  {Sanders}, {Stockdale}, {Turatto}, {Valenti}, {Fesen}, {Maza}, {Nomoto},
  {Phillips}, \& {Soderberg}}]{Bufano14}
{Bufano}, F., {Pignata}, G., {Bersten}, M., {et~al.} 2014, \mnras, 439, 1807,
  \dodoi{10.1093/mnras/stu065}

\bibitem[{{Castor}(2004)}]{Castor04}
{Castor}, J.~I. 2004, {Radiation Hydrodynamics}

\bibitem[{{Claeys} {et~al.}(2011){Claeys}, {de Mink}, {Pols}, {Eldridge}, \&
  {Baes}}]{Claeys11}
{Claeys}, J.~S.~W., {de Mink}, S.~E., {Pols}, O.~R., {Eldridge}, J.~J., \&
  {Baes}, M. 2011, \aap, 528, A131, \dodoi{10.1051/0004-6361/201015410}

\bibitem[{{DeLaney} {et~al.}(2010){DeLaney}, {Rudnick}, {Stage}, {Smith},
  {Isensee}, {Rho}, {Allen}, {Gomez}, {Kozasa}, {Reach}, {Davis}, \&
  {Houck}}]{DeLaney10}
{DeLaney}, T., {Rudnick}, L., {Stage}, M.~D., {et~al.} 2010, \apj, 725, 2038,
  \dodoi{10.1088/0004-637X/725/2/2038}

\bibitem[{{Dessart} {et~al.}(2021){Dessart}, {Hillier}, {Sukhbold}, {Woosley},
  \& {Janka}}]{Dessart21}
{Dessart}, L., {Hillier}, D.~J., {Sukhbold}, T., {Woosley}, S.~E., \& {Janka},
  H.~T. 2021, \aap, 652, A64, \dodoi{10.1051/0004-6361/202140839}

\bibitem[{{Dessart} {et~al.}(2013){Dessart}, {Hillier}, {Waldman}, \&
  {Livne}}]{Dessart13}
{Dessart}, L., {Hillier}, D.~J., {Waldman}, R., \& {Livne}, E. 2013, \mnras,
  433, 1745, \dodoi{10.1093/mnras/stt861}

\bibitem[{{Dessart} {et~al.}(2018{\natexlab{a}}){Dessart}, {Hillier}, \&
  {Wilk}}]{Dessart18clumping}
{Dessart}, L., {Hillier}, D.~J., \& {Wilk}, K.~D. 2018{\natexlab{a}}, \aap,
  619, A30, \dodoi{10.1051/0004-6361/201833278}

\bibitem[{{Dessart} {et~al.}(2018{\natexlab{b}}){Dessart}, {Yoon}, {Livne}, \&
  {Waldman}}]{Dessart18}
{Dessart}, L., {Yoon}, S.-C., {Livne}, E., \& {Waldman}, R. 2018{\natexlab{b}},
  \aap, 612, A61, \dodoi{10.1051/0004-6361/201732363}

\bibitem[{{Duffell}(2016)}]{Duffell16}
{Duffell}, P.~C. 2016, \apj, 821, 76, \dodoi{10.3847/0004-637X/821/2/76}

\bibitem[{{Durrell} {et~al.}(2010){Durrell}, {Sarajedini}, \&
  {Chandar}}]{Durrell10}
{Durrell}, P.~R., {Sarajedini}, A., \& {Chandar}, R. 2010, \apj, 718, 1118,
  \dodoi{10.1088/0004-637X/718/2/1118}

\bibitem[{{Eastman} \& {Pinto}(1993)}]{Eastman93}
{Eastman}, R.~G., \& {Pinto}, P.~A. 1993, \apj, 412, 731,
  \dodoi{10.1086/172957}

\bibitem[{{Eldridge} {et~al.}(2018){Eldridge}, {Xiao}, {Stanway}, {Rodrigues},
  \& {Guo}}]{Eldridge18}
{Eldridge}, J.~J., {Xiao}, L., {Stanway}, E.~R., {Rodrigues}, N., \& {Guo},
  N.~Y. 2018, \pasa, 35, e049, \dodoi{10.1017/pasa.2018.47}

\bibitem[{{Ensman} \& {Burrows}(1992)}]{Ensman92}
{Ensman}, L., \& {Burrows}, A. 1992, \apj, 393, 742, \dodoi{10.1086/171542}

\bibitem[{{Ergon} \& {Fransson}(2022)}]{Ergon22}
{Ergon}, M., \& {Fransson}, C. 2022, \aap, 666, A104,
  \dodoi{10.1051/0004-6361/202243448}

\bibitem[{{Ergon} {et~al.}(2015){Ergon}, {Jerkstrand}, {Sollerman},
  {Elias-Rosa}, {Fransson}, {Fraser}, {Pastorello}, {Kotak}, {Taubenberger},
  {Tomasella}, {Valenti}, {Benetti}, {Helou}, {Kasliwal}, {Maund}, {Smartt}, \&
  {Spyromilio}}]{Ergon15}
{Ergon}, M., {Jerkstrand}, A., {Sollerman}, J., {et~al.} 2015, \aap, 580, A142,
  \dodoi{10.1051/0004-6361/201424592}

\bibitem[{{Ergon} {et~al.}(2023){Ergon}, {Lundqvist}, {Fransson},
  {Kuncarayakti}, {Das}, {De}, {Ferrari}, {Fremling}, {Medler}, {Maeda},
  {Pastorello}, {Sollerman}, \& {Stritzinger}}]{Ergon23}
{Ergon}, M., {Lundqvist}, P., {Fransson}, C., {et~al.} 2023, arXiv e-prints,
  arXiv:2308.07158, \dodoi{10.48550/arXiv.2308.07158}

\bibitem[{{F{\"o}rster} {et~al.}(2018){F{\"o}rster}, {Moriya}, {Maureira},
  {Anderson}, {Blinnikov}, {Bufano}, {Cabrera-Vives}, {Clocchiatti}, {de
  Jaeger}, {Est{\'e}vez}, {Galbany}, {Gonz{\'a}lez-Gait{\'a}n}, {Gr{\"a}fener},
  {Hamuy}, {Hsiao}, {Huentelemu}, {Huijse}, {Kuncarayakti}, {Mart{\'\i}nez},
  {Medina}, {Olivares E.}, {Pignata}, {Razza}, {Reyes}, {San Mart{\'\i}n},
  {Smith}, {Vera}, {Vivas}, {de Ugarte Postigo}, {Yoon}, {Ashall}, {Fraser},
  {Gal-Yam}, {Kankare}, {Le Guillou}, {Mazzali}, {Walton}, \&
  {Young}}]{Forster18}
{F{\"o}rster}, F., {Moriya}, T.~J., {Maureira}, J.~C., {et~al.} 2018, Nature
  Astronomy, 2, 808, \dodoi{10.1038/s41550-018-0563-4}

\bibitem[{{Fransson} \& {Chevalier}(1989)}]{Fransson89}
{Fransson}, C., \& {Chevalier}, R.~A. 1989, \apj, 343, 323,
  \dodoi{10.1086/167707}

\bibitem[{{Fremling} {et~al.}(2019){Fremling}, {Ko}, {Dugas}, {Ergon},
  {Sollerman}, {Bagdasaryan}, {Barbarino}, {Belicki}, {Bellm}, {Blagorodnova},
  {De}, {Dekany}, {Frederick}, {Gal-Yam}, {Goldstein}, {Golkhou}, {Graham},
  {Kasliwal}, {Kowalski}, {Kulkarni}, {Kupfer}, {Laher}, {Masci}, {Miller},
  {Neill}, {Perley}, {Rebbapragada}, {Riddle}, {Rusholme}, {Schulze}, {Smith},
  {Tartaglia}, {Yan}, \& {Yao}}]{Fremling19}
{Fremling}, C., {Ko}, H., {Dugas}, A., {et~al.} 2019, \apjl, 878, L5,
  \dodoi{10.3847/2041-8213/ab218f}

\bibitem[{{Friend} \& {Castor}(1983)}]{Friend83}
{Friend}, D.~B., \& {Castor}, J.~I. 1983, \apj, 272, 259,
  \dodoi{10.1086/161289}

\bibitem[{{Gonz{\'a}lez-Gait{\'a}n} {et~al.}(2015){Gonz{\'a}lez-Gait{\'a}n},
  {Tominaga}, {Molina}, {Galbany}, {Bufano}, {Anderson}, {Gutierrez},
  {F{\"o}rster}, {Pignata}, {Bersten}, {Howell}, {Sullivan}, {Carlberg}, {de
  Jaeger}, {Hamuy}, {Baklanov}, \& {Blinnikov}}]{GonzoGaitan15}
{Gonz{\'a}lez-Gait{\'a}n}, S., {Tominaga}, N., {Molina}, J., {et~al.} 2015,
  \mnras, 451, 2212, \dodoi{10.1093/mnras/stv1097}

\bibitem[{{Guillochon} {et~al.}(2017){Guillochon}, {Parrent}, {Kelley}, \&
  {Margutti}}]{Guillochon17}
{Guillochon}, J., {Parrent}, J., {Kelley}, L.~Z., \& {Margutti}, R. 2017, \apj,
  835, 64, \dodoi{10.3847/1538-4357/835/1/64}

\bibitem[{{Hammer} {et~al.}(2010){Hammer}, {Janka}, \& {M{\"u}ller}}]{Hammer10}
{Hammer}, N.~J., {Janka}, H.~T., \& {M{\"u}ller}, E. 2010, \apj, 714, 1371,
  \dodoi{10.1088/0004-637X/714/2/1371}

\bibitem[{{Iwamoto} {et~al.}(1997){Iwamoto}, {Young}, {Nakasato}, {Shigeyama},
  {Nomoto}, {Hachisu}, \& {Saio}}]{Iwamoto97}
{Iwamoto}, K., {Young}, T.~R., {Nakasato}, N., {et~al.} 1997, \apj, 477, 865,
  \dodoi{10.1086/303729}

\bibitem[{{Jerkstrand} {et~al.}(2015){Jerkstrand}, {Ergon}, {Smartt},
  {Fransson}, {Sollerman}, {Taubenberger}, {Bersten}, \&
  {Spyromilio}}]{Jerkstrand15}
{Jerkstrand}, A., {Ergon}, M., {Smartt}, S.~J., {et~al.} 2015, \aap, 573, A12,
  \dodoi{10.1051/0004-6361/201423983}

\bibitem[{{Jin} {et~al.}(2023){Jin}, {Yoon}, \& {Blinnikov}}]{Jin23}
{Jin}, H., {Yoon}, S.-C., \& {Blinnikov}, S. 2023, \apj, 950, 44,
  \dodoi{10.3847/1538-4357/accf0d}

\bibitem[{{Kasen} \& {Woosley}(2009)}]{Kasen09}
{Kasen}, D., \& {Woosley}, S.~E. 2009, \apj, 703, 2205,
  \dodoi{10.1088/0004-637X/703/2/2205}

\bibitem[{{Kifonidis} {et~al.}(2003){Kifonidis}, {Plewa}, {Janka}, \&
  {M{\"u}ller}}]{Kifonidis03}
{Kifonidis}, K., {Plewa}, T., {Janka}, H.~T., \& {M{\"u}ller}, E. 2003, \aap,
  408, 621, \dodoi{10.1051/0004-6361:20030863}

\bibitem[{{Kozyreva} {et~al.}(2022){Kozyreva}, {Klencki}, {Filippenko},
  {Baklanov}, {Mironov}, {Justham}, \& {Chiavassa}}]{Kozyreva22}
{Kozyreva}, A., {Klencki}, J., {Filippenko}, A.~V., {et~al.} 2022, \apjl, 934,
  L31, \dodoi{10.3847/2041-8213/ac835a}

\bibitem[{{Kozyreva} {et~al.}(2020){Kozyreva}, {Shingles}, {Mironov},
  {Baklanov}, \& {Blinnikov}}]{Kozyreva20}
{Kozyreva}, A., {Shingles}, L., {Mironov}, A., {Baklanov}, P., \& {Blinnikov},
  S. 2020, \mnras, 499, 4312, \dodoi{10.1093/mnras/staa2704}

\bibitem[{{Kromer} \& {Sim}(2009)}]{Kromer09}
{Kromer}, M., \& {Sim}, S.~A. 2009, \mnras, 398, 1809,
  \dodoi{10.1111/j.1365-2966.2009.15256.x}

\bibitem[{{Lawrence} {et~al.}(1995){Lawrence}, {MacAlpine}, {Uomoto},
  {Woodgate}, {Brown}, {Oliversen}, {Lowenthal}, \& {Liu}}]{Lawrence95}
{Lawrence}, S.~S., {MacAlpine}, G.~M., {Uomoto}, A., {et~al.} 1995, \aj, 109,
  2635, \dodoi{10.1086/117477}

\bibitem[{{Maund} {et~al.}(2004){Maund}, {Smartt}, {Kudritzki},
  {Podsiadlowski}, \& {Gilmore}}]{Maund04}
{Maund}, J.~R., {Smartt}, S.~J., {Kudritzki}, R.~P., {Podsiadlowski}, P., \&
  {Gilmore}, G.~F. 2004, \nat, 427, 129, \dodoi{10.1038/nature02161}

\bibitem[{{Maund} {et~al.}(2011){Maund}, {Fraser}, {Ergon}, {Pastorello},
  {Smartt}, {Sollerman}, {Benetti}, {Botticella}, {Bufano}, {Danziger},
  {Kotak}, {Magill}, {Stephens}, \& {Valenti}}]{Maund11}
{Maund}, J.~R., {Fraser}, M., {Ergon}, M., {et~al.} 2011, \apjl, 739, L37,
  \dodoi{10.1088/2041-8205/739/2/L37}

\bibitem[{{Meynet} {et~al.}(2015){Meynet}, {Chomienne}, {Ekstr{\"o}m},
  {Georgy}, {Granada}, {Groh}, {Maeder}, {Eggenberger}, {Levesque}, \&
  {Massey}}]{Meynet15}
{Meynet}, G., {Chomienne}, V., {Ekstr{\"o}m}, S., {et~al.} 2015, \aap, 575,
  A60, \dodoi{10.1051/0004-6361/201424671}

\bibitem[{{Mihalas}(1978)}]{Mihalas78}
{Mihalas}, D. 1978, {Stellar atmospheres}

\bibitem[{{Mihalas} \& {Mihalas}(1984)}]{Mihalas84}
{Mihalas}, D., \& {Mihalas}, B.~W. 1984, {Foundations of radiation
  hydrodynamics}

\bibitem[{{Milisavljevic} \& {Fesen}(2013)}]{Milisavljevic13CasA}
{Milisavljevic}, D., \& {Fesen}, R.~A. 2013, \apj, 772, 134,
  \dodoi{10.1088/0004-637X/772/2/134}

\bibitem[{{Morales-Garoffolo} {et~al.}(2014){Morales-Garoffolo}, {Elias-Rosa},
  {Benetti}, {Taubenberger}, {Cappellaro}, {Pastorello}, {Klauser}, {Valenti},
  {Howerton}, {Ochner}, {Schramm}, {Siviero}, {Tartaglia}, \&
  {Tomasella}}]{MoralesGaroffolo14}
{Morales-Garoffolo}, A., {Elias-Rosa}, N., {Benetti}, S., {et~al.} 2014,
  \mnras, 445, 1647, \dodoi{10.1093/mnras/stu1837}

\bibitem[{{Moriya} {et~al.}(2011){Moriya}, {Tominaga}, {Blinnikov}, {Baklanov},
  \& {Sorokina}}]{Moriya11}
{Moriya}, T., {Tominaga}, N., {Blinnikov}, S.~I., {Baklanov}, P.~V., \&
  {Sorokina}, E.~I. 2011, \mnras, 415, 199,
  \dodoi{10.1111/j.1365-2966.2011.18689.x}

\bibitem[{{Moriya} {et~al.}(2018){Moriya}, {F{\"o}rster}, {Yoon},
  {Gr{\"a}fener}, \& {Blinnikov}}]{Moriya18}
{Moriya}, T.~J., {F{\"o}rster}, F., {Yoon}, S.-C., {Gr{\"a}fener}, G., \&
  {Blinnikov}, S.~I. 2018, \mnras, 476, 2840, \dodoi{10.1093/mnras/sty475}

\bibitem[{{Moriya} {et~al.}(2016){Moriya}, {Pruzhinskaya}, {Ergon}, \&
  {Blinnikov}}]{Moriya16}
{Moriya}, T.~J., {Pruzhinskaya}, M.~V., {Ergon}, M., \& {Blinnikov}, S.~I.
  2016, \mnras, 455, 423, \dodoi{10.1093/mnras/stv2336}

\bibitem[{{Morozova} {et~al.}(2015){Morozova}, {Piro}, {Renzo}, {Ott},
  {Clausen}, {Couch}, {Ellis}, \& {Roberts}}]{Morozova15}
{Morozova}, V., {Piro}, A.~L., {Renzo}, M., {et~al.} 2015, \apj, 814, 63,
  \dodoi{10.1088/0004-637X/814/1/63}

\bibitem[{{Morozova} {et~al.}(2017){Morozova}, {Piro}, \&
  {Valenti}}]{Morozova17}
{Morozova}, V., {Piro}, A.~L., \& {Valenti}, S. 2017, \apj, 838, 28,
  \dodoi{10.3847/1538-4357/aa6251}

\bibitem[{{Nagy} \& {Vink{\'o}}(2016)}]{NagyVinko16}
{Nagy}, A.~P., \& {Vink{\'o}}, J. 2016, \aap, 589, A53,
  \dodoi{10.1051/0004-6361/201527931}

\bibitem[{{Nakaoka} {et~al.}(2019){Nakaoka}, {Moriya}, {Tanaka}, {Yamanaka},
  {Kawabata}, {Maeda}, {Kawabata}, {Kawahara}, {Itagaki}, {Ouchi}, {Blinnikov},
  {Tominaga}, \& {Uemura}}]{Nakaoka19}
{Nakaoka}, T., {Moriya}, T.~J., {Tanaka}, M., {et~al.} 2019, \apj, 875, 76,
  \dodoi{10.3847/1538-4357/ab0dfe}

\bibitem[{{Nakar} \& {Piro}(2014)}]{NakarPiro14}
{Nakar}, E., \& {Piro}, A.~L. 2014, \apj, 788, 193,
  \dodoi{10.1088/0004-637X/788/2/193}

\bibitem[{Nomoto {et~al.}(1995)Nomoto, Iwamoto, \& Suzuki}]{NOMOTO1995173}
Nomoto, K., Iwamoto, K., \& Suzuki, T. 1995, Physics Reports, 256, 173,
  \dodoi{https://doi.org/10.1016/0370-1573(94)00107-E}

\bibitem[{{Paxton} {et~al.}(2011){Paxton}, {Bildsten}, {Dotter}, {Herwig},
  {Lesaffre}, \& {Timmes}}]{Paxton11}
{Paxton}, B., {Bildsten}, L., {Dotter}, A., {et~al.} 2011, \apjs, 192, 3,
  \dodoi{10.1088/0067-0049/192/1/3}

\bibitem[{{Paxton} {et~al.}(2013){Paxton}, {Cantiello}, {Arras}, {Bildsten},
  {Brown}, {Dotter}, {Mankovich}, {Montgomery}, {Stello}, {Timmes}, \&
  {Townsend}}]{Paxton13}
{Paxton}, B., {Cantiello}, M., {Arras}, P., {et~al.} 2013, \apjs, 208, 4,
  \dodoi{10.1088/0067-0049/208/1/4}

\bibitem[{{Paxton} {et~al.}(2015){Paxton}, {Marchant}, {Schwab}, {Bauer},
  {Bildsten}, {Cantiello}, {Dessart}, {Farmer}, {Hu}, {Langer}, {Townsend},
  {Townsley}, \& {Timmes}}]{Paxton15}
{Paxton}, B., {Marchant}, P., {Schwab}, J., {et~al.} 2015, \apjs, 220, 15,
  \dodoi{10.1088/0067-0049/220/1/15}

\bibitem[{{Paxton} {et~al.}(2018){Paxton}, {Schwab}, {Bauer}, {Bildsten},
  {Blinnikov}, {Duffell}, {Farmer}, {Goldberg}, {Marchant}, {Sorokina},
  {Thoul}, {Townsend}, \& {Timmes}}]{Paxton18}
{Paxton}, B., {Schwab}, J., {Bauer}, E.~B., {et~al.} 2018, \apjs, 234, 34,
  \dodoi{10.3847/1538-4365/aaa5a8}

\bibitem[{{Paxton} {et~al.}(2019){Paxton}, {Smolec}, {Schwab}, {Gautschy},
  {Bildsten}, {Cantiello}, {Dotter}, {Farmer}, {Goldberg}, {Jermyn}, {Kanbur},
  {Marchant}, {Thoul}, {Townsend}, {Wolf}, {Zhang}, \& {Timmes}}]{Paxton19}
{Paxton}, B., {Smolec}, R., {Schwab}, J., {et~al.} 2019, \apjs, 243, 10,
  \dodoi{10.3847/1538-4365/ab2241}

\bibitem[{Pellegrino {et~al.}(2023)Pellegrino, Hiramatsu, Arcavi, Howell,
  Bostroem, Brown, Burke, Elias-Rosa, Itagaki, Kaneda, McCully, Modjaz,
  Gonzalez, \& Pritchard}]{Pellegrino23}
Pellegrino, C., Hiramatsu, D., Arcavi, I., {et~al.} 2023, SN 2020bio: A
  Double-peaked Type IIb Supernova with Evidence of Early-time Circumstellar
  Interaction,  arXiv, \dodoi{10.48550/ARXIV.2301.04662}

\bibitem[{{Pinto} \& {Eastman}(2000)}]{Pinto00}
{Pinto}, P.~A., \& {Eastman}, R.~G. 2000, \apj, 530, 757,
  \dodoi{10.1086/308380}

\bibitem[{{Piro} {et~al.}(2017){Piro}, {Muhleisen}, {Arcavi}, {Sand},
  {Tartaglia}, \& {Valenti}}]{Piro17}
{Piro}, A.~L., {Muhleisen}, M., {Arcavi}, I., {et~al.} 2017, \apj, 846, 94,
  \dodoi{10.3847/1538-4357/aa8595}

\bibitem[{{Podsiadlowski} {et~al.}(1993){Podsiadlowski}, {Hsu}, {Joss}, \&
  {Ross}}]{Podsiadlowski93}
{Podsiadlowski}, P., {Hsu}, J.~J.~L., {Joss}, P.~C., \& {Ross}, R.~R. 1993,
  \nat, 364, 509, \dodoi{10.1038/364509a0}

\bibitem[{{Podsiadlowski} {et~al.}(1992){Podsiadlowski}, {Joss}, \&
  {Hsu}}]{Podsiadlowski92}
{Podsiadlowski}, P., {Joss}, P.~C., \& {Hsu}, J.~J.~L. 1992, \apj, 391, 246,
  \dodoi{10.1086/171341}

\bibitem[{{Potashov} {et~al.}(2021){Potashov}, {Blinnikov}, \&
  {Sorokina}}]{Potashov21}
{Potashov}, M.~S., {Blinnikov}, S.~I., \& {Sorokina}, E.~I. 2021, Astronomy
  Letters, 47, 204, \dodoi{10.1134/S1063773721030051}

\bibitem[{{Reed} {et~al.}(1995){Reed}, {Hester}, {Fabian}, \&
  {Winkler}}]{Reed95}
{Reed}, J.~E., {Hester}, J.~J., {Fabian}, A.~C., \& {Winkler}, P.~F. 1995,
  \apj, 440, 706, \dodoi{10.1086/175308}

\bibitem[{{Rest} {et~al.}(2011){Rest}, {Foley}, {Sinnott}, {Welch}, {Badenes},
  {Filippenko}, {Bergmann}, {Bhatti}, {Blondin}, {Challis}, {Damke}, {Finley},
  {Huber}, {Kasen}, {Kirshner}, {Matheson}, {Mazzali}, {Minniti}, {Nakajima},
  {Narayan}, {Olsen}, {Sauer}, {Smith}, \& {Suntzeff}}]{Rest11}
{Rest}, A., {Foley}, R.~J., {Sinnott}, B., {et~al.} 2011, \apj, 732, 3,
  \dodoi{10.1088/0004-637X/732/1/3}

\bibitem[{{Richmond} {et~al.}(1994){Richmond}, {Treffers}, {Filippenko},
  {Paik}, {Leibundgut}, {Schulman}, \& {Cox}}]{Richmond94}
{Richmond}, M.~W., {Treffers}, R.~R., {Filippenko}, A.~V., {et~al.} 1994, \aj,
  107, 1022, \dodoi{10.1086/116915}

\bibitem[{{Schlafly} \& {Finkbeiner}(2011)}]{Schlafly11}
{Schlafly}, E.~F., \& {Finkbeiner}, D.~P. 2011, \apj, 737, 103,
  \dodoi{10.1088/0004-637X/737/2/103}

\bibitem[{{Shigeyama} \& {Nomoto}(1990)}]{Shigeyama90}
{Shigeyama}, T., \& {Nomoto}, K. 1990, \apj, 360, 242, \dodoi{10.1086/169114}

\bibitem[{{Smith}(2014)}]{Smith14}
{Smith}, N. 2014, \araa, 52, 487, \dodoi{10.1146/annurev-astro-081913-040025}

\bibitem[{{Stancliffe} \& {Eldridge}(2009)}]{Stancliffe09}
{Stancliffe}, R.~J., \& {Eldridge}, J.~J. 2009, \mnras, 396, 1699,
  \dodoi{10.1111/j.1365-2966.2009.14849.x}

\bibitem[{{Swartz} {et~al.}(1991){Swartz}, {Wheeler}, \& {Harkness}}]{Swartz91}
{Swartz}, D.~A., {Wheeler}, J.~C., \& {Harkness}, R.~P. 1991, \apj, 374, 266,
  \dodoi{10.1086/170115}

\bibitem[{{Taddia} {et~al.}(2018){Taddia}, {Stritzinger}, {Bersten}, {Baron},
  {Burns}, {Contreras}, {Holmbo}, {Hsiao}, {Morrell}, {Phillips}, {Sollerman},
  \& {Suntzeff}}]{Taddia18}
{Taddia}, F., {Stritzinger}, M.~D., {Bersten}, M., {et~al.} 2018, \aap, 609,
  A136, \dodoi{10.1051/0004-6361/201730844}

\bibitem[{{Tartaglia} {et~al.}(2017){Tartaglia}, {Fraser}, {Sand}, {Valenti},
  {Smartt}, {McCully}, {Anderson}, {Arcavi}, {Elias-Rosa}, {Galbany},
  {Gal-Yam}, {Haislip}, {Hosseinzadeh}, {Howell}, {Inserra}, {Jha}, {Kankare},
  {Lundqvist}, {Maguire}, {Mattila}, {Reichart}, {Smith}, {Smith},
  {Stritzinger}, {Sullivan}, {Taddia}, \& {Tomasella}}]{Tartaglia17}
{Tartaglia}, L., {Fraser}, M., {Sand}, D.~J., {et~al.} 2017, \apjl, 836, L12,
  \dodoi{10.3847/2041-8213/aa5c7f}

\bibitem[{{Tsvetkov} {et~al.}(2009){Tsvetkov}, {Volkov}, {Baklanov},
  {Blinnikov}, \& {Tuchin}}]{Tsvetkov09}
{Tsvetkov}, D.~Y., {Volkov}, I.~M., {Baklanov}, P., {Blinnikov}, S., \&
  {Tuchin}, O. 2009, Peremennye Zvezdy, 29, 2.
\newblock \doarXiv{0910.4242}

\bibitem[{{Tsvetkov} {et~al.}(2012){Tsvetkov}, {Volkov}, {Sorokina},
  {Blinnikov}, {Pavlyuk}, \& {Borisov}}]{Tsvetkov12}
{Tsvetkov}, D.~Y., {Volkov}, I.~M., {Sorokina}, E., {et~al.} 2012, Peremennye
  Zvezdy, 32, 6.
\newblock \doarXiv{1207.2241}

\bibitem[{{Utrobin} {et~al.}(2021){Utrobin}, {Wongwathanarat}, {Janka},
  {M{\"u}ller}, {Ertl}, {Menon}, \& {Heger}}]{Utrobin21}
{Utrobin}, V.~P., {Wongwathanarat}, A., {Janka}, H.~T., {et~al.} 2021, \apj,
  914, 4, \dodoi{10.3847/1538-4357/abf4c5}

\bibitem[{{Van Dyk} {et~al.}(2011){Van Dyk}, {Li}, {Cenko}, {Kasliwal},
  {Horesh}, {Ofek}, {Kraus}, {Silverman}, {Arcavi}, {Filippenko}, {Gal-Yam},
  {Quimby}, {Kulkarni}, {Yaron}, \& {Polishook}}]{VanDyk11}
{Van Dyk}, S.~D., {Li}, W., {Cenko}, S.~B., {et~al.} 2011, \apjl, 741, L28,
  \dodoi{10.1088/2041-8205/741/2/L28}

\bibitem[{{Van Dyk} {et~al.}(2014){Van Dyk}, {Zheng}, {Fox}, {Cenko}, {Clubb},
  {Filippenko}, {Foley}, {Miller}, {Smith}, {Kelly}, {Lee}, {Ben-Ami}, \&
  {Gal-Yam}}]{VanDyk14}
{Van Dyk}, S.~D., {Zheng}, W., {Fox}, O.~D., {et~al.} 2014, \aj, 147, 37,
  \dodoi{10.1088/0004-6256/147/2/37}

\bibitem[{{Wongwathanarat} {et~al.}(2017){Wongwathanarat}, {Janka},
  {M{\"u}ller}, {Pllumbi}, \& {Wanajo}}]{Wongwathanarat17}
{Wongwathanarat}, A., {Janka}, H.-T., {M{\"u}ller}, E., {Pllumbi}, E., \&
  {Wanajo}, S. 2017, \apj, 842, 13, \dodoi{10.3847/1538-4357/aa72de}

\bibitem[{{Woosley} {et~al.}(1994){Woosley}, {Eastman}, {Weaver}, \&
  {Pinto}}]{Woosley94}
{Woosley}, S.~E., {Eastman}, R.~G., {Weaver}, T.~A., \& {Pinto}, P.~A. 1994,
  \apj, 429, 300, \dodoi{10.1086/174319}

\bibitem[{{Woosley} \& {Weaver}(1995)}]{Woosley95}
{Woosley}, S.~E., \& {Weaver}, T.~A. 1995, \apjs, 101, 181,
  \dodoi{10.1086/192237}

\bibitem[{{Yoon} \& {Cantiello}(2010)}]{Yoon10}
{Yoon}, S.-C., \& {Cantiello}, M. 2010, \apjl, 717, L62,
  \dodoi{10.1088/2041-8205/717/1/L62}

\bibitem[{{Yoon} {et~al.}(2019){Yoon}, {Chun}, {Tolstov}, {Blinnikov}, \&
  {Dessart}}]{Yoon19}
{Yoon}, S.-C., {Chun}, W., {Tolstov}, A., {Blinnikov}, S., \& {Dessart}, L.
  2019, \apj, 872, 174, \dodoi{10.3847/1538-4357/ab0020}

\bibitem[{{Yoon} {et~al.}(2017){Yoon}, {Dessart}, \& {Clocchiatti}}]{Yoon17}
{Yoon}, S.-C., {Dessart}, L., \& {Clocchiatti}, A. 2017, \apj, 840, 10,
  \dodoi{10.3847/1538-4357/aa6afe}

\end{thebibliography}
\bibliographystyle{aasjournal}

\end{document}